\documentclass[aip,graphicx]{revtex4-1}

\draft 

\usepackage{amsmath, amssymb}

\usepackage[english]{babel}
\usepackage[utf8x]{inputenc}
\usepackage{tikz}

\newcommand{\be}{\begin{equation}}
\newcommand{\ee}{\end{equation}}

\newcounter{unnumber}

\newtheorem{thrm}{Theorem}

\newtheorem{cor}{Corollary}
\newtheorem{prop}{Proposition}

\newtheorem{prf}[unnumber]{Proof}
\newenvironment{pf}{\begin{prf} \rm}{\end{prf}}

 \font\tenscr=rsfs10 scaled1100
\font\sevenscr=rsfs7 
\font\fivescr=rsfs5 
\skewchar\tenscr='177 \skewchar\sevenscr='177
\skewchar\fivescr='177
\newfam\scrfam
\textfont\scrfam=\tenscr \scriptfont\scrfam=\sevenscr
\scriptscriptfont\scrfam=\fivescr

\begin{document}

\title{On the Liouvillian
solutions to the
perturbation equations of the Schwarzschild black hole} 

\author{Evangelos Melas}
\affiliation{}

\date{\today}

\begin{abstract}
It is well known  that the equations governing the evolution of
scalar, electromagnetic and gravitational  perturbations of the
background geometry of a Schwarzschild black hole can be reduced
to a single master equation.
We  use Kovacic's algorithm  to obtain all Liouvillian solutions, i.e., essentially all solutions  in terms of
quadratures, of this master equation.
We show that all solutions in quadratures of this equation
contain a polynomial solution
to an associated ordinary differential equation (ODE).
In the case of the gravitational perturbations only, for a Liouvillian solution
$\chi$ to the master equation,
this ODE
reduces to an
algebraic equation of the first order. In all the
other cases, apart from a few trivial cases which do not give  any Liouvillian solutions,  this
ODE
falls into the confluent Heun class.
In the case of the gravitational perturbations,
for the
Liouvillian solution
$\chi \int \frac {{\rm
d}r_{\!\ast}}{\chi^{2}}$,
we find in ``closed form''  the polynomial
solution P
to the associated confluent Heun ODE.
We prove that
the Liouvillian solution $\chi \int \frac {{\rm
d}r_{\!\ast}}{\chi^{2}}$ is a product of
elementary functions, one of them being the
 polynomial P.
We extend previous results by Hautot and use the extended results we derive
in order
to prove that P
admits a finite expansion in terms of
truncated confluent hypergeometric functions of the first kind. We also prove, by using the
extended results we derive, that P
 admits also a finite expansion in terms of
 associated Laguerre polynomials.
Remarkably both expansions entail not
constant coefficients but appropriate
function coefficients instead.
We highlight the relation of these results
with inspiring new developments.
We prove, save for two unresolved cases,
that the Liouvillian solutions $\chi$ and
$\chi \int \frac {{\rm
d}r_{\!\ast}}{\chi^{2}}$, initially found by Chandrasekhar, are the only
Liouvillian solutions to the
master equation.
We improve previous results
in the literature on this problem and compare
our results with them.
Comments are made for a more efficient
implementation of Kovacic's algorithm 
to any second order ODE with rational function coefficients.
Our results set the stage for deriving similar results in other black hole geometries 4$-$dim and higher.


\end{abstract}


\maketitle 

\section{Introduction}
\label{nnbddfserhnbcv}
\noindent

It is well known (see for instance \cite{Leaver}) that the
equations governing the evolution of external sourceless test
fields (be it scalar, electromagnetic, or a perturbation of the
gravitational field itself)  on the background geometry of a
Schwarzschild black hole can be reduced to a single second order
ordinary differential equation (ODE) with rational function coefficients.
This master equation, equation (\ref{adamsmith}),
assumes formally the ${\rm Schr} \ddot{ {\rm o} } {\rm dinger} $
form, with an effective curvature potential. The functional form
of this potential is such that it is difficult, in general, to
obtain exact solutions to the
master equation.
For that
reason, most of the work done so far has used approximation
methods.

For instance, in the gravitational case, the master
equation is the Regge$-$Wheeler equation
\cite{Regge,Edelstein,Vish} (RWE), and a number of numerical,
semianalytic and analytic approximation techniques have been
applied over the years for calculating frequencies of the
quasinormal modes of Schwarzschild black holes (see for example
\cite{Det,Leaver1,Andersson,Nollert,Schutz,Liu, Onozawa} and
references therein). The quasinormal modes are solutions of the
Regge$-$Wheeler and Zerilli equations \cite{Zerilli,Zerilli1} (both
to be discussed in section 2) that fulfil the boundary conditions:
the waves must be {\it outgoing} at spatial infinity and {\it
ingoing} across the event horizon.

However, Chandrasekhar \cite{Chandr}
 found solutions in terms of quadratures for
algebraically special perturbations of the Schwarzschild, ${\rm
Reissner}\!\!-\!\!{\rm Nordstr} \ddot{{\rm o}} {\rm m} $ and Kerr
black holes. These algebraically special perturbations excite
gravitational waves that are either purely ingoing or purely outgoing.
Such algebraically special perturbations of the
Kerr
space$-$time
were first considered in passing  by Wald
\cite{Wald}, who showed that  necessary and sufficient condition
for the existence of these algebraically special perturbations is the vanishing of the Starobinsky constant,
though the explicit solutions in quadratures of the perturbation
equations were not given by him. Wald's result generalized an earlier one
by Couch \& Newman \cite{Couch} in the context of the Schwarzschild black hole.

The important point to be noted
here
is that these algebraically special perturbations solutions can occur only for
special values of the frequency of the perturbations as first noticed by
Couch \& Newman \cite{Couch} in the context of the Schwarzschild black hole.
The vanishing of the Starobinsky constant, first noticed by Wald \cite{Wald} in the case of the Kerr black hole,
and then studied by Chandrasekhar \cite{Chandr} in the case of Kerr, Reissner$-$$\bf {{\rm Nordstr}\ddot{ {\rm o} } {\rm m} }$,
and Schwarzschild black holes,
is precisely the condition that the algebraically special perturbations occur only for special values of the frequencies in all black hole geometries.
In the
Schwarzschild case, Qi and Schutz \cite{Qi} have shown that the
algebraically special solution is the linear approximation to the
Robinson$-$Trautman solution, when that solution tends to the
Schwarzschild solution.


In this paper we reexamine the issue of obtaining all solutions
in quadratures of the perturbation equations of the Schwarzshild geometry by using
 Kovacic's algorithm \cite{Kova,Duva,Duval1,UW,SU1,SU2,HR}.
 The employment of Kovacic's algorithm to find all possible Liouvillian solutions
of the master equation which governs the evolution of the perturbations of the Schwarzshild geometry was initiated in \cite{M} and the only published work on the subject is,
to the best of our knowledge \cite{C}, by Couch \& Holder (2007; henceforth C.H.).

In our work we address problems
unnoticed in C.H.
and in related work \cite{r1,r2,r3,r4,r5,r6,F1,F2,F3,F4,F5,F6,r8,r9}.
These problems  are described in subsections
\ref{pairs}, \ref{ic}, and \ref{id}.   In a nutshell, these problems are:

\begin{enumerate}

\item{The search for pairs of Liouvillian solutions $\eta$ and  $\eta \int \frac { e^{-\int a}}{\eta ^{2}}$ in Kovacic's algorithm.}
\item{
The expression of the second Liouvillian
solution $\eta \int \frac { e^{-\int a}}{\eta ^{2}}$, initially found by \newline Chandrasekhar \cite{Chandr}, in the case of the gravitational
perturbations of the Schwarzschild geometry,
as a product of elementary functions, one of them being a polynomial $\Pi$, which is a solution to an associated confluent Heun equation.}
\item{The determination of $\Pi$ in ``closed$-$form''.}
\item{The expansion of $\Pi$ in terms of confluent hypergeometric functions of the first kind.}

\item{The expansion of $\Pi$ in terms of associated Laguerre polynomials.}

\end{enumerate}

We comment on problem 1 in subsection \ref{pairs} and solve it in section \ref{gl}. We comment on problems 2 and 3 in subsection \ref{ic} and solve them in section \ref{gl}. We comment on problems 4 and 5  in subsection
\ref{id} and solve them in section \ref{summ2}
(Theorem \ref{Th2} and Theorem \ref{Th3}).









Our results are apt to generalization to other black hole geometries, 4$-$dim and higher \cite{Ida}.


\subsection{Kovacic's algorithm}
\label{kovalg}

Kovacic's algorithm will find all possible ``closed$-$form'' solutions
(i.e., essentially, all solutions in terms of quadratures) of
linear second order homogeneous ODEs with complex rational
function coefficients of the form
\be \label{ant}  y''+ay'+by=0, \ee
where $a \, {\rm and} \, b$ are rational functions of a complex
variable $x$.
The ``closed$-$form'' solution means a
Liouvillian solution, i.e. one that can be expressed in terms of
algebraic functions, exponentials and indefinite integrals.
Let $\eta$ be a (non$-$zero) Liouvillian solution of the
differential equation (\ref{ant}). The method of
reduction of order produces a second Liouvillian solution, namely
$$ \eta \int \frac { e^{-\int a}}{\eta ^{2}}. $$

Kovacic's algorithm reduces the problem of
finding Liouvillian solutions to   equation (\ref{ant}) to the problem of finding
polynomial solutions to  an associated set $\mathcal D$  of linear ODEs
whose coefficients
are rational functions over the field of complex numbers.
The algorithm is so constructed
that if no polynomial solutions exist to any of the differential equations in $ \mathcal D$
then equation (\ref{ant}) has no liouvillian solutions, whereas,
if  a polynomial solution exists to a differential equation in $\mathcal D$ then this polynomial
solution will give rise to  a Liouvillian solution to equation (\ref{ant}), and, different
polynomial solutions to the  differential equations in
$\mathcal D$   give rise to  different Liouvillian solutions to equation (\ref{ant}).

The most difficult aspects of applying Kovacic's algorithm are reduced to the solution of the following three problems:
\begin{enumerate}

\item{In general, it is not true, that if the polynomial solution to one of the differential equations in $\mathcal D$
gives rise to a Liouvillian solution  $\eta$ to
equation (\ref{ant}) then the polynomial solution to another differential equation in $\mathcal D$
will give rise to the second solution
 $\eta \int \frac { e^{-\int a}}{\eta ^{2}}$ of
equation (\ref{ant}), which is also Liouvillian.
This is precisely the case in the problem we are
examining here; it turns out, as we show in subsection \ref{1l}, that in the case
of the gravitational perturbations of the Schwarzschild geometry there exists Liouvillian solution $\eta$, and it takes us considerable effort  in subsection \ref{G7} to prove that
Kovacic's algorithm does also give rise to the second Liouvillian solution
 $\eta \int \frac { e^{-\int a}}{\eta ^{2}}$.}

\item{In general, it is not easy, to find polynomial solutions, if they exist at  all, to the differential equations in
    $\mathcal D$. This difficulty becomes more apparent when the degree of the sought polynomial solution is not fixed but it depends on a parameter  which appears in the coefficients of the differential equation in $\mathcal D$.
    When a differential equation in $\mathcal D$ leads to a 2$-$term recurrence relation for the coefficients of its Frobenius solutions the problem of finding a polynomial solution to it can be solved in a systematic way, either its degree is fixed or depends on a parameter  which appears in the coefficients of the differential equation in $\mathcal D$.
However, the difficulty becomes even more prominent
    when the differential equation in $\mathcal D$ leads to a $d$$-$term recurrence relation, $d=3,4,5,...,$ for the coefficients of its Frobenius solutions, since in general, $d$$-$term  recurrences do not admit ``closed$-$form''
solutions. This is precisely the case in the problem we are examining here; as we show in subsection \ref{G7}, one of the differential equations in $\mathcal D$  associated to equation
(\ref{ant}), in the case of the gravitational
perturbations of the Schwarzschild geometry,
is equation (\ref{73}) which falls into the confluent Heun class, and as such,
leads to a 3$-$term recurrence relation for the coefficients of its Frobenius solutions.
Moreover the degree of the sought polynomial solution  to equation (\ref{73})
is not fixed but it depends on a parameter  which appears in the coefficients of  (\ref{73}). It is the structure of Kovacic’s algorithm in combination with the method
we follow in subsection \ref{G7} which allow us to find in ``closed$-$form'' a polynomial solution to equation (\ref{73}).    }

\item{As we said Kovacic's algorithm reduces the problem of
finding Liouvillian solutions to   equation (\ref{ant}) to the problem of finding
polynomial solutions to  an associated set $\mathcal D$  of linear ODEs. Let $\mathcal M$ be a member of $\mathcal D$. Kovacic's algorithm determines the degree $d$ of a
polynomial solution to $\mathcal M$, if such a
polynomial solution exists. The degree $d$
can be easily determined independently of the determination of $d$ by Kovacic's algorithm.
For each $\mathcal M$ the following problem
must be solved:

\it{Find all polynomial solutions of degree $d$ to $\mathcal M$, or else, prove that $\mathcal M$ does not admit a polynomial solution. } \normalfont

The difficulty of the problem of finding 
all polynomial solutions of degree $d$
to $\mathcal M$
constitutes the second difficult aspect of
applying Kovacic's algorithm. Here we restrict
attention to the problem of proving that
$\mathcal M$ does not admit polynomial solution, when this is the case.
We note that when $\mathcal M$ leads to a 2$-$term recurrence relation for the coefficients of its Frobenius series solutions the problem of proving that the
Frobenius series does not terminate,
when this is the case,
can be solved in a systematic way, either its degree is fixed or it depends on a parameter  which appears in the coefficients of $\mathcal M$. However when $\mathcal M$ leads to a $q$$-$term recurrence relation,  $q=3,4,5,...,$ for the coefficients of its Frobenius series solutions the problem of proving that the
Frobenius series does not terminate,
is by no means a trivial problem and a
variety of approaches can be employed to
solve the problem. These methods, in the case $q=3$, are elaborated in subsection \ref{evidence}.






}

\end{enumerate}

Problems 1 and 2 are respectively problems 1 and 3 in the list of problems in section \ref{nnbddfserhnbcv}. C.H.\cite{C} address and solve partially only  problem 3. We address all problems 1, 2, and 3. We comment on problem 1 in subsection \ref{pairs} and solve it in section \ref{gl}. We comment on problem 2 in subsection \ref{ic} and solve it in section \ref{gl}. We
solve partially problem 3 in sections \ref{gl} and \ref{rem}.

We improve the results derived by
C.H.\cite{C} on the solution of
problem 3. Namely, the results
derived in subsections
\ref{confluent}, \ref{elementarysym},
\ref{sections3},
\ref{evidence},  \ref{reduction}, and
\ref{comments}
are new.
We compare our results on the solution of problem 3 with the results of C.H. on the solution of problem 3 in subsection \ref{comparison}.
This is as far as the application of Kovacic's algorithm is concerned.

In subsection \ref{elementary} we prove that
the second Liouvillian
solution $\eta \int \frac { e^{-\int a}}{\eta ^{2}}$, initially found by Chandrasekhar \cite{Chandr}, in the case of the gravitational
perturbations of the Schwarzschild geometry,
can be expressed as a product of  elementary functions,
one of these elementary functions  being a polynomial solution to a confluent Heun equation.

Moreover, we address
and solve a problem which is not addressed at all by
C.H.\cite{C}; this is the union of problems 4 and 5 in the list of problems in section \ref{nnbddfserhnbcv}.
Motivated by previous work on the confluent
Heun equation and the previous knowledge and experience acquired from
non$-$relativistic quantum mechanics \cite{Lan}, we expect that this polynomial
can be expressed as a linear combination of (confluent) hypergeometric functions.

We comment on this problem in subsection
\ref{id}
 and solve it in section \ref{summ2},
where we prove that the polynomial in question
can be expressed as a sum, with appropriate function coefficients, of truncated confluent hypergeometric functions of the first kind
(Theorem \ref{Th2}). Furthermore we prove that the polynomial in question can be expressed as a sum, with appropriate function coefficients, of associated Laguerre polynomials (Theorem \ref{Th3}).

\subsection{The search for pairs $\eta$ and  $\eta \int \frac { e^{-\int a}}{\eta ^{2}}$}

\label{pairs}

It is not generally true \cite{Duval1} that
if the polynomial solution to one of the differential equations in $\mathcal D$
gives rise to a Liouvillian solution  $\eta$ to
equation (\ref{ant}) then the polynomial solution to another differential equation in $\mathcal D$
will give rise
to $\eta \int \frac { e^{-\int a}}{\eta ^{2}}$.
In fact establishing when this holds, as we said in subsection \ref{kovalg}, is one of the most difficult aspects of applying Kovacic's algorithm
and it can take considerable amount of time and effort to find the pair(s)
of equations in $\mathcal D$, \it{if they exist at all}, \normalfont
which yield the Liouvillian solutions $\eta$ and $\eta \int \frac { e^{-\int a}}{\eta ^{2}}$
to   equation (\ref{ant}).

C.H.  do not even mention the possibility that
Kovacic's algorithm, besides the Liouvillian solution $\eta$  may also gives rise to the Liouvillian solution
$\eta \int \frac { e^{-\int a}}{\eta ^{2}}$.
In fact in section \ref{gl} we prove that this is indeed the  case for the
gravitational perturbations
of the Schwarzschild geometry.
In particular in subsection \ref{1l} we prove that Kovacic's algorithm gives rise to the
algebraically special Liouvillian  perturbation $\eta$ of Schwarzschild black hole, found by Chandrasekhar \cite{Chandr},  which excites gravitational
waves that are either purely ingoing or purely outgoing.

This is the easy part of the proof and has been given in a slightly different form by C.H..
We include it here  because it is short
and for completeness of the presentation.
However, the difficult part of the proof
in section \ref{gl}, not given by C.H., is given in subsection  \ref{G7}.
In subsection  \ref{G7} we prove that Kovacic's algorithm,
in the case of the gravitational perturbations of the Schwarzschild geometry,
gives also rise to the Liouvillian solution $\eta \int \frac { e^{-\int a}}{\eta ^{2}}$.

Both  algebraically special Liouvillian solutions $\eta $ and $\eta \int \frac { e^{-\int a}}{\eta ^{2}}$ to the
RWE equation were found by Chandrasekhar \cite{Chandr}.
However this does not render subsection VB
superfluous. The significance of this subsection does not result from the discovery  of the second Liouvillian solution
$\eta \int \frac { e^{-\int a}}{\eta ^{2}}$ but rather from the proof itself that Kovacic's algorithm gives also rise to the
second Liouvillian solution
$\eta \int \frac { e^{-\int a}}{\eta ^{2}}$ and from other results which are obtained in the course of the proof.

In particular, in subsections \ref{G7},
\ref{elementary},
we prove that $\eta \int \frac { e^{-\int a}}{\eta ^{2}}$ can be expressed via elementary functions,
one of them being a polynomial solution to a  confluent Heun equation.
This result remained unnoticed by
Chandrasekhar \cite{Chandr}.
The proof in subsection \ref{G7} is also
an indispensable part of the proof that there
are no other Liouvillian solutions to the master
equation (\ref{adamsmith}) apart from those
found by Chandrasekhar\cite{Chandr}.

We note in passing that it has been erroneously claimed \cite{Chee} that Chandrasekhar  found \cite{Chandr} only the Liouvillian solution $\eta \int \frac { e^{-\int a}}{\eta ^{2}}$
and that he missed the other Liouvillian solution   $\eta$.
A careful reading of  Chandrasekhar's paper\cite{Chandr} proves that this is false.

Most likely the proof in subsection \ref{G7}
that Kovacic's algorithm
gives also rise to the Liouvillian solution $\eta \int \frac { e^{-\int a}}{\eta ^{2}}$
is going to be useful to future applications of the algorithm.
Indeed,
the
logical structure of the proof does not depend on the particular problem at hand, namely the linear
perturbations of the Schwarzschild geometry, but it is tailored for identifying Liouvillian pairs
$\eta$ and $\eta \int \frac { e^{-\int a}}{\eta ^{2}}$ when the order of one of the polynomial solutions
associated to the  Liouvillian solutions  $\eta$, $\eta \int \frac { e^{-\int a}}{\eta ^{2}}$ is not fixed but it
depends on a parameter which appears in the
coefficients
of the polynomial solution.

This situation appears quite frequently in many problems of mathematical physics and in particular
on the perturbation theory of the black$-$hole geometries in 4$-$dim and higher where this parameter
comes
from the
master equation which governs the evolution of the perturbations.
As the relevant analysis in section \ref{FF} shows
this parameter is
the frequency of the monochromatic waves to which the
perturbation is Fourier decomposed.
Therefore it is highly likely that analogues and/or generalizations of the  proof in subsection \ref{G7} are
going to be useful in the search for
pairs of Liouvillian solutions
$\eta$ and $\eta \int \frac { e^{-\int a}}{\eta ^{2}}$ in subsequent applications of the algorithm;
either these applications refer to black hole geometries, or otherwise.


The absence of search for pairs $\eta$ and $\eta \int \frac { e^{-\int a}}{\eta ^{2}}$,
which is one of the most difficult aspects of
applying Kovacic's algorithm
is not unique with C.H.,
but it is also observed in other applications of Kovacic's algorithm (e.g. \cite{r9}).
The cause of this negligence can  be traced back to the original article by Kovacic
\cite{Kova} in which no mention is made of the efficacy of the algorithm to
capture the second Liouvillian solution $\eta \int \frac { e^{-\int a}}{\eta ^{2}}$ when a Liouvillian solution  $\eta$ is identified.

Subsequent works on  Kovacic's algorithm \cite{Duva,Duval1,UW,SU1,SU2,HR}  remain insistently silent
on this point; the only exception being \cite{Duval1} where on page we 214 we read:
``When one liouvillian solution is found, another one, linearly independent, is $\eta \int \frac { e^{-\int a}}{\eta ^{2}}$.
This second solution is not always detected by the algorithm;
''.
This scarce and inconclusive mention in the literature on the existence of pairs $\eta$ and
$\eta \int \frac { e^{-\int a}}{\eta ^{2}}$ in the Liouvillian solutions found by Kovacic's algorithm
probably explains the lack of search for such pairs in the applications and extensions of the algorithm so far \cite{Kova,Duva,Duval1,UW,SU1,SU2,HR,C,r9}.


\subsection{Liouvillian solutions, confluent
Heun equation and its polynomial solutions}

\label{ic}

  In subsections \ref{1l} and \ref{G7}
we prove that RWE equation,
which governs the gravitational
perturbations of the Schwarzschild
geometry, admits two
 Liouvillian solutions,
$\eta$ and $\eta \int \frac { e^{-\int a}}{\eta ^{2}}$, initially found by Chandrasekhar \cite{Chandr}.

However, Chandrasekhar
did not obtain the
results of subsection \ref{G7} and \ref{elementary}:
As we prove in subsection \ref{G7},
the Liouvillian solution
 $\eta \int \frac { e^{-\int a}}{\eta ^{2}}$, initially found by Chandrasekhar \cite{Chandr},  is the product of  functions (equation (\ref{grasssop})), one of them being a polynomial solution to an associated confluent Heun equation,
 equation (\ref{73});
 equation (\ref{73}) falls into the confluent Heun class after a trivial change of the independent variable, however, this does not  affect in any way the results and conclusions stated here.

An immediate corollary of this, as we show
in subsection \ref{elementary},
is that $\eta \int \frac { e^{-\int a}}{\eta ^{2}}$
is a product of elementary functions.
This   result,
 was  given for the first time by Araujo and MacCallum \cite{Ara} in an unpublished work;  however, these authors do not give the proof, which we give in subsection \ref{G7},  which leads to it.

It is difficult to imagine how we could have obtained the results of subsections \ref{G7}
and \ref{elementary} without applying Kovacic's algorithm to the master
equation (\ref{adamsmith}), RWE equation is a special case of (\ref{adamsmith}):
It is the structure of Kovacic's algorithm
in combination
with the   method we follow in subsection \ref{G7} which allow
to find in ``closed$-$form'' the polynomial factor in $\eta \int \frac { e^{-\int a}}{\eta ^{2}}$.

Remarkably, we obtain in ``closed$-$form'' the polynomial
factor in $\eta \int \frac { e^{-\int a}}{\eta ^{2}}$, despite the fact that the associated confluent Heun equation
(\ref{73}), leads to a 3$-$term recurrence relation for the coefficients of its Frobenius solutions, and in general, 3$-$term recurrences do not admit``closed$-$form''
solutions.


Regarding the associated confluent Heun equation (\ref{73}) we note the following:
The degree of the sought polynomial solution to equation (\ref{73}), if such a polynomial solution exists at all,
is found either
by using Kovacic's algorithm or by a straightforward calculation. The crucial feature
of the sought polynomial solution to
equation (\ref{73})
is that  its
degree is not fixed but it depends on a parameter  which appears in the coefficients of  (\ref{73}).

It is precisely this feature which renders difficult to find polynomial solutions to (\ref{73}) if we
proceed with brutal force with any of the textbook methods, eg., by solving the associated linear homogeneous system,
by considering the  continued fraction associated to the 3$-$term  recurrence relation \cite{Gau} to which the  confluent Heun equation leads,
or by applying the relevant knowledge and experience
from non$-$relativistic quantum mechanics \cite{Lan} in relation with the  hypergeometric function solutions to 3$-$term recurrences arising from the
time$-$independent  Schr\"{o}dinger equation.

The difficulties persist even if we apply not textbook methods but results and methods which have appeared  in the relevant research on the subject, e.g.,
a new method of solving linear differential equations \cite{Gur}, of arbitrary order, and the subsequent technique to find polynomial
solutions to linear differential equations \cite{Gur1}, or, expansion of the solution of (\ref{73}) in terms of confluent hypergeometric functions and the like \cite{Ish,Ish1}.

Since previous experience and knowledge and other results on the subject do not offer any clues
for finding polynomial solutions to equation (\ref{73}), we follow instead another method in subsection \ref{G7}
which allows us to find a polynomial solution to equation (\ref{73}) and at the same time to prove that  Kovacic's
algorithm does give rise to a pair of Liouvillian solutions $\eta$ and  $\eta \int \frac { e^{-\int a}}{\eta ^{2}}$.

The results
 in subsections \ref{1l}, \ref{G7}, and \ref{rem}, show that \it all \normalfont Liouvillian solutions  to the master equation (\ref{adamsmith})
contain a polynomial; in \it all \normalfont cases,
this polynomial
is solution to an associated
confluent Heun differential equation \cite{He,Bat,Dec1,Dec2,He1,Sla,Ma}.
The only  exception is studied
in subsection \ref{1l} and  occurs  in the case
of the gravitational   perturbations of the Schwarzschild geometry. In this case we do find
a Liouvillian solution $\chi$, given in equation (\ref{vannav}), to the master equation (\ref{adamsmith}), and the polynomial which is
contained in  $\chi$  is of first order and it
satisfies an ODE of second order, which reduces to an
algebraic equation of first order for the constant term of the polynomial.

A precursor of this result appeared in \cite{Ga,Su,Ba} where it was shown that the
master equation which governs the evolution of various
perturbations on a class of black hole geometries
falls into the Heun class after a change of the
independent variable and after a subsequent change of the dependent variable.
However a systematic, exhaustive construction of all
Liouvillian solutions of the master equation
(\ref{adamsmith}) which governs the evolution of perturbations of the Schwarzschild geometry
is out of the reach of these studies, this can only be accomplished by using
Kovacic's algorithm.

\subsection{
Liouvillian solutions,
truncated hypergeometric functions, and associated Laguerre polynomials}

\label{id}


The master equation (\ref{adamsmith}),
after the change of the dependent variable
(\ref{pupul}),
becomes a time$-$independent Schr\"{o}dinger   type equation (equation (\ref{kooala})) with an effective curvature potential.
For many potentials, e.g. for
the harmonic oscillator potential,
the Kratzer potential, the Morse potential, the  Bargmann potential, the P\"{o}schl$-$Teller potential, to
mention a few,
time$-$independent Schr\"{o}dinger    equation admits
Liouvillian solutions in terms of truncated (confluent) hypergeometric functions \cite{Lan}.

This suggests, given the general form of the
Liouvillian solution obtained by applying Kovacic's algorithm to the master equation
(\ref{adamsmith}) (see e.g. equation (\ref{grasssop})),
that
there may exist polynomial solutions to the 
confluent Heun equations associated with the master equation
(\ref{adamsmith}) which are expressed in terms of (confluent) hypergeometric functions.
As it is stated in subsections \ref{kovalg}
and \ref{ic}         Kovacic's algorithm
reduces the problem of
finding Liouvillian solutions to   master equation
(\ref{adamsmith}) to the problem of finding
polynomial solutions to  an associated set $\mathcal D$  of confluent Heun equations.



The suggestion
in turn
that there may exist polynomial solutions to the
confluent Heun equations associated with the master equation
(\ref{adamsmith}) which are expressed in terms of (confluent) hypergeometric functions
makes contact with the very extensive
literature  on the Heun equation and its
various confluent forms, and in particular, with
Hautot's results \cite{Hautot,Hautot1,Hautot2,Hautot3},
who proved that when certain sufficient conditions are satisfied, then the confluent
Heun equation admits polynomial solutions
which can be expressed as a finite sum
of truncated confluent hypergeometric
functions of the first kind.

What makes relevant Hautot's work
with our work, is as we show in subsection
\ref{sufficient}, Hautot's sufficient conditions are satisfied
in the case of the algebraically special gravitational
perturbations of the Schwarzschild
geometry, and in particular, are satisfied for the Liouvillian solution
$\eta \int \frac { e^{-\int a}}{\eta ^{2}}$
initially found by Chanadrasekhar
\cite{Chandr}. As we show in subsection
\ref{sufficient}, in this case,
Hautot's sufficient conditions
are reduced to the condition that
a  determinant  of fixed order is zero.

Interestingly  enough,
the vanishing
of this determinant
is equivalent to the vanishing of the Starobinsky constant,
i.e., it is equivalent to the fact that the frequencies of
the algebraically  special gravitational perturbations of the Schwarzschild geometry take the values firstly found by Couch and Newman\cite{Couch}, and then subsequently rediscovered by Wald\cite{Wald} and by  Chandrasekhar\cite{Chandr}.

We note in passing that the importance of Hautot's result lies in an apparently
curious fact:
Confluent Heun equation, an equation with three singular points,
when certain sufficient conditions are satisfied, admits
polynomial solution which can be expressed as a sum of a fixed number,
the number being determined by the confluent Heun equation at hand,
of truncated confluent  hypergeometric functions of the first  kind, which
satisfy the hypergeometric confluent differential equation, an equation
with two  singular points!

So it seems, that when Hautot's sufficient conditions are satisfied, a singularity
disappears! Hautot does not give an explanation of the phenomenon. To the best of our knowledge the
first systematic study and explanation of this phenomenon was given by Craster and Shanin in \cite{Shanin}. Their work is
illuminating since their explanation involves
isomonodromy mappings of the solution space of the
confluent Heun equation. An inspiring  recent study\cite{Gr}
sets the stage for a group theoretical explanation
of this phenomenon with the use of tridiagonalization.

 The results derived in this
paper, and in particular the extension of
Hautot's results we give in subsections \ref{extension}, \ref{truncon} and \ref{Laguerre1}, provide a test bed for
a verification of the results in\cite{Shanin,Gr}, and even more, for extending these results and for unifying them in a coherent framework.
We elaborate more on these issues in section \ref{D}.

Interestingly enough, in the problem under consideration, Hautot's sufficient conditions
are nothing but the requirement that the frequencies of the gravitational perturbations
of the Schwarzschild geometry are the frequencies (equation   (\ref{nutnuta1}))
of the algebraically special perturbations!


However, when we try to form Hautot's
finite sum of truncated confluent of
hypergeometric functions of the first
kind, we immediately encounter an
obstruction, since, as
 we show in subsection
\ref{extension},
two of the four
terms of the sum are not defined
in the case of the  Liouvillian solution
$\eta \int \frac { e^{-\int a}}{\eta ^{2}}$
initially found by Chanadrasekhar
\cite{Chandr}.

So it appears, that oddly enough,
although
in the case of the
the Liouvillian solution
$\eta \int \frac { e^{-\int a}}{\eta ^{2}}$,
initially found by Chanadrasekhar
\cite{Chandr},
Hautot's
sufficient conditions are satisfied, Hautot's  polynomial solution
to the associated confluent Heun equation does not arise in this case.

We resolve this apparent paradox
in subsections
\ref{extension}, \ref{truncon} and \ref{Laguerre1}, where
we introduce an extension of
Hautot's results \cite{Hautot,Hautot1,Hautot2,Hautot3}, and
use this extension
in order
to prove that the confluent Heun
equation, associated with Chandrasekhar's
Liouvillian solution
$\eta \int \frac { e^{-\int a}}{\eta ^{2}}$,
does admit polynomial solution which can be expressed both as a sum of truncated confluent hypergeometric functions of the first kind and as a sum of associated Laguerre polynomials, with appropriate function coefficients in each case
(Theorem \ref{Th2}
and Theorem
\ref{Th3} respectively).

The Euler$-$Gauss   hypergeometric function, the confluent hypergeometric function, and generalizations thereof, play  a prominent role in many branches of
physics and mathematics  \cite{Nik}.
The relation of the Liouvillian solutions of the RWE, and in particular of the Chandrasekhar's algebraic special solution
$\eta \int \frac { e^{-\int a}}{\eta ^{2}}$, to the hypergeometric functions
has not been investigated so far. It is an enigma that even publications devoted to the analytic treatment\cite{r8} of the
Chandrasekhar's algebraic special solution
$\eta \int \frac { e^{-\int a}}{\eta ^{2}}$ do not even address the problem of the relation of
this solution with the hypergeometric functions.



The effective curvature potential (\ref{virginet}) of the RWE  is complicated, so the existence of Liouvillian solutions
in the case of the gravitational perturbations of the Schwarzschild geometry
comes as a surprise, and their expression as a
product of elementary functions,
one of them being a polynomial
which admits an expansion both
in terms of
 truncated confluent
hypergeometric functions of the first
kind (Theorem \ref{Th2}), and
in terms of
associated Laguerre polynomials
(Theorem \ref{Th3}), even more so.
This opens a new chapter in the ever active area of solvability
of Schr\"{o}dinger  potentials \cite{Ish2,Ish2a,Ish3,Der}
and further study  is needed
in order to find out how the results
we obtain in this paper are generalized to other black hole geometries, 4$-$dim and higher \cite{Ida}.



This paper is organised as follows.
In section \ref{FF} some fundamental formulae are presented.
In section \ref{KA}  Kovacic's algorithm is outlined. In section \ref{AKA}
Kovacic's algorithm is applied to the master equation (\ref{adamsmith}) which governs the evolution
of the first$-$order perturbations of the Schwarzschild geometry. In section \ref{gl}
two Liouvillian solutions of the master equation (\ref{adamsmith}) are found in the gravitational case, and it is shown that both of them
can be written as product of elementary
functions.
In section \ref{Hautotre} Hautot's results are summarized
and their relation to the perturbations of the Schwarzschild geometry is given.
In section \ref{summ2} Hautot's results are extended and the extended results are used
in order to express one of the Liouvillian solutions found in section
\ref{gl}
as a product of elementary functions,
one of them being a polynomial which
can be expressed
 as a sum of truncated confluent hypergeometric functions of the first kind (Theorem \ref{Th2}), and also,
it can be expressed as a sum of associated Laguerre polynomials (Theorem \ref{Th3}), with appropriate function coefficients in both cases.
In section \ref{rem}
we prove, save for two
unresolved cases,
that the Liouvillian solutions given, in the case of the gravitational perturbations, in section \ref{gl},
are the only
Liouvillian solutions to the
master equation (\ref{adamsmith}).
We also compare our results with
the results of C.H.\cite{C} and
by using the results derived in this
paper we make suggestions for a more
efficient implementation of
Kovacic's algorithm to any second
order ODE with rational function
coefficients.
Finally in section \ref{D} we state interesting open problems related to the results obtained in this paper and highlight connections of these results with inspiring new developments.



\section{Fundamental Formulae}
\label{FF}
\noindent

It has been found (see, for instance, \cite{Leaver} and references
therein) that the equations describing scalar as well as
electromagnetic or gravitational radiative perturbations on the
static part of the Schwarzschild space$-$time can be reduced to the
master equation that follows, by separating out the dependence on
angles and assuming that the time dependence on the perturbations
is ${\rm exp} \left[-{\rm i} \sigma t \right]$.

This assumption
corresponds to a Fourier analysis of the perturbing field
$\Phi(t,r,\theta,\phi)$:
\be \label{Fourier} \Phi(t,r,\theta,\phi)=\frac {1}{2\pi}
\int_{-\infty}^{\infty} {\rm exp} \left[-{\rm i} \sigma t \right]
\left ( \sum_{l} \frac {1}{r} \psi_{l}(r, \sigma){\rm
Y}_{l0}(\theta,\phi) \right ) {\rm d}\sigma. \ee
$\Phi(t,r,\theta,\phi)$ is such that from it {\it all} components of
the perturbed metric tensor can be reconstructed.

Separation of the dependence on
angles leads (see, for instance, \cite{Leaver}) to the
equation
\be
\label{radiative} \frac {{\rm d}^{2}}{{\rm d}r^{2}_{\! \ast}}
\psi_{l}(r)+ \left [ \sigma^{2}- \left ( 1- \frac{2 M}{r} \right )
\left ( \frac{l(l+1)} {r^{2}} +\frac {2\beta M}{r^{3}} \right)
\right] \psi_{l}(r)=0. \ee \noindent
Equation (\ref{radiative}) is the master equation
which governs the evolution of first order
scalar,
electromagnetic and
gravitational radiative perturbations on the static part of the Schwarzschild space$-$time.
In equation (\ref{radiative}) and throughout in this paper we use geometric unit system in which c=G=1.

${\rm Y}_{lm}$
are the spherical harmonics, $r$ is the Schwarzschild radial
coordinate, $r_{\!\ast} = r+2 M \,{\rm log} \left ( \frac{r}{2 M}-1
\right)$ is the Regge$-$Wheeler ``tortoise coordinate'', $\sigma$ is
the frequency, and $\beta=-3,0,1$ and distinguishes the various
types of external perturbations, gravitational, electromagnetic or
massless scalar fields respectively. When a function ${\rm
f}(r_{\!\ast})$ is written in terms of the Schwarzschild
coordinate r, the inverse function, $r(r_{\!\ast})$, is to be
assumed.

The origins of these equations are reviewed by Leaver
\cite{Leaver}. Let us make a few remarks relevant to our work.
Regge and Wheeler \cite{Regge} showed how odd$-$parity (axial)
gravitational perturbations to the geometry could be expressed in
terms of the solutions of equation (\ref{radiative}) when
$\beta=-3.$ A very similar equation obeyed by even$-$parity (polar)
gravitational perturbations, but with a different potential, was
derived by Zerilli \cite{Zerilli1}. Chandrasekhar
\cite{Chandr1,Chandr2} and Chandrasekhar and Detweiler \cite{Det}
subsequently showed that solutions to Zerilli's even$-$parity
equation could be expressed in terms of the Regge$-$Wheeler
odd$-$parity solutions and vice versa. In particular, the following
relations hold \be \label{diatodas} \left [
\mu^{2}(\mu^{2}+2)+12{\rm i}\sigma M \right ] {\rm Z}^{(+)}= \left [
\mu^{2}(\mu^{2}+2)+72 M^{2} \frac{r-2 M}{r^{2}(\mu^{2}r+6 M)}   \right ]
{\rm Z}^{(-)} +12 M \frac {{\rm d}}{{\rm d}r_{\!\ast}} {\rm Z}^{(-)},
\ee \be \label{diatodas1} \left [ \mu^{2}(\mu^{2}+2)-12{\rm
i}\sigma M \right ] {\rm Z}^{(-)}= \left [ \mu^{2}(\mu^{2}+2)+72 M^{2}
\frac{r-2 M}{r^{2}(\mu^{2}r+6 M)}   \right ] {\rm Z}^{(+)} -12 M \frac
{{\rm d}}{{\rm d}r_{\!\ast}} {\rm Z}^{(+)}, \ee (\cite{Det},
equation (61); see also \cite{Chandr2}, page 162, equations
(152),(153)), where $ \mu^{2}=(l-1)(l+2), \,{\rm Z}^{(+)}$  are
the solutions to Zerilli's even$-$parity equation, and ${\rm
Z}^{(-)}\equiv \psi_{l}(r)$ when $\beta=-3$ and $\frac {{\rm
d}}{{\rm d}r_{\!\ast}}=\frac {r-2 M}{r} \frac {{\rm d}}{{\rm d}r} $.

Therefore (see equation (\ref{diatodas})), a solution  ${\rm
Z}^{(-)}$ to Regge$-$Wheeler's equation will yield a solution  ${\rm
Z}^{(+)}$ to Zerilli's equation and vice versa (see equation
(\ref{diatodas1})). Due to the particular functional dependence of
${\rm Z}^{(-)}$ and  ${\rm Z}^{(+)}$, if ${\rm Z}^{(-)}$ is
expressed in terms of quadratures so is  ${\rm Z}^{(+)}$ (equation
(\ref{diatodas})) and vice versa (equation (\ref{diatodas1})).
Thus it is not necessary to consider Zerilli's equation
separately. The $\beta=0$ equation for electromagnetic
perturbations was first derived by Wheeler \cite{Wheeler}, while
Ruffini, Tiomno and Vishveshwara \cite{Ruffini} showed how the two
equations governing the radial parts of even$-$parity and odd$-$parity
electromagnetic perturbations with a source are reduced to a
single differential equation,namely, Wheeler's equation, when the
source is switched off.

The lowest multipoles which can radiate are the gravitational
quadrupole, the electromagnetic dipole, and the scalar monopole
\cite{Price,Bardeen}. The master equation (\ref{radiative}) is not
applicable for non$-$radiative multipoles, since the field equations
are not the same in this case
\cite{Regge,Edelstein,Vish,Zerilli,Wheeler,Ruffini,Price}. In the
gravitational case, exact solutions both of the odd$-$parity and
even$-$parity perturbation equations when $l=0,1$ can be given
\cite{Zerilli}. In the electromagnetic case and when $l=0$, the
smooth allowed transition from a Schwarzschild to a
Reisner$-$Nordstrom geometry, which takes place through the capture
of a charge particle in a given Schwarzschild background, has been
studied by Hanni and Ruffini \cite{Hanni}. Thus, in our
investigation, the angular harmonic index $l$ will take the values
2,3,... in the gravitational case, 1,2,... in the electromagnetic
case, and 0,1,2,... in the scalar case.

For the purposes of our study we transform
the master equation
(\ref{radiative}) into the standard ${\rm Schr} \ddot{ {\rm o} }
{\rm dinger} $ form, with the Schwarzschild coordinate r as the
independent variable. To accomplish this, first we change
the independent variable from
$r_{\! \ast}$ to $r$, and then we eliminate the
 term
containing the first derivative
by a standard transformation
of the dependent variable. The details are as follows.

By changing the independent variable from
$r_{\! \ast}$ to $r$, the master equation (\ref{radiative}) is
rewritten as follows
\be \label{adamsmith}
\frac {{\rm
d}^{2}}{{\rm d}r^{2}} \psi_{l}(r)+ \frac{2 M}{r(r-2 M)}\frac{{\rm
d}}{{\rm d}r} \psi_{l}(r)- \left[\frac
{s^{2}r^{4}}{4r^{2}(r-2 M)^{2}} + \frac {l(l+1)}{r(r-2 M)}+ \frac
{2\beta M}{r^{2}(r-2 M)} \right] \psi_{l}(r)=0, \ee
where
\be
\hspace{-1cm} s^{2}=-4\sigma^{2} \ee will prove a convenient substitution later
on. When $\beta=-3$, equation (\ref{adamsmith}),
or equivalently equation (\ref{radiative}),
is known as the Regge$-$Wheeler equation \cite{Regge}
(equation (\ref{adamsmith}) is identical to equation
(23), page 144 \cite{Chandr2})). Equation (\ref{adamsmith})  does not change if $s$ is
replaced with $-$$s$. Thus, in case of real $s$, only
non$-$negative values of  $s$ need to be considered.

The term
containing the first derivative of the dependent variable in
(\ref{adamsmith}) can be eliminated by a standard transformation
of the dependent variable. To obtain the ${\rm Schr} \ddot{ {\rm
o} } {\rm dinger}-{\rm like}$ diffential equation in r, we perform
this transformation  in equation (\ref{adamsmith}) \be
\label{pupul} \psi'_{l}(r)= \left ( 1- \frac{2 M}{r} \right
)^{\frac{1}{2}} \psi_{l}(r) \ee The result is \be \label{kooala}
\frac {{\rm d}^{2}}{{\rm d}r^{2}} \psi'_{l}+ {\rm
R}_{l}(r)\psi'_{l}=0 \ee where \be \label{virginet} {\rm
R}_{l}(r)= \frac {r^{2}}{(r-2 M)^{2}} \left [ -\frac {s^{2}}{4}-{\rm
V} _{l}(r)+\frac{2 M}{r^{3}}-\frac {3 M^{2}}{r^{4}} \right ] \ee and \be
{\rm V}_{l}(r)= \left ( 1-\frac{2 M}{r} \right ) \left [ \frac
{l(l+1)}{r^{2}} + \frac {2 \beta M}{r^{3}} \right ]. \ee By using
Kovacic's algorithm on the equation (\ref{kooala}), we will
investigate all its solutions in terms of quadratures, and hence,
due to (\ref{pupul}), all the Liouvillian solutions of the master
equation (\ref{adamsmith}).

For convenience, we rewrite equation (\ref{kooala}) in the form
\be
\mathcal O_{(r,s,M)} \psi'_{l}=0,
\ee
where $\mathcal O_{(r,s,M)}$ is the linear operator
\be
\mathcal O_{(r,s,M)}=\frac{{\rm d}^{2}}{{\rm d}r^{2}}+{\rm
R}_{l}(r).
\ee
The subscript $(r,s,M)$ in $\mathcal O_{(r,s,M)}$
emphasizes the dependence of $\mathcal O_{(r,s,M)}$
on the parameters $r,s$ and $M$.

Regarding  the parameter $M$, the mass of the black hole,
which appears in equation (\ref{kooala}) we note the
following: The transformation
\be
\label{tran}
r = \rho k M,   \quad
s = \frac{\tau}{k M},
\ee
where $k$ is a positive integer, leaves equation
(\ref{kooala}) quasi$-$invariant, it replaces
the parameter $M$ with $\frac{1}{k}.$
Thus we have
\be
\label{equivalence}
\mathcal O_{(r,s,M)} \psi'_{l}=0 \Leftrightarrow \mathcal O_{(\rho,\tau,\frac{1}{k})} \psi'_{l}=0.
\ee

We conclude that by making use of the transformation
(\ref{tran}) we can set $M$, without loss of generality, to be any  positive number.
For simplicity, henceforth, we choose $k$ in (\ref{tran}) to be equal to one, i.e., for simplicity, throughout in this paper, we set $M=1$. The same remark, regarding the
parameter $M$, applies also to equations (\ref{radiative}) and (\ref{adamsmith}). The transformation (\ref{tran}) and the subsequent equivalence (\ref{equivalence}) are
of crucial importance in subsections \ref{comparison} and \ref{comments}.
In subsection \ref{comparison}
we compare our results with the results of C.H.\cite{C} on the solution of  problem 3, mentioned in subsection \ref{kovalg}.
In subsection \ref{comments} we make
some comments on the implementation of
Kovacic's algorithm.

In section \ref{KA} we give Kovacic's algorithm in some detail and in section \ref{AKA} we give in detail
the application of Kovacic's algorithm. This is
useful and necessary for a number of reasons:

\begin{enumerate}
\item{It facilitates the clear and detailed exposition of the results presented in this paper.}

\item{It allows and makes easier the comparison of our results with the results of other researchers on the problem
of finding all  Liouvillian solutions to the perturbation equations of the Schwarzschild geometry which we examine in this paper.}

\item{It facilitates  the comparison of our results
    with other results on the problem of finding all  Liouvillian solutions to the perturbation equations of other black hole geometries 4$-$dim and higher \cite{Ida}.}

\end{enumerate}

Regarding the second reason we note that
C.H.\cite{C} present the version of Kovacic's algorithm they use, its application, and their results
 on the problem
of finding all  Liouvillian solutions to the perturbation equations of the Schwarzschild geometry
in a succinct form. This makes difficult to trace the origin of some differences in our and their results
on the solution of  problem 3, mentioned in subsection \ref{kovalg}. These differences are stated in subsection
\ref{comparison}. We have checked our calculations several times and we have found them correct. On the other hand the
succinct form of  presentation of the results in
\cite{C} has prohibited us from
repeating all their calculations and so the origin of the differences remains obscure.

\section{Kovacic's algorithm}
\label{KA}
\noindent

Kovacic's algorithm \cite{Kova} finds a ``closed$-$form'' solution
of the differential equation (\ref{ant}) $$  y''+ay'+by=0 $$
where $a \, {\rm and} \, b$ are rational functions of a complex
variable $x$, provided a ``closed$-$form'' solution exists. The
algorithm is so arranged that if no solution is found, then no
solution can exist. The ``closed$-$form'' solution means a
Liouvillian solution, i.e. one that can be expressed in terms of
algebraic functions, exponentials and indefinite integrals. (As
functions of a complex variable are considered, trigonometric
functions need not be mentioned explicitly, as they can be written
in terms of exponentials. Logarithms are indefinite integrals and
hence are allowed). A more precise definition involves the notion
of Liouvillian field \cite{Kova}.

Let $\eta$ be a (non$-$zero) Liouvillian solution of the
differential equation (\ref{ant}). It follows that every solution
of this differential equation is Liouvillian. Indeed the method of
reduction of order produces a second solution, namely \be
\label{kritskrits} \eta \int \frac { e^{-\int a}}{\eta ^{2}}. \ee
This second solution is evidently Liouvillian and the two
solutions are linearly independent. Thus any solution, being a
linear combination of these two, is Liouvillian.

A well$-$known
change of dependent variable may be used to eliminate the term
involving $y'$ from the differential equation (\ref {ant}). Let
\be \label{ursuz1} z=e^{\frac {1}{2} \int a} y \ee (this is
nothing else but the transformation (\ref {pupul})). Then \be
\label{ursuz} z''+ \left ( b- \frac{1}{4} a^{2}- \frac {1}{2}a'
\right )z=0 \ee (thus  ${\rm R}_{l}(r)$  in equation (\ref
{virginet}) equals $ b- \frac{1}{4} a^{2}- \frac {1}{2}a'$ where
$a=\frac{2}{r(r-2)}$ and $
\;\;\;\;\;\;\;\;\;\;
b=-\left [ \frac
{s^{2}r^{4}}{4r^{2}(r-2)^{2}} + \frac {l(l+1)}{r(r-2)}+ \frac
{2\beta}{r^{2}(r-2)} \right ] $ ). ~Equation (\ref{ursuz}) still
has rational function coefficients and evidently (see equation
(\ref {ursuz1})) $y$ is Liouvillian if and only if $z$ is
Liouvillian. Thus no generality is lost by assuming that the term
involving $y'$ is missing from the differential equation
(\ref{ant}).

Before giving the main result obtained by Kovacic
\cite{Kova} we first introduce some notation and some terminology.
$C$ denotes the complex numbers and $C(x)$ the rational functions
over $C$. An algebraic function $\omega$ of $x$ is of degree k,
where k is a positive integer, when $\omega$ solves an irreducible
algebraic equation \be \label{doladola} \Pi(\omega,x)=\sum_{{\rm
i}=0}^{{\rm k}} \frac{{\rm P}_{{\rm i}}(x)} {({\rm k}-{\rm i})!}
\omega^{{\rm i}}=0 \ee where ${\rm P}_{{\rm i}}(x)$ are rational
functions of $x$. Let $\nu \in C(x)$ (to avoid triviality, $\nu
\in \!\!\!\!\!| \,\,C$).

Then the following holds\cite{Kova}
\begin{thrm}
\label{kovacic}
Equation \be \label{katrouba} y''=\nu y, \, \, \, ,\, \, \, \,
\nu \in C(x) \ee has a Liouvillian solution if and only if it has
a solution of the form \be y=e^{\int \omega {\rm d}x} \ee where
$\omega$ is an algebraic function of x of degree 1,2,4,6 or 12.
\end{thrm}

The search of Kovacic's algorithm for $\Pi(\omega,x)$ is based on
the knowledge of the poles of $\nu$ and consists in constructing
and testing a finite number of possible candidates for
$\Pi(\omega,x)$. If no $\Pi(\omega,x)$ is found then the
differential equation (\ref{katrouba}) has no Liouvillian
solutions. If such a
 $\Pi(\omega,x)$ is found and $\omega$ is a solution of the equation
(\ref{doladola}) then the function $\eta=e^{\int\omega{\rm d}x}$
is a Liouvillian solution of (\ref {katrouba}).

If \be
\label{katrouba1} \nu(x)=\frac{s(x)}{t(x)}, \ee
 with $ s,t \in C[x]$,
($C[x]$ denotes the polynomials over C), relatively prime, then
the poles of $\nu$ are the zeros of $t(x)$ and the order of the
pole is the multiplicity of the zero of $t$. The order of $\nu$ at
$\infty$ , ${\rm o}(\infty)$, is defined as \be {\rm
o}(\infty)={\rm max}(0, 4+{\rm d}^{{\rm o}}s- {\rm d}^{{\rm o}}t)
\ee where ${\rm d}^{{\rm o}}s$ and  ${\rm d}^{{\rm o}}t$ denote
the leading degree of $s$ and $t$ respectively (for a
justification of this definition see \cite{Duva} page 10; Kovacic
originally gave a different definition, see \cite{Kova} page 8).
We give now an outline of Kovacic's algorithm. This is the outline
of an improved version of the algorithm given by Duval and
Loday$-$Richaud \cite{Duva}.




\subsection{The algorithm}

\it{Notations.} \normalfont Let ${\rm
L}_{\rm{max}}=\{1,2,4,6,12\}$ and let $h$ be the function defined
on ${\rm L}_{\rm{max}}$ by \be
h(1)=1\,\,,\,\,h(2)=4,\,\,\,\,h(4)=h(6)=h(12)=12. \ee

\noindent \bf{Input:} \normalfont A rational function
$$ \nu(x)=\frac{s(x)}{t(x)} \qquad ({\rm equation} \\ (\ref{katrouba1})). $$
\noindent The polynomials $ s,t \in C[x]$ are supposed to be
relatively prime. \newline The differential equation under consideration is
$$ y''- \nu y= 0 \qquad ({\rm equation} \\ (\ref{katrouba})).  $$

\noindent \bf {First step:} \normalfont { \it The set L of
possible degrees of $ \omega .$ } \normalfont

\noindent We are interested in equation (\ref{katrouba}) where
$\nu(x)$ is given by (\ref{katrouba1}).

\noindent \textbf{1a}. \hspace{0.5 cm} If $t(x)=1$, set $m=0$.
Otherwise, factorize $t(x)$ \be
t=t_{1}t_{2}^{2}t_{3}^{3}...t_{m}^{m} \ee where the $t_{{\rm
i}}\,,\,{\rm i}=1,2,...,m,$ \,are relatively prime two by two and
each $t_{{\rm i}}$ either is equal to one or has simple zeros. Let
\be \Gamma'=\{ c\in C\,,\,t(c)=0 \} \,\,\,\, {\rm and} \,\,\,\,
\Gamma=\Gamma'\cup \{\infty\}, \ee where $\cup$ denotes
set$-$theoretic union. Associate orders with the elements of
$\Gamma$: \be {\rm o}(c)={\rm i} \ee for all $c\in \Gamma'$ ,where
i is such that $t_{{\rm i}}(c)=0$, and
$$
{\rm o}(\infty)={\rm max}(0, 4+{\rm d}^{{\rm o}}s- {\rm d}^{{\rm
o}}t).
$$
Let \be m^{+}={\rm max}(m,{\rm o}(\infty)). \ee For \,\,
$0\leq{\rm i}\leq m^{+}$ \,\, let \be \Gamma_{{\rm i}}=\{ c \in
\Gamma|\,{\rm o}(c)={\rm i} \}. \ee
\noindent \textbf{1b}. \hspace{0.5 cm} If \, $m^{+}\geq 2$ \,
define \,$\gamma_{2}$\, and \,$\gamma$ \,by \be \gamma_{2}=\left
|\Gamma_{2} \right|\,\,{\rm and}\,\, \gamma=\gamma_{2}+ \left |
\cup \Gamma_{\rm k} \right |, \,\, {\rm k}\,\,{\rm is}\,\,{\rm
odd},\,\,{\rm and}\,\, 3\leq{\rm k}\leq{\rm m}^{+} \ee
respectively, where, if S is a set then $|{\rm S}|$ denotes the
number of elements of S.

 $\!\!\!\!\!\!\!\!\!$ \textbf{1c}. \hspace{0.5 cm} For each $c\in
\Gamma_{2}$, if $c\neq \infty$, calculate the number ${\rm a}_{c}$
so that \be \label{droud} \nu(x)=\frac{{\rm
a}_{c}}{(x-c)^{2}}+{\rm O}\left ( \frac{1}{x-c} \right), \ee where
the O symbol has its usual meaning, $ f(x)={\rm O}(g(x))$ if
$|f(x)|\leq{\rm M}g(x)$ for some positive constant M; $g(x)$ is
assumed to be positive. The coefficient  ${{\rm a}_{c}}$ in
(\ref{droud}) is the coefficient of $1/(x-c)^{2}$ in the Laurent
series expansion of $\nu(x)$ about $c$.

 $\!\!\!\!\!\!\!\!\!$ \textbf{1d}. \hspace{0.5 cm} Construct L, a
subset of
${\rm L}_{\rm{max}},$
as follows
\be
1\in {\rm L}\Longleftrightarrow \gamma=\gamma_{2}, \ee \be 2\in
{\rm L}\Longleftrightarrow \gamma\geq2\,\,,\,\,{\rm and}, \ee \be
4,6\,\,{\rm and}\,\,12 \in {\rm L}\Longleftrightarrow m^{+}\leq2.
\ee
\textbf{1e}. \hspace{0.5 cm} If \,${\rm L}=\emptyset$ \,go to the
stage of the algorithm END. Otherwise, let n  be equal to the
smallest element of L.

\noindent  \bf {Second step:}  \normalfont { \it
The  sets ${\rm E}_{c}$ associated to the singular points.}
\normalfont

\noindent  Construction of the sets ${\rm E}_{c}$, \,$c \in
\Gamma$.

$\!\!\!\!\!$\textbf{2a}. \hspace{0.5 cm} When n=1, for each
$c\in\Gamma_{2}$, define \,${\rm E}_{c}$ as follows \be
\label{ladiladi} {\rm E}_{c}=\left\{ \frac{1}{2} \left ( 1
\underline{+} \sqrt{1+4{\rm a}_{c}} \right ) \right\}. \ee
$\;$ \textbf{2b}. \hspace{0.5 cm} When ${\rm n}\geq 2$, for each
$c\in \Gamma_{2}$, define ${\rm E}_{c}$ as follows \be
\label{gyretaate} {\rm E}_{\rm c}=\left\{ \frac{h({\rm n})}{2}
\left(1 - \sqrt{1+4{\rm a}_{c}}\right )+ \frac{h({\rm n})}{\rm
n}{\rm k}\sqrt{1+4{\rm a}_{c}} | {\rm k}=0,1,...,{\rm n}
\right\}\cap Z, \ee where $\cap$ denotes set-theoretic
intersection and $Z$ denotes the set of integers.

$\!\!\!\!\!\!$ \textbf{2c}. \hspace{0.5 cm} When
n=1, for each $c \in \Gamma_{2{\rm q}}$ with ${\rm q} \geq 2$,
calculate one of the two ``square roots'' $ \left [ \sqrt \nu
\,\right ] _c$ of $\nu$ defined as follows:

\noindent If $c=\infty$, \be \label{MilkyWay} \left [ \sqrt \nu
\,\right ] _{\infty}={\rm a}_{\infty} \  x^{{\rm q}-2} +
\sum_{{\rm i}=0}^{{\rm q}-3}\delta_{{\rm i},\infty} \ x^{{\rm i}},
\ee
and \be \label{MilkyWay1} \nu-\left [ \sqrt \nu \,\right ]
_{\infty}^{2} =-{\rm \beta}_{\infty} \  x^{{\rm q}-3} + {\rm O}
\left (  x^{{\rm q}-4} \right ).\ee

\noindent Let \be \label{sauvage} {\rm E}_{\infty}=\left \{ \frac
{1}{2} \left ( {\rm q}+\epsilon \frac{\beta_{\infty}}{{\rm
a}_{\infty}} \right ) | \ \epsilon= \underline{+} 1 \right \}. \ee

\noindent Define a function ``sign'' S with domain ${\rm
E}_{\infty}$ as follows \be \label{blooo} {\rm S} \left (
\frac{1}{2} \left ( {\rm q}+ \epsilon \frac{\beta_{\infty}}{{\rm
a}_{\infty}} \right ) \right )=
 \left \{ \begin{array}{lr}
\epsilon & {\rm if} \ \beta_{\infty} \neq 0 \\
1 & {\rm otherwise}
\end{array}
\right . . \ee

$\!\!\!\!\!\!\!\!\!\!\!\!$ \textbf{2d}. \hspace{0.5 cm} When n=2,
for each $c \in \Gamma_{\rm q}$ with ${\rm q} \geq 3$, define
${\rm E}_{c}$ as follows \be \label{nugehamm} {\rm E}_{c}=\{ {\rm
q} \}. \ee In equation (\ref{MilkyWay}) $\left [ \sqrt \nu \,
\right ] _{\infty}$ is the sum of terms involving $x^{{\rm i}}$
for $0 \leq {\rm i} \leq {\rm q}-2$ in the Laurent series for
$\sqrt \nu$ at $\infty$. In practice, one would not form the
Laurent series for $\sqrt \nu$, but rather would determine $ \left
[ \sqrt \nu \right ]_{\infty}$ by using undetermined coefficients,
i.e. by equating $\left ( \left [ \sqrt \nu \right ]_{\infty}
\right )^{2}= \left ( {\rm a}_{\infty} x^{{\rm q }-2} +
\delta_{\nu-3, \infty} x^{{\rm q}-3} + ... + \delta _{0,\infty}
\right )^{2}$ with the corresponding part of the Laurent series
expansion of $\nu$ at $\infty$. There are two possibilities for $
\left [ \sqrt \nu \right ]_{\infty}$, one being the negative of
the other, and any of these may be chosen. In equation
(\ref{MilkyWay1}), $\nu$ denotes its Laurent series expansion at
$\infty$. This equation defines $\beta_{\infty}$.

\noindent \bf {Third step:} \normalfont {\it Possible degrees for
P and possible values for $\theta$.} \normalfont

\noindent  $\!$ \textbf{3a}. \hspace{0.5 cm} For each family
$\underline{{\rm e}}= \left ({\rm e}_{c} \right ) _{c \in \Gamma}$
of elements ${\rm e}_{c} \in {\rm E}_{c}$ calculate \be {\rm d}
\left ( \underline{{\rm e}} \right )= {\rm n}- \frac{{\rm n}}{h
({\rm n} )} \sum_{c \in \Gamma} {\rm e}_{c} \ee \noindent
$\!$ \textbf{3b}. \hspace{0.5 cm} Retain the families
$\underline{{\rm e}}$ which are such that
$$
{\rm i})\,\,\, {\rm d}(\underline{{\rm e}}) \in N \,\, {\rm where}
\, \,N \, \, {\rm \, \, is \, \, the \, \, set \, \, of \, \,
non-negative \, \, integers \, \, and}
$$
$$
\;\;\;\;\;{\rm ii})\,\,\,{\rm when}\,\,{\rm n}=2\,\,{\rm
or}\,\,4\,\, {\rm at\,\, least \,\,two\,\,{\rm
e}_{c}\,\,are\,\,odd\,\,numbers\,\,in\,\,each\,\,family}.
$$
If none of the families $\underline{{\rm e}}$ is retained, go to
the stage of the algorithm $\underline{{\rm CONTINUATION}}$.

\noindent
$\!$ \textbf{3c}. \hspace{0.6 cm} For each family $\underline{{\rm
e}}$ retained from step \textbf{3b}, form the rational function
\be \label{krookraa} \theta=\frac{{\rm n}}{h({\rm n})}\sum_{c\in
\Gamma'} \frac{{\rm e}_{c}}{x-c} \,+ \, \delta_{{\rm n}}^{1}
\!\!\!\!\!\!\!\!\! \sum_{\begin{array}{c}
c \in \cup \Gamma_{2{\rm q}} \\
{\rm q} \geq 2
\end{array}}
\!\!\!\!\!\!\!\!\!{\rm S} \left ({\rm e}_{c} \right ) \left [
\sqrt r \right ]_{c}, \ee where $\delta_{{\rm n}}^{1}$ is the
Kronecker symbol.

\noindent  {\bf Fourth Step:} \normalfont {\it Tentative
computation of P.} \normalfont


\noindent Search for a polynomial P of degree d (as defined in
step \textbf{3a}) such that \be \label{karokaro}
\begin{array}{l}
{\rm P}_{{\rm n}}=-{\rm P} \\
\cdot \, \cdot \, \cdot \\
{\rm P}_{{\rm i}-1}=-{\rm P}^{'}_{{\rm i}} -\theta{\rm P}_{{\rm
i}}-
({\rm n}-{\rm i})({\rm i}+1)\nu{\rm P}_{{\rm i}+1} \\
\cdot \, \cdot \, \cdot \\
{\rm P}_{-1}=0,
\end{array}
\ee where the prime in ${\rm P}^{'}(x)$ denotes differentiation
with respect to the independent variable $x$.

\noindent \bf {Output:} \normalfont

\noindent
OUTPUT1: If such a polynomial is found and $\omega$ is a solution
of the irreducible algebraic  equation
$$
\sum_{{\rm i}=0}^{{\rm n}} \frac{{\rm P}_{{\rm i}}(x)} {({\rm
n}-{\rm i})!} \omega^{{\rm i}}=0 \,\,\,\,\,\,\,\,\,\, \left ( {\rm
equation} \,\,\,\,(\ref{doladola}) \right ),
$$
where the rational functions  ${\rm P}_{{\rm i}}(x)$ are defined
in (\ref{karokaro}), then the function $\eta=e^{\int \omega }$ is
a Liouvillian solution of the equation under consideration
$$y''=\nu y  \,\,\,\,\,\,\,\,\,\, ({\rm equation}(\ref{katrouba})).$$
If no such polynomial is found for any family retained from step
\textbf{3b}, go to the stage of algorithm $\underline{{\rm
CONTINUATION}}.$

\noindent $\underline{{\rm CONTINUATION}}:$\,\, If ${\rm n}$ is
different from ${\rm the } \,\, {\rm largest}\,\, {\rm
element}\,\,{\rm of}\,\,{\rm L}$ then set n equal to the next (in
increasing order) element of L and go to \textbf{Step 2}.


\noindent  OUTPUT2: The equation $y''=\nu y$ has no Liouvillian
solutions.

\indent We sketched Kovacic's algorithm in the form needed in
the subsequent section.
Now, besides
the original formulation of this algorithm \cite{Kova} we have its several versions,
improvements \cite{Duva,Duval1,UW},
and extensions to higher order equations \cite{SU1,SU2,HR}.
For example,
the formulation of
Kovacic's algorithm given
in \cite{UW} is
different both from its
original formulation
\cite{Kova} and
from the formulation of Kovacic's algorithm
given in this paper.
It seems that the formulation in \cite{UW}
is much more convenient for computer implementation and in fact
has been implemented in Maple.

However, it seems that for differential equations with simple structure of singularities,
and
depending on
parameters,
the
formulation of the algorithm given in this paper, which is an outline of the formulation given in \cite{Duva}, is well suited.
Moreover, the original formulation of the algorithm \cite{Kova} consists in fact of three separated algorithms
each of them repeating similar steps. In \cite{Duva} one can find a
modification of the original formulation unifying and improving these three algorithms in one. This
form is very convenient for applications and it is the one employed here.

\section{Application of Kovacic's algorithm}
\label{AKA}
 \noindent \bf {Input:} \normalfont Equation (\ref{kooala}) is
the same as equation (\ref{katrouba}) with

\begin{eqnarray}
\label{gorgor} \nu & = &  -{\rm R}_{l}(r)  =
\frac{r^{2}}{(r-2)^{2}} \left [ \frac {s^{2}}{4} + \left (
1-\frac{2}{r} \right ) \left ( \frac{l(l+1)}{r^{2}} + \frac {2
\beta}{r^{3}} \right )
-\frac {2}{r^{3}} + \frac{3}{r^{4}} \right ]   \nonumber \\
& &   \nonumber \\
&  = &  \frac { \frac{s^{2}}{4}r^{4} + l(l+1)r^{2} +
2[\beta-l(l+1)-1]r+3-4\beta}{r^{2}(r-2)^{2}}
  \equiv  \frac {s(r)}{t(r)}.
\end{eqnarray}

\noindent The partial fraction expansion for $\nu$ is the
following
\begin{eqnarray}
\label{poustpoust1} \nu(r) & = & \frac{s^{2}}{4} +
\frac{1}{4}(3-4\beta)\frac{1}{r^{2}} +
\frac{\frac{1}{4}(-2\beta-2l(l+1)+1)}{r} \nonumber \\
& & \nonumber \\
& & + \; \frac{ \frac{1}{4}(4s^{2}-1)}{(r-2)^{2}}+
\frac{\frac{1}{4}(4s^{2}+2l(l+1)+2\beta-1)}{r-2}.
\end{eqnarray}

\noindent We now apply Kovacic's algorithm to equation
(\ref{kooala}):


\noindent
\bf{First step:} \normalfont


\noindent $\textbf{1a.}$ \hspace{0.1cm} ${\rm From \: \;
(\ref{gorgor})  \; \;we  \;  \; obtain}$
\begin{eqnarray}
\label{poustpoust} t(r)=r^{2}(r-2)^{2}.\;{\rm Hence}\; m=2 \; {\rm
and} \; &
\Gamma'=\{0,2\}, & \Gamma=\{0,2,\infty\} \nonumber \\
{\rm o}(0)= & {\rm o}(2)= & 2 \nonumber \\
{\rm d}^{\rm o}s={\rm d}^{\rm o}t=4, & {\rm o}(\infty)=4 \; \;{\rm
and} &
m^{+}=4 \nonumber \\
\label{kolossss} \Gamma_{\rm o}=\Gamma_{1}=\emptyset, &
\Gamma_{2}=\{0,2\}, \;   \Gamma_{3}=\emptyset \; {\rm and} \; &
\Gamma_{4}=\{\infty\}.
\end{eqnarray}


\noindent $\textbf{1b.}$ \hspace{0.5cm} ${\rm Equations  \: \;
(\ref{poustpoust})  \; \;give  }$ \be \label{holoholo}
\gamma_{2}=2 \; \; \; \; \; {\rm and} \; \; \; \; \;
\gamma=\gamma_{2}=2. \ee


\noindent $\textbf{1c.}$ \hspace{0.5cm} ${\rm From   \; \;
(\ref{poustpoust1})  \; \;we \; \; obtain  }$ \be
\label{koloskalos} {\rm a}_{\rm
o}=\frac{1}{4}(3-4\beta)\;\;\;\;\;{\rm and} \;\;\;\;\;{\rm
a}_{2}=\frac{1}{4}(4s^{2}-1). \ee


\noindent $\textbf{1d.}$ \hspace{0.5cm} ${\rm Equations    \; \;
(\ref{holoholo})  \; \; imply    }$ \be {\rm L}=\{1,2\}. \ee

\noindent $\textbf{1e.}$ \hspace{0.5cm} \be \label{toratora} {\rm
n}=1. \ee


\noindent
$\!\!\!$ \bf  {Second step:} \normalfont


\noindent {\bfseries 2a.} \hspace{0.5cm} \normalfont We are
considering the case n=1 .
The roots $0,2  \in
 \Gamma_{2} $
and therefore by using equations
(\ref{ladiladi}),(\ref{koloskalos}) one finds that \be
\label{katra1} {\rm E}_{0}=\left \{ \frac{1}{2}(1 \underline{+}
\sqrt{1+4{\rm a}_{\rm o}} ) \right \}=\left \{ \frac{1}{2}
\underline{+} \sqrt{1-\beta} \right \} . \ee Equation
(\ref{katra1}) implies
\begin{eqnarray}
\label{katra2} {\rm E}_{0}=\left \{ \frac{5}{2},-\frac{3}{2}
\right \} \; &  {\rm when} & \;
 \beta=-3, \\
& & \nonumber \\
\label{katra3} {\rm E}_{0}=\left \{ \frac{3}{2},-\frac{1}{2}
\right \} \; &  {\rm when} & \;
 \beta=0, \\
& & \nonumber \\
\label{katra4} {\rm E}_{0}=\left \{ \frac{1}{2} \right \} \; &  {\rm
when} & \;
 \beta=1.
\end{eqnarray}
Combining equations (\ref{ladiladi}) and (\ref{koloskalos}) yields
\be \label{belbel} {\rm E}_{2}=\left \{ \frac {1}{2} \left ( 1
\underline{+} \sqrt{1+4{\rm a}_{2}} \right ) \right \} = \left \{
\frac{1}{2}+s,\frac{1}{2}-s \right \}. \ee


\noindent {\bfseries 2b.} \hspace{0.5cm} \normalfont It is not
applicable  since n=1.

\noindent {\bfseries 2c.} \hspace{0.5cm} \normalfont $ \infty \;
\in \Gamma_{4}$
and therefore \be \label{abelonas} {\rm q}=2. \ee Equations
(\ref{MilkyWay}) and       (\ref{abelonas}) give \be
\label{kolkol} \left [ \sqrt \nu \right ]_{\infty}={\rm
a}_{\infty}. \ee

\noindent From equations (\ref{gorgor}) and (\ref{kolkol}) we
obtain


\be \label{kolkal} \left ( \left [ \sqrt \nu \right ]_{\infty}
\right )^{2}= {\rm a}^{2}_{\infty}=\frac{s^{2}}{4}. \ee
We choose \be \label{jolyy} {\rm a}_{\infty}=\frac{s}{2}. \ee


\noindent Combining equations (\ref{MilkyWay1}), (\ref{gorgor})
and (\ref{kolkal})
yields \be \label{kotkotkto} \beta_{\infty}=-s^{2}. \ee Making use
of equations (\ref{sauvage}), (\ref{abelonas}), (\ref{jolyy}),  and
(\ref{kotkotkto}), one readily finds that \be
\label{hohohoo} {\rm E}_{\infty}=\left \{ 1-s, 1+s \right \}. \ee
Combining equations   (\ref{blooo}), (\ref{abelonas}), (\ref{jolyy}),
and (\ref{hohohoo}), yields \be \label{tirara} {\rm S}(1-s)=1 \; \;
{\rm and} \; \; {\rm S}(1+s)=-1. \ee

\noindent {\bfseries 2d.} \hspace{0.5cm} \normalfont It is not
applicable since n=1.


\noindent
\bf  {Third Step:} \normalfont


\noindent {\bfseries 3a.} \hspace{0.5cm} \normalfont For each
family $\underline{{\rm e}}= \left ({\rm e}_{c} \right ) _{c \in
\Gamma}$ of elements ${\rm e}_{c} \in {\rm E}_{c}$ we calculate
the degree ${\rm d} \left ( \underline{{\rm e}} \right )$ of the
corresponding, prospective polynomial P. The sets ${\rm E}_{c}$
are given by equations (\ref{katra1}),(\ref{belbel}) and
(\ref{hohohoo}). Since ${\rm E}_{\rm o}$ depends on the value of
$\beta$
we distinguish three cases:

\vspace{0.2cm}

\centerline{\boldmath${\beta=-3}\;\;\;$  \qquad $\bf
(Gravitational \; \; case)$}
\begin{tabular}{llllc}
\boldmath ${\rm e}_{0}\;\;\;
\qquad $ &\boldmath${\rm
e}_{2}\;\;\;
\qquad
$ &\boldmath${\rm e}_{\infty}\;\;\;
\qquad
$ &\boldmath${\rm d}=1-\displaystyle\sum_{c\in
\Gamma}
{\rm e}_{c}\;\;\;
\qquad
$ & ${\bf Families}$ \\
$\; \frac{5}{2}$ &  $\! \! \! \! \ \! \frac{1}{2}+s $ & $\! \! \!
\  1-s$
& $
\qquad
-3 $ & G1 \\
$\; \frac{5}{2}$ &  $\! \! \! \! \ \! \frac{1}{2}+s $ & $\! \! \!
\  1+s$
& $ \! \! \! \! \qquad -3-2s $ & G2 \\
$\; \frac{5}{2}$ &  $\! \! \! \! \ \! \frac{1}{2}-s $ & $\! \! \!
\  1-s$
& $ \! \! \! \! \qquad -3+2s $ & G3 \\
$\; \frac{5}{2}$ &  $\! \! \! \! \ \! \frac{1}{2}-s $ & $\! \! \!
\  1+s$
& $ \qquad -3 $ & G4 \\
$\! \! \! -\frac{3}{2}$ &  $\! \! \! \! \ \! \frac{1}{2}+s $ & $\!
\! \! \  1-s$
& $ \qquad \; \; \;1 $ & G5 \\
$\! \! \! -\frac{3}{2}$ &  $\! \! \! \! \ \! \frac{1}{2}+s $ & $\!
\! \! \  1+s$
& $   \qquad \; 1-2s $ & G6 \\
$\! \! \!  -\frac{3}{2}$ &  $\! \! \! \! \ \! \frac{1}{2}-s $ &
$\! \! \! \  1-s$
& $ \; \qquad 1+2s $ & G7 \\
$\! \! \! -\frac{3}{2}$ &  $\! \! \! \! \ \! \frac{1}{2}-s $ & $\!
\! \! \  1+s$ & $ \qquad \; \; \;1 $ & G8
\end{tabular}
\normalfont

\centerline{\boldmath ${\beta=1} \;\;\;$ \qquad  $\bf (Scalar \;
\; case)$}
\begin{tabular}{llllc}
\boldmath ${\rm e}_{0}\;\;\;
\qquad $ &\boldmath${\rm
e}_{2}\;\;\;
\qquad $ &\boldmath${\rm e}_{\infty}\;\;\;
\qquad $ &\boldmath${\rm d}=1-\displaystyle\sum_{c\in
\Gamma}
{\rm e}_{c}\;\;\;
\qquad $ & ${\bf Families}$ \\
$\; \frac{1}{2}$ &  $\! \! \! \! \ \! \frac{1}{2}+s $ & $\! \! \!
\  1-s$
& $ \qquad -1$ & S1 \\
$\; \frac{1}{2}$ &  $\! \! \! \! \ \! \frac{1}{2}+s $ & $\! \! \!
\  1+s$
& $ \! \! \! \! \qquad -1-2s $ & S2 \\
$\; \frac{1}{2}$ &  $\! \! \! \! \ \! \frac{1}{2}-s $ & $\! \! \!
\  1-s$ & $ \! \! \! \! \; \; \;
\qquad 2s-1  $ & S3 \\
$\; \frac{1}{2}$ &  $\! \! \! \! \ \! \frac{1}{2}-s $ & $\! \! \!
\  1+s$ & $ \qquad -1 $ & S4
\end{tabular}
\vspace{0.2cm}

\normalfont \centerline{\boldmath${\beta=0}\;\;\;$ \qquad  $\bf
(Electromagnetic  \; \; case)$}

\begin{tabular}{llllc}
\boldmath ${\rm e}_{0}\;\;\;
\qquad $ &\boldmath${\rm
e}_{2}\;\;\;
\qquad $ &\boldmath${\rm e}_{\infty}\;\;\;
\qquad $ &\boldmath${\rm d}=1-\displaystyle\sum_{c\in
\Gamma}
{\rm e}_{c}\;\;\;
\qquad $ & ${\bf Families}$ \\
$\; \frac{3}{2}$ &  $\! \! \! \! \ \! \frac{1}{2}+s $ & $\! \! \!
\  1-s$ & $ \qquad -2 $ & E1 \\
$\; \frac{3}{2}$ &  $\! \! \! \! \ \! \frac{1}{2}+s $ & $\! \! \!
\  1+s$ & $ \! \! \! \! \qquad -2-2s $ & E2 \\
$\; \frac{3}{2}$ &  $\! \! \! \! \ \! \frac{1}{2}-s $ & $\! \! \!
\  1-s$ & $ \! \! \! \! \qquad -2+2s $ & E3 \\
$\; \frac{3}{2}$ &  $\! \! \! \! \ \! \frac{1}{2}-s $ & $\! \! \!
\  1+s$ & $ \qquad -2 $ & E4 \\



$\! \! \! -\frac{1}{2}$ &  $\! \! \! \! \ \! \frac{1}{2}+s $ & $\!
\! \! \  1-s$
& $ \qquad \; \; \;0 $ & E5 \\
$\! \! \! -\frac{1}{2}$ &  $\! \! \! \! \ \! \frac{1}{2}+s $ & $\!
\! \! \  1+s$
& $   \qquad -2s $ & E6 \\
$\! \! \!  -\frac{1}{2}$ &  $\! \! \! \! \ \! \frac{1}{2}-s $ &
$\! \! \! \  1-s$
& $ \; \qquad \; \ 2s $ & E7 \\
$\! \! \! -\frac{1}{2}$ &  $\! \! \! \! \ \! \frac{1}{2}-s $ & $\!
\! \! \  1+s$ & $ \qquad \; \; \;0 $ & E8
\end{tabular}
\unboldmath

\normalfont \noindent In the last column we
enumerated the different families arising within each of the three
cases.


\noindent {\bfseries 3b.} \hspace{0.5cm} $\mathbf  i) \; \; \; \;
\; $\ \ \ If d is a non$-$negative integer, the family should be
retained, otherwise the family is discarded. Thus the families
G1, G4, E1, E4, S1 and S4 are discarded. It is assumed that $s$ is
non$-$negative
and therefore the families G2, E2 and S2 are also discarded since
for all of them ${\rm d}$ becomes negative for the
non$-$negative values of $s$. For the family G6 there are only two
non$-$negative values of $s$ for which ${\rm d}$ becomes
non$-$negative,
${\rm d}=0$ when $s=1/2$ and ${\rm d}=1$ when $s=0$,
and, for the family E6 there is only one,
${\rm d}=0$ when
 $s=0$
 .
We easily find that for  these values of $s$ the families  G6 and
E6 do not give rise to any Liouvillian solutions. A quick check
shows that the same result holds for the families G5, E5 and E8.
Therefore the families which remain for consideration are the
following: G3, G7, G8, E3, E7 and S3.

\hspace{0.cm} $\bf   i i ) \; \; \; \; \; $\ \ \ It
is not applicable since n=1.
\noindent

{\bfseries 3c}. \hspace{0.6 cm} For each family  retained from
step  \textbf{3b}, we form the rational function $\theta$ given by
equation (\ref{krookraa}). Since n=1, $\Gamma_{2}=\{0,2\}$ and
$\Gamma_{4}=\{\infty\}$
equation (\ref{krookraa})
implies \be \label{kliklii} \theta=\frac{{\rm e}_{0}}{r} +
\frac{{\rm e}_{2}}{r-2} + {\rm S}({\rm e}_{\infty}) \left [ \sqrt
\nu \right ]_{\infty}, \ee where ${\rm e}_{c}$ denotes any element
of ${\rm E}_{c}$
,
${\rm S}({\rm e}_{\infty})$ are given by
equations  (\ref{tirara}) and $ \left [ \sqrt \nu \right
]_{\infty}$ is given by equation (\ref{jolyy}). By making use of
equation (\ref{kliklii}) for each of the retained families we
obtain \newline
\centerline{\boldmath${\beta=-3}\;\;\;$  \qquad $\bf
(Gravitational \; \; case)$}
\begin{eqnarray}
\boldmath
\qquad
\qquad
{\bf \theta}
\qquad
\qquad
\qquad
\qquad
\qquad
\qquad
\qquad
&
&
{\bf Families}
\nonumber \\
\label{ygflygflygflty} \frac{5/2}{r} +
\frac{\frac{1}{2}-s}{r-2}+\frac{s}{2}
\quad \qquad \qquad \qquad
\qquad \qquad
&
&
\! \!\qquad {\rm G3}
\\
\label{gravity7} \frac{-3/2}{r} +
\frac{\frac{1}{2}-s}{r-2}+\frac{s}{2}
\quad \qquad \qquad \qquad
\qquad \qquad
&
&
\! \!\qquad {\rm G7}
\\
\label{gravity8} \frac{-3/2}{r} +
\frac{\frac{1}{2}-s}{r-2}-\frac{s}{2}
\quad \qquad \qquad \qquad
\qquad \qquad
&
&
\! \!\qquad {\rm G8}
\end{eqnarray}

\centerline{\boldmath${\beta=1}\;\;\;$  \qquad $\bf
(Scalar \; \; case)$} \vspace{-1.0cm}
\begin{eqnarray}
\boldmath  \qquad \qquad  {\bf \theta}
\qquad
\qquad
\qquad
\qquad
\qquad
\qquad
\qquad
&
& \ \  {\bf Families} \nonumber \\
\label{scalar11} \frac{1/2}{r} +
\frac{\frac{1}{2}-s}{r-2}+\frac{s}{2} \quad \qquad \qquad \qquad
\qquad \qquad
& & \qquad \ \ {\rm S3}
\end{eqnarray}


\centerline{\boldmath${\beta=0}\;\;\;$  \qquad  $\bf
(Electromagnetic\; \; case)$} \vspace{-1.0cm}
\begin{eqnarray}
\boldmath  \qquad \qquad  {\bf \theta}
\qquad
\qquad
\qquad
\qquad
\qquad
\qquad
\qquad
\quad
&
& {\bf Families} \nonumber \\
\label{ugh.ujhb.b.b,b,b.}
\frac{3/2}{r} +
\frac{\frac{1}{2}-s}{r-2}+\frac{s}{2} \quad \qquad \qquad \qquad \quad
\qquad \qquad
&
& \! \!\qquad {\rm E3} \\
\label{yumkfcmufmkfcym} \frac{-1/2}{r} +
\frac{\frac{1}{2}-s}{r-2}+\frac{s}{2} \qquad \qquad \qquad \qquad
\qquad \qquad
&
& \! \!\qquad {\rm E7}  \\
\nonumber
\end{eqnarray}
\noindent \bf {Fourth step - Output:} \normalfont

\noindent For n=1 equations (\ref{karokaro}) imply \be
\label{asaasa} {\rm P}_{1}=-{\rm P}, \ee \be \label{asaasa1} {\rm
P}_{0}={\rm P}^{'}+\theta{\rm P}, \ee and \be
\label{gfkkukku} {\rm P}_{-1}=0={\rm P}^{''}+2\theta{\rm P}^{'}+ (
\theta^{2}+\theta^{'}-\nu  ) {\rm P}. \ee Combining equations
(\ref{doladola}), (\ref{asaasa}) and (\ref{asaasa1}) yields \be
\label{asaasa3} \omega=\frac{{\rm P}^{'}}{{\rm P}} + \theta. \ee
For each of the retained families we search for a polynomial P of
degree ${\rm d}$ (as defined in step \textbf{3a}) such that
equation (\ref{gfkkukku}) is satisfied.
If such a
polynomial ${\rm P}$ is found then $\omega$ is given by equation
(\ref{asaasa3}) and the function \be \label{dramdram}
\eta=y=e^{\int \omega {\rm d}r}= e^{\int  ( \frac {{\rm
P}^{'}}{{\rm P}}+\theta  ){\rm d}r}= {\rm P} e^{\int \theta{\rm
d}r} \ee is a Liouvillian solution of equation (\ref{kooala}).

\noindent In
section \ref{gl} it will be shown that the
families G7 and G8 give the solutions found by Chandrasekhar
\cite{Chandr}. Finally, the families G3, E3, E7 and S3 are
considered in section \ref{rem}.
Now the
algorithm can be considered complete when n=1. \newline
\noindent $\underline{{\rm CONTINUATION}}$: The analysis of the
case n=2 is given in  Appendix \ref{app}.


\vspace{-0.457cm}

\section{
The families G7 and G8
}
\label{gl}
\noindent

In this section we prove that the family
G8 gives the Liouvillian solution $\chi$
and more importantly we also prove that the
family
G7 gives the Liouvillian solution $\chi \int \frac {{\rm
d}r_{\!\ast}}{\chi^{2}}$. As a spin off, in subsection \ref{elementary},    we prove
that the integral $\int \frac {{\rm
d}r_{\!\ast}}{\chi^{2}}$ has an elementary
functions answer
for \it{every} \normalfont value of the angular harmonic index $l$, $l$=2,3,... . Consequently,
given the functional form of $\chi$ (equations
(\ref{xirxirr}) and (\ref{nutnuta})),
the Liouvillian solution $\chi \int \frac {{\rm
d}r_{\!\ast}}{\chi^{2}}$
has an elementary
functions answer
for \it{every} \normalfont value of the angular harmonic index $l$, $l$=2,3,... .
The Liouvillian solutions $\chi$ and $\chi \int \frac {{\rm
d}r_{\!\ast}}{\chi^{2}}$ were firstly found by
Chandrasekhar  \cite{Chandr} in the case of the
first order gravitational perturbations
of the Schwarzschild geometry.
We start by describing Chandrasekhar's
Liouvillian solutions.

\subsection{Chandrasekhar's
Liouvillian solutions}

Chandrasekhar found \cite{Chandr} ``closed$-$form'' solutions to the
Regge$-$Wheeler and Zerilli equations. These solutions correspond to
algebraically special perturbations, i.e. perturbations in which
we have only incoming $(\Psi_{0}\neq 0)$ or only outgoing
$(\Psi_{4}\neq 0)$ waves. The vanishing of the Starobinsky
constant, which is the necessary and sufficient condition for the
existence of these solutions, implies that the frequency $\sigma$
of the perturbations takes pure imaginary values.

The
algebraically special solutions \cite{Chee}, initially found by Chandrasekhar, to the RWE, 
which describe outgoing waves, are the following \be
\label{xirxirr} \chi=\frac{\mu^{2}r+6}{r}{\rm e}^{-{\rm i}\sigma
r_{\!\ast}} \ee and \be \label{xirxirr1} \chi \int \frac {{\rm
d}r_{\!\ast}}{\chi^{2}} \ee where, \be \label{frogfrog}
\mu^{2}=(l-1)(l+2) \qquad \qquad l=2,3,..., \ee \be
\label{nutnuta} r_{\!\ast} = r+2\,{\rm log} \left ( \frac{r}{2}-1
\right) \; \; {\rm or} \; \; {\rm equivalently} \; \; \frac {{\rm
d}}{{\rm d}r_{\!\ast}}=\left ( 1-\frac{2}{r} \right ) \frac {{\rm
d}}{{\rm d}r}, \ee and, \be \label{nutnuta1} \sigma_{0}\equiv {\rm
i}\sigma=\frac{\mu^{2}(\mu^{2}+2)}{12}=
\frac{l(l-1)(l+1)(l+2)}{12}. \ee Once the solution (\ref{xirxirr})
is derived the method of reduction of order produces the solution
(\ref{xirxirr1}) (cf. equation (\ref{kritskrits}) with $a=0$). Now
we prove the following:

\subsection{The family G8 gives
the Liouvillian solution $\chi$
}
\label{1l}
\begin{prop}
The family G8 gives rise to the solution
$\chi=\frac{\mu^{2}r+6}{r}{\rm e}^{-{\rm i}\sigma r_{\!\ast}}.$
\end{prop}

\begin{pf}
\noindent We need to find a polynomial ${\rm P}=r+ {\rm k}$ such
that equation
(\ref{gfkkukku}) is satisfied,
where, $\theta$ now is given by equation (\ref{gravity8}).
Combination
of equations (\ref{gorgor}), (\ref{gravity8}) and
(\ref{gfkkukku})
yields \be \label{grosgroos} (-{\rm k}s + l(l+1))r +
(l+2)(l-1){\rm k}-6=0. \ee From equation (\ref{grosgroos}) it
follows that \be \label{toutrou} {\rm k}=\frac{6}{(l+2)(l-1)}
\qquad {\rm and} \qquad s=\frac{l(l-1)(l+1)(l+2)}{6}. \ee

Since a
polynomial solution $ {\rm P}=r +{\rm k}$
to equation
(\ref{gfkkukku}) is found the family
G8 does give rise to a Liouvillian solution to equation
(\ref{kooala}). This solution is given by equation
(\ref{dramdram}):
\be \label{drk1drk1}
\eta=\frac{e^{-\frac{s}{2}r}}{r^{\frac{3}{2}}(r-2)^{s-\frac{1}{2}}}
(r+{\rm k}), \ee where $s$ and k are given by equations
(\ref{toutrou}). By combining equations (\ref{pupul})  and
(\ref{drk1drk1}) one obtains the corresponding Liouvillian
solution $\eta'$ to
equation (\ref{adamsmith}):
\be \label{roocck}
\eta'=\frac{(r+{\rm k})}{r(r-2)^{s}}e^{-\frac{s}{2}r}. \ee

The solution $\chi$ found by Chandrasekhar (equation
(\ref{xirxirr})) is
the same as
the solution $\eta'$. Indeed, by
combining equations (\ref{xirxirr}),(\ref{nutnuta}) and
(\ref{nutnuta1}) we obtain
\be \label{vannav}
\chi=2^{2\sigma_{0}}\frac{\mu^{2}r+6}{r(r-2)^{2\sigma_{0}}}
e^{-\sigma_{0}r}. \ee \noindent On the other hand, from equations
(\ref{frogfrog}), (\ref{nutnuta1}), (\ref{toutrou}) and
(\ref{roocck}) one finds that \be \label{kotaak}
\eta'=\frac{1}{\mu^{2}} \frac{\mu^{2}r+6}{r(r-2)^{2\sigma_{0}}}
e^{-\sigma_{0}r}. \ee From equations (\ref{vannav}) and
(\ref{kotaak}) it follows that the solution $\eta'$ is
the same as the solution
$\chi$; the solution to an homogeneous equation is determined up to an arbitrary multiplicative constant.
This completes the proof. \end{pf}

\subsection{The family G7 gives
the
Liouvillian solution $\chi \int \frac {{\rm
d}r_{\!\ast}}{\chi^{2}}$
}
\label{G7}

\noindent

The
significance
of this subsection does not
stem from
the discovery of the Liouvillian solution
$\chi \int \frac {{\rm d}r_{\!\ast}}{\chi^{2}}$; a Liouvillian solution
$\chi$ to the master equation (\ref{radiative}) was found in the previous subsection
and it is well known    that  a second linearly independent solution to
 equation (\ref{radiative}) is  $\chi \int \frac {{\rm
d}r_{\!\ast}}{\chi^{2}}$ which is also Liouvillian. Moreover, both solutions $\chi$ and
$\chi \int \frac {{\rm
d}r_{\!\ast}}{\chi^{2}}$
were found by Chandrasekhar a long time
ago\cite{Chandr}.

It is the proof itself, that the family G7 gives rise to the Liouvillian solution
$\chi \int \frac {{\rm d}r_{\!\ast}}{\chi^{2}}$, and other results which are obtained in the course of the
proof that constitute the new findings of this subsection and contribute to its significance.
In particular
\begin{itemize}
\item{It is verified that Kovacic's algorithm may capture the second Liouvillian solution $\eta \int \frac { e^{-\int a}}{\eta ^{2}}$
when a Liouvillian solution  $\eta$ is identified. So far in the applications of Kovacic's algorithm the search for
pairs $\eta$ and $\eta \int \frac { e^{-\int a}}{\eta ^{2}}$, which
as it is stated in
subsections \ref{kovalg} and \ref{pairs}
is one of the most difficult aspects of the application of the
algorithm,
has been neglected (e.g. \cite{C}, \cite{r9}). The cause of this negligence can  be traced back to the original article by Kovacic
\cite{Kova}. }

\item{The structure and the logic of the proof does not depend on the particular problem at hand, namely the linear
perturbations of the Schwarzschild geometry, but it is tailored for identifying Liouvillian pairs
$\eta$ and $\eta \int \frac { e^{-\int a}}{\eta ^{2}}$. Therefore it is highly likely that analogues and/or generalizations of the  proof
are
going to be useful in the search for
pairs of Liouvillian solutions
$\eta$ and $\eta \int \frac { e^{-\int a}}{\eta ^{2}}$ in subsequent applications of the algorithm;
either these applications refer to black hole geometries, or otherwise. }

\item{ Equation (\ref{73}) after a change of the
independent variable falls into the confluent Heun class.
Therefore in this case Kovacic's algorithm reduces the problem of finding
Liouvillian solutions to the RWE equation to the problem of finding polynomial solutions
to equation  (\ref{73}) which falls into the confluent Heun class and one may hope
that the literature on the solutions of the confluent Heun equation,
which  is rather extensive (see e.g. \cite{Gur,Gur1,Ish,Ish1}), may provide some clues
or even provide the answer. However the literature is not helpful
and does not provide any clues.
The polynomial solution we find
here (equations (\ref{hutyytuh}), (\ref{hutyytuh1}) and (\ref{nugyygun})) has not appeared in the  literature so far. We find it by combining
Kovacic's algorithm with a new approach.   }


\item{In general the coefficients of the series solutions of ODEs with rational function coefficients
cannot be calculated in ``closed$-$form'' when the  ODEs lead to
three (or four, five...)$-$term recurrence relations for these coefficients.
Equation (\ref{73}) is constructed by Kovacic's algorithm and plays a prominent role: Polynomial solutions
to equation (\ref{73}) lead to Liouvillian solutions to the RWE equation (\ref{adamsmith}).
We find a ``closed$-$form'' polynomial solution (equations (\ref{hutyytuh}), (\ref{hutyytuh1}), (\ref{nugyygun}), and (\ref{cobi}))
to equation (\ref{73}) at odds with what generally happens with ODEs
which lead to
three$-$term recurrence relations for the coefficients of their series solutions.  }

\item{The second Liouvillian solution $\eta \int \frac { e^{-\int a}}{\eta ^{2}}$, initially found by  Chandrasekhar\cite{Chandr},
can be expressed via elementary functions
for \it{all} \normalfont values of the angular harmonic index $l$, $l$=2,3,... .
This result was  given for the first time by Araujo and MacCallum \cite{Ara} in an unpublished work;
however, these authors do not give the proof which leads to it.
This  result was not derived  by Chandrasekhar \cite{Chandr}
and it is difficult to
imagine how one could obtain it without  applying   Kovacic's algorithm to the RWE.
}

\item{
It is a nice and unexpected surprise that we can combine
the result (equations (\ref{hutyytuh}), (\ref{hutyytuh1}), (\ref{nugyygun}))   regarding the polynomial solution to equation (\ref{73}) in this subsection,       with
the
 extension of Hautot's results we give in subsection
\ref{extension}, and express, in subsection \ref{truncon}, the polynomial solution
to equation (\ref{73}) as a finite sum of truncated confluent hypergeometric functions of the first kind (equation (\ref{tel1})), and also express, in subsection \ref{Laguerre1}, this
polynomial as a finite sum of associated Laguerre polynomials (equation (\ref{tel2})).
Both sums hold for \it{every} \normalfont  value of the angular harmonic index $l$, $l$=2,3,... .

}

\item{

As a result, in subsection \ref{truncon}, we
express in equation (\ref{interms}),
the second Liouvillian solution $\eta \int \frac { e^{-\int a}}{\eta ^{2}}$ as a finite sum of truncated confluent hypergeometric functions multiplied by a factor
involving elementary functions,
and also, in subsection \ref{Laguerre1},
we
express in equation (\ref{interms1}),
the second Liouvillian solution $\eta \int \frac { e^{-\int a}}{\eta ^{2}}$ as a finite sum of associated Laguerre polynomials
multiplied by the same
factor of
elementary functions as before.
It is again difficult to imagine
how we could obtain
this
result
without
combining the application of  Kovacic's algorithm to the RWE equation with the extension of Hautot's results we give in subsection \ref{extension}.
}.

\item{Last but not least, the proof
we give in this subsection, showing that
the family G7 gives rise to
the Liouvillian solution $\chi \int \frac {{\rm
d}r_{\!\ast}}{\chi^{2}}$, is
an indispensable part of the proof that there
are no other Liouvillian solutions to the master
equation (\ref{adamsmith}) apart from those
found by Chandrasekhar\cite{Chandr}. The proof continues in section
\ref{rem}.
The proof
we give in this subsection
is totally absent in the consideration
of the problem by C.H.\cite{C}. What is even more curious is that  in \cite{C}
there is no family, in the case of the gravitational perturbations of the Schwarzschild geometry, which gives rise to the Liouvillian solution
$\chi \int \frac {{\rm
d}r_{\!\ast}}{\chi^{2}}$; we
elaborate more on this issue
in subsection \ref{comparison}.
However, in \cite{C},
 the family which leads to the Liouvillian solution
 $\chi$, we examine this family in subsection \ref{1l}, does appear.
 Due to the succinct form of the presentation of the results in \cite{C} we have not been able to trace the origin of this
discrepancy.
}

\end{itemize}

It is not generally true \cite{Duval1} that if one of the
retained families gives rise to a Liouvillian solution  $\eta$ to
equation (\ref{ant}) then another retained family will give rise
to $\eta \int \frac { e^{-\int a}}{\eta ^{2}}$.
In fact, as it is stated in subsections \ref{kovalg} and \ref{pairs}, establishing when this holds is one of the most difficult aspects of applying Kovacic's algorithm
and it can take considerable amount of time and effort to find the pair(s)
of families, \it{if they exist at all}, \normalfont  which yield the Liouvillian solutions $\eta$ and $\eta \int \frac { e^{-\int a}}{\eta ^{2}}$.
We prove now that in the particular
problem under consideration

\begin{thrm}
\label{prop2}
The family G7 gives rise to the solution $\chi \int \frac {{\rm
d}r_{\!\ast}}{\chi^{2}}$.
\end{thrm}
\noindent \textbf{Proof} \hspace{0.1cm}
In this case, where  $\theta(r)$ and $\nu(r)$ are given respectively  by equations (\ref{gravity7})
and  (\ref{gorgor}), with $\beta=-3$, equation (\ref{gfkkukku}) reads
\be
\label{73}
  {\rm P}^{''}(r)    + \frac {  6-2r-4rs+r^{2}s }  {r(r-2)} {\rm P}^{'}(r)    +  \frac    { 2-l(l+1)+6s-r s(1+2s) } {r(r-2)}   {\rm P}(r)=0.
\ee
The last equation has two regular singular points at $r=0$,
$r=2$, and,  one irregular singular point of Poincare rank 1 at $r=
\infty .$ The
roots of the associated indicial equations are the following
$$
\begin{array}{ccccc}
Singular \; points & & & &  \! \!     \! \!      \! \!   \! \! \!
\!    \! \!    \! \!    \! \!   \! \!   Roots
\\
\! \! \! r=0 & & & & \! \! \! \! \! \!  \! \! \! \! \!   \! \! \!
\!  \! \! \!  \! \!  \!
\rho_{1}=0 \ , \ \rho_{2}=4,   \\
\! \! r=2 & & & & \! \! \! \! \! \!  \! \! \!  \! \! \! \! \! \!  \! \! \! \rho_{1}=0 \ , \ \rho_{2}=
 2s,
\\
\;  \;     r=
\infty & & & &
     \rho= 1+2s.
\end{array}
$$

One can easily check that
if   equation  (\ref{73}) admits a polynomial solution at all then
this solution will have degree $1 + 2 s$, which is precisely the degree
determined by  Kovacic's algorithm.
One can easily check that equation  (\ref{73})
cannot admit polynomial solution of zero or first
degree. Thus in this proof we will assume hereafter that $2 s$
is an integer which takes values in the set
$ \{1,2,3,...    \}$.




According to Fuchs' Theorem
a basis of the space of solutions of equation (\ref{73}) in
the punctured neighbourhood $(0,2)\cup (2,4)$  of 2, when $\rho_{2}-\rho_{1}=2 s - 0= 2 s \in Z^{+}, \ Z^{+}=\{1,2,3,...    \},$
is
given by
\begin{eqnarray}
\label{iuhij;uhn} \hspace{1.5cm} p_{1}(r) & = &
(r-2)^{2 s} \sum_{{\rm n}=0}^{\infty}{\rm P}_{\rm n}(r-2)^{\rm n}, \quad \rm P_{0} \neq 0,
\quad {\rm and,}
\\
\label{iuhij;uhn1} \hspace{1.5cm} p_{2}(r) & = &
\sum_{{\rm n}=0}^{\infty}{\rm a}_{\rm n}(r-2)^{\rm n} +{\rm C}
p_{1}{\rm ln}|r-2|, \quad {\rm a}_{\rm 0} \neq 0.
\end{eqnarray}
The constant $\rm C$ may turn out to be zero.

A remark regarding the constant $\rm C$ is here
in order.
The constant $\rm C$ may be determined
by
substituting $p_{1}(r)  =
(r-2)^{2 s} \sum_{{\rm n}=0}^{\infty}{\rm P}_{\rm n}(r-2)^{\rm n}, \rm P_{0} \neq 0,$ (equation (\ref{iuhij;uhn})) into equation (\ref{73}),
in order to find
the coefficients ${\rm P}_{\rm n}$, and then by substituting $p_{2}(r)  =
\sum_{{\rm n}=0}^{\infty}{\rm a}_{\rm n}(r-2)^{\rm n} +{\rm C}
p_{1}{\rm ln}|r-2|,  {\rm a}_{\rm 0} \neq 0,
$  (equation (\ref{iuhij;uhn1})) into equation (\ref{73})  in order to find the coefficients
${\rm a}_{\rm n}$ and the constant $\rm C.$
The constant $\rm C$ may turn out to be zero.

Whenever in this paper,
 power series solutions of the form
(\ref{iuhij;uhn}) and (\ref{iuhij;uhn1})
 are used in the proofs of Lemmas, Corollaries and Theorems, and in these proofs it is stated that
``The constant $\rm C$ may turn out to be zero''
this will  mean that  the proof will not necessitate the evaluation of $\rm C.$  If the
value of $\rm C$ is required from the proof, this
will be explicitly stated, $\rm C$ will be evaluated with the method described above, and its
value will be given.

By substituting ${\rm P}(r)=(r-2)^{\rho} \sum_{\rm n = 0}^{+ \infty} {\rm P}_{\rm n} (r-2)^{\rm n}$
into equation (\ref{73}), where $\rho=0$ or
$\rho=2 s$, we find that the coefficients
${\rm P}_{\rm n}$ satisfy the following
three$-$term recurrence relation
\begin{eqnarray}
\label{rrr10}
& - & s ( 2 - {\rm n}  - \rho + 2 s       ) {\rm P}_{{\rm n}-1}
+
(2-l(l+1)+{\rm n}^{2}
+ {\rm n}(2 \rho   -   3)
+ \rho (\rho - 3)
+ 4 s (1-s)
){\rm P}_{{\rm n}} \nonumber \\
& +  & 2(1+{\rm n}+\rho)(1+{\rm n}+\rho-2 s) {\rm P}_{{\rm n}+1}=
 0,
\quad \rm{where}, \ \ \  {\rm P}_{-1}=0.
\end{eqnarray}

We note that when $\rho=2 s$ equation
(\ref{rrr10}) gives the three$-$term recurrence relation satisfied by the coefficients ${\rm P}_{\rm n}$
of the power series solution
$p_{1}(r)  =
(r-2)^{2 s} \sum_{{\rm n}=0}^{\infty}{\rm P}_{\rm n}(r-2)^{\rm n},  \rm P_{0} \neq 0,$ given by equation (\ref{iuhij;uhn}).
When $\rho=0$ equation
(\ref{rrr10}) gives the three$-$term recurrence relation satisfied by the coefficients ${\rm a}_{\rm n}$
of the
solution
$p_{2}(r)  =
\sum_{{\rm n}=0}^{\infty}{\rm a}_{\rm n}(r-2)^{\rm n} +{\rm C}
p_{1}{\rm ln}|r-2|,  {\rm a}_{\rm 0} \neq 0,$ given by equation (\ref{iuhij;uhn1}), \it
only when $\rm C = 0.$ \normalfont When
$\rm C \neq 0$ the coefficients
${\rm a}_{\rm n}$
of the power series
$\sum_{{\rm n}=0}^{\infty}{\rm a}_{\rm n}(r-2)^{\rm n}$
appearing in the solution
$p_{2}(r)$,
given by equation (\ref{iuhij;uhn1}),
are determined by substituting  $p_{2}(r)$
into equation (\ref{73}).


From the three$-$term recurrence relation (\ref{rrr10}) we easily
verify, by examining the system of equations resulting from (\ref{rrr10}) when $\rho=2 s$
and $\rm n=0,1,2$,
that the series solution $p_{1}$
given by equation (\ref{iuhij;uhn}),
can in principle
terminate
to give a polynomial solution of degree $2 s + 1$,
but this can only happen when $s= - \frac{l(l-1)(l+1)(l+2)}{6},$ $l=2,3,... \ .$
Consequently this case is dismissed since the
only accepted values of $s$ are
$s= 2 \sigma_{0} =\frac{l(l-1)(l+1)(l+2)}{6},$
$l=2,3,...,$ (equation (\ref{nutnuta1})).

However,
as we easily verify from the three$-$term recurrence relation
(\ref{rrr10}),
by examining the system of equations resulting from (\ref{rrr10}) when $\rho=0$
and $\rm n= 2 \it s + \rm 1, \rm 2 \it s + \rm 2$,
the series solution $p_{2}$
given by equation (\ref{iuhij;uhn1}),
can, in principle, terminate and give a polynomial solution
of degree $2 s + 1$, when
$s=  \frac{l(l-1)(l+1)(l+2)}{6},$ $l=2,3,...,$ if C is zero.

If this is indeed the case the coefficients ${\rm P}_{\rm n}$ of
the polynomial solution
${\rm P}(r)=\sum_{\rm n = 0}^{2 \it s + \rm 1} {\rm P}_{\rm n} (r-2)^{\rm n}$
to equation (\ref{73}),
satisfy the
3$-$term
recurrence relation
\begin{eqnarray}
\label{rr10}
s ( {\rm n} - 2( s+1)) {\rm P}_{{\rm n}-1}+(2-l(l+1)+{\rm n}({\rm n}-3)-4 s (s-1)){\rm P}_{{\rm n}}+2(1+{\rm n})(1+{\rm n}-2 s) {\rm P}_{{\rm n}+1}=  0,&& \nonumber \\
{\rm P}_{-1}=0, \quad \quad
\end{eqnarray}
which results from the
3$-$term
recurrence relation (\ref{rrr10}) when $\rho=0.$


We search now for a polynomial solution
${\rm P}(r)$
of degree $1+2s$
to equation
(\ref{73}).
Proceeding with brutal force and trying to
find a polynomial solution to equation (\ref{73}) is not effective because
the degree of the polynomial solution sought for the equation (\ref{73}) is not fixed but
it depends on the  parameter $s$ which appears in the coefficients of ${\rm P}^{'}$ and of ${\rm P}$
in equation (\ref{73}).
Remarkably equation  (\ref{73}) after a change of the independent variable from $r$ to $z=\frac{r}{2}$
falls into the confluent Heun class. The extensive literature on the solutions of the confluent Heun equation
(see e.g. \cite{Gur,Gur1,Ish,Ish1}) does not provide any clues either for obtaining polynomial solutions to equation (\ref{73}).

For this reason, we follow instead another approach:  We \it{assume} \normalfont that a
polynomial solution ${\rm P}(r)$ to equation (\ref{gfkkukku}) of
degree $1+2s$ exists, i.e., we assume that the family G7 gives
rise to a Liouvillian solution $y$. By making use of equations
(\ref{gravity7}) and (\ref{dramdram}) we find that \be
\label{drsdrachmas} y=\frac{{\rm
P}(r)e^{\frac{s}{2}r}}{r^{\frac{3}{2}}(r-2)^{s-\frac{1}{2}}}. \ee
By combining equations (\ref{pupul}) and (\ref{drsdrachmas}) we
find the corresponding Liouvillian solution $y'$ to
equation (\ref{adamsmith}) : \be \label{grasssop} y'=\frac{{\rm
P}(r)e^{\frac{s}{2}r}}{r(r-2)^{s}}. \ee

We also assume that $y'$
is equal to the solution $\chi \int \frac {{\rm
d}r_{\!\ast}}{\chi^{2}}$ given by equation (\ref{xirxirr1}) and
that $s$ in equation
({\ref{grasssop})
takes the values given by
(equation (\ref{nutnuta1})),
i.e., we assume that \be
\label{btrkoala} s=2\sigma_{0}=\frac{\mu^{2}(\mu^{2}+2)}{6}=
\frac{l(l-1)(l+1)(l+2)}{6}, \ee where $l=2,3,...$ . By making use
of equations (\ref{xirxirr}), (\ref{nutnuta}), (\ref{grasssop})
and (\ref{btrkoala}) the last assumption reads \be
\label{grougroo} \int \frac {{\rm d}r_{\!\ast}}{\chi^{2}}=
\int\frac{r^{3}(r-2)^{4\sigma_{0}-1}e^{2\sigma_{0}r}}{(\mu^{2}r+6)^{2}}=
\frac{{\rm P}(r)e^{2\sigma_{0}r}}{\mu^{2}r+6}. \ee

Equation
(\ref{grougroo}) implies \be \label{grougroo1} ({\rm
P}^{'}(r)+2\sigma_{0}{\rm P}(r))(\mu^{2}r+6)-\mu^{2}{\rm P}(r)=
r^{3}(r-2)^{4\sigma_{0}-1} . \ee By
making the change of the independent variable $r=w+{\rm 2}$
in equation
(\ref{grougroo1}) we obtain
\be \label{troustrouss} ({\rm
P}^{'}(w)+2\sigma_{0}{\rm P}(w))(\mu^{2}w+2\mu^{2}+6)-\mu^{2}{\rm P}(w)=
w^{4\sigma_{0}-1}(w+2)^{3}. \ee P($w$) is a polynomial of degree
$1+2s=1+4\sigma_{0}$ .
By setting
\be \label{voklaal} {\rm P}(w)=\sum_{{\rm n}=0}^{1+4\sigma_{0}} {\rm
P}_{{\rm n}}w^{n} \ee in (\ref{troustrouss}) we find
\begin{eqnarray}
\hspace{-1.2cm} \sum_{{\rm n}=0}^{1+4\sigma_{0}} \left \{ 2\sigma_{0}\mu^{2}{\rm
P}_{{\rm n}-1} + [\mu^{2}{\rm n} +
2\sigma_{0}(6+2\mu^{2})-\mu^{2}]{\rm P}_{\rm n}+
(6+2\mu^{2})({\rm n}+1){\rm P}_{{\rm n}+1} \right \}w^{\rm n}+ & & \nonumber \\
\hspace{-0.65cm} \label{oddddo} \!\!\!\!\!\! 2\sigma_{0}\mu^{2}{\rm
P}_{4\sigma_{0}+1}w^{4\sigma_{0}+2}=
8w^{4\sigma_{0}-1}+12w^{4\sigma_{0}}+6w^{4\sigma_{0}+1}+w^{4\sigma_{0}+2},
\end{eqnarray}
 where, ${\rm P}_{-1}={\rm P}_{4\sigma_{0}+2}=0$.
From equation (\ref{oddddo}) all the coefficients ${\rm P}_{\rm
n}$ can be found.

This derivation of the coefficients ${\rm
P}_{\rm n}$ depends  on the validity of the initial assumptions.
In order to verify the correctness of the result, we derive again
the values of the coefficients ${\rm P}_{\rm n}$ by using
in addition a theoretical input independent of any initial
assumption. In case of agreement, we do expect the coefficients determined by
equation (\ref{oddddo}) to be proportional to the new
coefficients, i.e., we expect the following relation to hold \be
\label{fradeedarf} {\rm P}_{{\rm n}}=c{\rm P}^{'}_{\rm n} \qquad
\forall \qquad 0\leq{\rm n}\leq 1+4\sigma_{0}, \ee where,  $ {\rm
P}^{'}_{\rm n}$ are the new coefficients and $c$ is a constant of
proportionality. We use the same symbol ${\rm P}_{\rm n}$ for both
derivations. This hopefully does not give rise to any confusion.

Now the second derivation of the coefficients ${\rm P}_{\rm n}$ is
as follows.
\noindent
The points $r$=${\rm 0}$ and $r$=${\rm 2}$ are regular singular points of
equation (\ref{adamsmith}).
By Fuchs' Theorem
a basis of the space of solutions of equation (\ref{adamsmith}), in
in
the punctured neighbourhood $(0,2)\cup (2,4)$  of 2,
is
given by
\begin{eqnarray}
\label{lotrolotro} \psi_{1} & = & (r-2)^{s}\sum_{{\rm
n}=0}^{\infty}{\rm b}_{\rm n}(r-2)^{\rm n}, \quad  {\rm b}_{\rm
0} \neq 0,  \quad {\rm and,} \qquad \qquad \qquad \!
\\
\label{lotrolotro1} \psi_{2} & = & (r-2)^{-s}\sum_{{\rm
n}=0}^{\infty}{\rm c}_{\rm n}(r-2)^{\rm n} +{\rm C}\psi_{1}{\rm
ln}|r-2|,  \quad  {\rm c}_{\rm 0} \neq 0.
\end{eqnarray}
The constant {\rm C}  may
turn out to be zero, and,
$s$ and $-s$ are the roots of the indicial equation
associated with the singular point $r=2$. Their difference
$s-(-s)=2s=4\sigma_{0}$ is a strictly positive integer (equation
(\ref{btrkoala})).

The two linearly independent solutions $\chi$
and $y'$ of equation (\ref{adamsmith}), given by equations
(\ref{vannav}) and (\ref{grasssop}) respectively, may be expressed
as linear combinations of $\psi_{1}$ and $\psi_{2}$, i.e., there are constants A, B, D, and F, such that
\begin{eqnarray}
\label{jolojoloi}
\frac{(\mu^{2}r+6)e^{-\sigma_{0}r}}{r(r-2)^{2\sigma_{0}}} & = &
{\rm A}\psi_{1}+{\rm B}\psi_{2}, \ \\
\label{tressert} \frac{\rm P(r)
e^{\sigma_{0}r}}{r(r-2)^{2\sigma_{0}}} & = & {\rm D}\psi_{1}+{\rm
F}\psi_{2}. \
\end{eqnarray}
\noindent

 By combining equations
(\ref{lotrolotro}), (\ref{lotrolotro1}), (\ref{jolojoloi}), and
(\ref{tressert}), we find that
 the functions $(\mu^{2}w+2\mu^{2}+6)e^{-\sigma_{0}w}$ and ${\rm
P}(w)e^{\sigma_{0}w}$ have proportional expansions up to order
$w^{4\sigma_{0}-1}$, i.e.,
we find that \be \label{ftysftys} {\rm
P}(w)\sim(\mu^{2}w+2\mu^{2}+6)e^{-2\sigma_{0}w}, \qquad
up
\quad
to
\quad
order
\quad w^{4\sigma_{0}-1}, \ee where, in order to simplify notation, we have used the same symbol
{\rm P} for the functions {\rm P($r$)} and {\rm p}{\rm(}$w${\rm
)}={\rm P}{\rm(}$w${\rm +2)}.
\noindent

 Now
$$(\mu^{2}w+2\mu^{2}+6)\sum_{{\rm n}=0}^{\infty}
\frac{(-2\sigma_{0}w)^{\rm n}}{{\rm n}!}$$ has ${\rm n}^{\rm th}$
coefficient
$$
\frac{(-2\sigma_{0})^{{\rm n}-1}}{{\rm n}!} \left [({\rm
n}-4\sigma_{0})\mu^{2}-12 \sigma_{0} \right ],
$$
and therefore equation (\ref{ftysftys}) gives \be \label{bgrnhy}
{\rm P}_{\rm n}\sim \frac{(-2\sigma_{0})^{{\rm n}-1}}{{\rm n}!}
\left [({\rm n}-4\sigma_{0})\mu^{2}-12 \sigma_{0} \right ] \qquad
{\rm for} \quad 0\leq{\rm n}\leq4\sigma_{0}-1. \ee

From equation
(\ref{bgrnhy}) it follows that \be
\label{afbmoradz} {\rm
P}_{4\sigma_{0}-1}\sim\frac
{(-2\sigma_{0})^{4\sigma_{0}-2}}{(4\sigma_{0}-1)!}(-\mu^{2}-12\sigma_{0})
\ , \qquad {\rm P}_{4\sigma_{0}-2}\sim\frac
{(-2\sigma_{0})^{4\sigma_{0}-3}}{(4\sigma_{0}-2)!}(-2\mu^{2}-12\sigma_{0}).
\ee Equation (\ref{afbmoradz}) gives \be \label{redderklo} \frac
{{\rm P}_{4\sigma_{0}-1}}{{\rm P}_{4\sigma_{0}-2}}=
\frac{\sigma_{0}}{1-4\sigma_{0}}\frac{\mu^{2}+12\sigma_{0}}
{\mu^{2}+6\sigma_{0}}. \ee By a straightforward calculation we
verify     
that the set of values of
the coefficients ${\rm P}_{{\rm n}}$ given by equation
(\ref{bgrnhy}) satisfy equation (\ref{oddddo}) for all powers up
to $w^{4\sigma_{0}-2}$.
This is a partial verification of the validity of our initial
assumptions. There was no reason to anticipate that the
coefficients given by equation  (\ref{bgrnhy}) should satisfy
equation (\ref{oddddo}) for all powers up to $w^{4\sigma_{0}-2}$,
unless our initial assumptions are valid.

To complete the
verification we need to show that these values, i.e. the values
(\ref{bgrnhy}),
can give a consistent solution to the four
equations which are obtained by equating the coefficients of
$w^{4\sigma_{0}+2}$, $w^{4\sigma_{0}+1}$, $w^{4\sigma_{0}}$, and
$w^{4\sigma_{0}-1}$, from both sides of (\ref{oddddo}). By equating
the coefficients of $w^{4\sigma_{0}+2}$ we find \be
\label{hutyytuh} {\rm
P}_{4\sigma_{0}+1}=\frac{1}{2\sigma_{0}\mu^{2}}. \ee By equating
the coefficients of $w^{4\sigma_{0}+1}$ and by making use of
equation ({\ref{hutyytuh}})
we have that \be \label{hutyytuh1} {\rm
P}_{4\sigma_{0}}=\frac{\mu^{2}-3}{\sigma_{0}\mu^{4}}. \ee Equating
the coefficients of $w^{4\sigma_{0}}$ and combining equations
(\ref{hutyytuh}) and (\ref{hutyytuh1}) yields \be \label{doreerod}
{\rm
P}_{4\sigma_{0}-1}=\frac{3(6\sigma_{0}-\mu^{2})}{\sigma_{0}^{2}\mu^{6}}.
\ee

Finally, by equating the coefficients of $w^{4\sigma_{0}-1}$
and by making use of equations (\ref{hutyytuh1}) and
(\ref{doreerod}) we obtain \be \label{jokliilkoj} {\rm
P}_{4\sigma_{0}-2}= \frac{3 \left
(-36\sigma_{0}^{2}-12\mu^{2}\sigma_{0}^{2}+12\mu^{2}\sigma_{0}
+4\mu^{4}\sigma_{0}-\mu^{4} \right ) } {\sigma_{0}^{3}\mu^{8}} \ .
\ee From equations (\ref{doreerod}) and (\ref{jokliilkoj}) it
follows that \be \label{bgiceromas} \frac{{\rm P}_{4\sigma_{0}-1}}
{{\rm P}_{4\sigma_{0}-2}}= \frac
{(\mu^{2}-6\sigma_{0})\sigma_{0}\mu^{2}}
{36\sigma_{0}^{2}+12\mu^{2}\sigma_{0}^{2}-12\mu^{2}\sigma_{0}
-4\mu^{4}\sigma_{0}+\mu^{4}} \ . \ee

The values of the
coefficients ${\rm P}_{{\rm n}}$ given by equation (\ref{bgrnhy})
agree with the values given by equations
(\ref{hutyytuh}), (\ref{hutyytuh1}), (\ref{doreerod}), and
(\ref{jokliilkoj}), if and only if the ratio (\ref{redderklo}) is
equal to the ratio (\ref{bgiceromas}). It is straightforward to
verify that the equality of the two ratios (\ref{redderklo}) and
(\ref{bgiceromas}) \be
\frac{\sigma_{0}}{1-4\sigma_{0}}\frac{\mu^{2}+12\sigma_{0}}
{\mu^{2}+6\sigma_{0}}=
\frac{(\mu^{2}-6\sigma_{0})\sigma_{0}\mu^{2}}
{36\sigma_{0}^{2}+12\mu^{2}\sigma_{0}^{2}-12\mu^{2}\sigma_{0}
-4\mu^{4}\sigma_{0}+\mu^{4}} \ee holds if and only if \be
\sigma_{0}=\frac{\mu^{2}(\mu^{2}+2)}{12}. \ee But this happens to
be one of our initial assumptions (equation (\ref{btrkoala})).
This completes the proof.

We adopt the values of the coefficients ${\rm P}_{{\rm
n}}$ determined by the first derivation (equations
(\ref{oddddo}), (\ref{redderklo}), (\ref{hutyytuh}), (\ref{hutyytuh1}),
and  (\ref{doreerod})). The values of the coefficient ${\rm
P}_{4\sigma_{0}-1}$ given by the two derivations are different
(equations (\ref{afbmoradz}) and (\ref{doreerod})). Their ratio is
equal to the costant of proportionality $c$ (cf.equation
(\ref{fradeedarf})) \be \label{folawwalof}
c=\frac{3(4\sigma_{0})!(\mu^{2}-6\sigma_{0})}
{(-2\sigma_{0})^{4\sigma_{0}}(12\sigma_{0}+\mu^{2})\sigma_{0}\mu^{2}}.
\ee

To make the values of the coefficients ${\rm P}_{\rm n}$ given
by equation (\ref{bgrnhy}) consistent with the adopted values
(equation (\ref{oddddo})) we multiply them by the constant of
proportionality $c$ given by equation (\ref{folawwalof}). By doing
so we obtain \be \label{nugyygun} \hspace{-1.9cm}   {\rm P}_{\rm n}=\frac
{3(-2\sigma_{0})^{{\rm
n}-4\sigma_{0}-1}(4\sigma_{0})!(\mu^{2}-6\sigma_{0}) \left [ ({\rm
n}-4\sigma_{0})\mu^{2}-12\sigma_{0} \right ] } {{\rm
n}!(\mu^{2}+12\sigma_{0})\sigma_{0}\mu^{6}},
\qquad 0\leq{\rm
n}\leq4\sigma_{0}-1. \ee \noindent Equations
(\ref{hutyytuh}), (\ref{hutyytuh1}) and (\ref{nugyygun}) give all
the coefficients ${\rm P}_{\rm n}$,
$0\leq{\rm
n}\leq1+4\sigma_{0}$,
of the polynomial ${\rm P}(w)$.
As expected, these coefficients also satisfy the 3$-$term recurrence relation (\ref{rr10}),
when $s = 2 \sigma_{0} =\frac{l(l-1)(l+1)(l+2)}{6}$.

${\rm
P}(w)$ is a polynomial of degree $1+4\sigma_{0}$ in $w=r-2$
related to the
polynomial ${\rm P}(r)$ appearing in the numerator of the
solution $y'$ (equation (\ref{grasssop})) by the 
relation \be
\label{derossored} {\rm P}(r)=\sum_{{\rm n}=0}^{1+4\sigma_{0}}{\rm
p}_{\rm n}r^{\rm n}= \sum_{{\rm n}=0}^{1+4\sigma_{0}}{\rm P}_{\rm
n}(r-2)^{\rm n},
\ee where ${\rm P}_{\rm n}$ are given by
equations (\ref{hutyytuh}), (\ref{hutyytuh1}) and
(\ref{nugyygun}). From equation (\ref{derossored}) one can find
the coefficients ${\rm p}_{\rm n}$ by using Newton's binomial
formula to expand $(r-2)^{{\rm n}}$.
We easily find
\be
\label{cobi}
{\rm
p}_{\rm n}=
\sum_{i=0}^{(1+4\sigma_{0}) - \rm n}
{\rm P}_{\rm n + \it i}
\binom{{\rm n + \it i}}{\it i}    (- \rm 2)^{\it i}, \quad  0 \leq \rm n  \leq 1+4\sigma_{0}.
\ee
This completes the proof.



\subsection{
$\chi \int \frac {{\rm
d}r_{\!\ast}}{\chi^{2}}$ has an elementary functions answer for all $l$
}

\label{elementary}

\noindent

The integral $\int \frac {{\rm d}r_{\!\ast}}{\chi^{2}}$
in the solution $\chi \int \frac {{\rm d}r_{\!\ast}}{\chi^{2}}$
(see equation (\ref{xirxirr1})) was left unevaluated by
Chandrasekhar. Equation (\ref{grougroo}) shows that it has an
elementary function answer for all $l$. For example, for $l=2 \:$
we have $\mu^{2}=4$ and $\sigma_{0}=2$ (equations (\ref{frogfrog})
and (\ref{nutnuta1})) and by combining equations
(\ref{grougroo}), (\ref{hutyytuh}), (\ref{hutyytuh1}), (\ref{nugyygun}),
and (\ref{derossored}), we find
\begin{eqnarray}
\int \frac {{\rm d}r_{\!\ast}}{\chi^{2}}= \int \frac
{r^{3}(r-2)^{7}e^{4r}}{(4r+6)^{2}}{\rm d}r=
\frac{{\rm P}(r)e^{4r}}{4r+6} & = & \nonumber \\
\left ( -\frac{1164765}{16384}+\frac{1941275}{8192}r
-\frac{388255}{1024}r^{2}+\frac{388255}{1024}r^{3}-\frac{132149}{512}r^{4}
\right.
& + & \nonumber \\
\left.
\frac{31345}{256}r^{5}-40r^{6}+\frac{275}{32}r^{7}-\frac{35}{32}r^{8}+
\frac{1}{16}r^{9} \right ) \frac{e^{4r}}{4r+6}\cdot & &
\end{eqnarray}

\section{
Hautot's results and
perturbations of the Schwarzschild geometry
}

\label{Hautotre}


In 1969 Hautot wrote a paper \cite{Hautot} on the polynomial solutions of the differential equations of the form
\be
\label{hautot}
z(z-1) {\rm P}^{''}(z) + (a z^{2}  + b z + c) {\rm P}^{'}(z) + (d + e z + f z^{2}){\rm P}(z)=0.
\ee
Kovacic's algorithm key result, stated in
subsection \ref{kovalg}  and in section
\ref{AKA}, is that there are  Liouvillian solutions to the master equation (\ref{adamsmith})
if and only if there are polynomial solutions
to equation (\ref{gfkkukku})which reads
$$
{\rm P}^{''}+2\theta{\rm P}^{'}+ ( \theta^{2}+\theta^{'}-\nu  )
{\rm P} =0$$.
Equation (\ref{hautot})
is of interest to us because
in the problem we are examining,
in all cases, apart from a few
cases examined in section
\ref{AKA} and a case examined in subsection
\ref{1l}, equation (\ref{gfkkukku}),
as it is shown in subsection
\ref{G7} and  in section \ref{rem},
falls into the class of equations
(\ref{hautot}).


Hautot starts his paper by finding the necessary and sufficient conditions for the existence of polynomial solutions to equation (\ref{hautot}).
These conditions
amount to
the requirement that the determinant of a tridiagonal matrix,
the matrix of the linear homogeneous system  associated with the differential equation (\ref{hautot}), must be zero.
The elements of this determinant are the coefficients of the three$-$term recurrence relation which results from equation (\ref{hautot})
when the polynomial solution ${\rm P}_{\rm n}(z)=\sum_{k=0}^{\rm n} \lambda_{k} z^{k}$ of degree $\rm n$ is substituted into it.

Hautot proceeds to
point out that the key difficulty in satisfying this requirement  is that the order  of the determinant increases
with the order of the polynomial solution.
In fact it is precisely this difficulty which has deterred up to this day the derivation of all
the polynomial solutions
to equation (\ref{hautot}) despite the numerous efforts (e.g. \cite{Gur,Gur1,Ish,Ish1}).

Hautot then makes a crucial remark upon which his whole paper is based: The key difficulty can be circumvented
at the expense of not obtaining the whole set of polynomial solutions but a subset of them
by arriving at
sufficient, but not necessary, conditions for the existence of polynomial solutions to equation (\ref{hautot}).

These new  sufficient conditions reduce
to the requirement that the determinant of a \it fixed \normalfont order $ (j+1) \times (j+1) $,
the fixed number $j$  being determined by equation (\ref{hautot}), must be zero.
So now one is only faced with the easy task of checking if a determinant of fixed order is zero.
If it is, and if in addition,
as the relevant analysis shows, $f=0$ and $e=-a \rm n,$
then equation (\ref{hautot}) admits polynomial solutions of degree n.


Hautot manages to obtain the necessary conditions
by using the structure of the three$-$term recurrence relation which results from equation (\ref{hautot}) and an elementary fact
about determinants which one can find in any textbook on linear algebra!
Hautot's work is a nice reminder that  significant results can be obtained with elementary methods.
What adds more to the significance
of Hautot's results  is that Hautot carries on and proves that when the sufficient
conditions are satisfied, i.e., when the aforementioned determinant of fixed order is zero, then   the resulting polynomial solutions of degree n
can be written as a sum of
$j+1$
truncated confluent hypergeometric functions of the first kind!

\subsection{Hautot's  results and Liouvillian solutions
}

\noindent

What makes Hautot's
results
relevant to our analysis is that the  sufficient
conditions derived by Hautot for the existence of polynomial solutions to equation (\ref{hautot})
are met precisely in the case of the Liouvillian solution $\chi \int \frac {{\rm
d}r_{\!\ast}}{\chi^{2}}$, initially found by Chandrasekhar\cite{Chandr},
in the case of the gravitational perturbations of the Schwarzschild geometry. In fact Hautot's sufficient conditions in the case of
the Liouvillian solution $\chi \int \frac {{\rm
d}r_{\!\ast}}{\chi^{2}}$  reduce to the requirement that the frequency of the gravitational
perturbations can only take the values of the frequencies of the algebraically special gravitational perturbations of the Schwarzschild geometry
initially found by Couch and Newman \cite{Couch}!



After the application, in section \ref{AKA}, of Kovacic's algorithm to the master equation (\ref{adamsmith}),
and the examination of the families
G8 and G7 in subsections \ref{1l} and
\ref{G7} respectively, there remain four families, namely the families
G3, E3, E7, and S3.

In section \ref{rem} the four remaining families G3, E3, E7, and S3 are reduced into two families G3 and E7.
There is strong evidence, given in subsection \ref{evidence},  that
the two families G3 and E7
 do not give rise to Liouvillian solutions, but  a proof of this is still missing.


More to the evidence given in section \ref{rem} adds the fact that Hautot's
sufficient conditions are not satisfied in cases G3, E3,  and S3 and therefore no Liouvillian solutions arise in these cases from Hautot's
sufficient conditions.

Moreover, in the case E7, which arises in the
electromagnetic perturbations of the Schwarzschild geometry,
Hautot's sufficient conditions are satisfied
but they give rise to Liouvillian solutions only when
the angular harmonic index $l$ takes the value 0. This solution describes the first order perturbation of the
Schwarzschild geometry due to the presence of a point charge located at the center of  the mass$-$source of the gravitational field,
and it is interesting in its own right. However, it does not describe gravitational waves travelling on the Schwarzschild background and as such it is dismissed.

Hautot's sufficient conditions are derived for and are tailored to equation (\ref{hautot}) which appears in the application of Kovacic's algorithm to the
Schwarzschild geometry in the form of equation (\ref{gfkkukku}). In section \ref{G7}
we prove that the application of Kovacic's algorithm to the Schwarzschild geometry leads to the result that
Chandrasekhar's Liouvillian solution  $\chi \int \frac {{\rm
d}r_{\!\ast}}{\chi^{2}}$, in the case of the gravitational perturbations of the Schwarzschild geometry,
is a product of elementary functions,
one of them being a polynomial
of degree $2 s + 1$, where $s=2 {\rm i} \sigma$ and $\sigma$ is the frequency
of the monochromatic waves to which the perturbation of the Schwarzschild geometry is Fourier decomposed.


By using an appropriate extension of
Hautot's results, which we derive
in subsection \ref{extension}, we prove  in subsection \ref{truncon} that
this
polynomial,
of degree $2 s + 1$,
admits a finite expansion
in terms of truncated confluent hypergeometric functions of the first kind (equation (\ref{tel1})). Moreover
in
subsection \ref{Laguerre1}, by using  the same
extension of Hautot's results, we prove that
this polynomial,
of degree $2 s + 1$,
admits also a finite expansion in terms of associated Laguerre polynomials (equation (\ref{tel2})).
Remarkably, both expansions hold
\it{for every value of the angular harmonic index l}, \normalfont $l$=2,3,... .

We expect that Hautot's analysis,
appropriately modified and extended,
is applicable to other black$-$hole geometries,
4$-$dim and higher, and will help, in conjunction with Kovacic's algorithm, to find Liouvillian solutions to the perturbation equations  of these other geometries, and express them in certain cases, as finite sums of truncated (confluent) hypergeometric functions.

A modification of Hautot's results
is necessary in order to apply them to other black hole geometries, because in other geometries Kovacic's algorithm
will lead in general to equations which will not fall into the
class (\ref{hautot}), and an extension of Hautot's results is needed, for the
reasons explained in subsection \ref{extension}, in order to apply them to the Schwarzschild geometry in 4$-$dim,
and also in order to apply them,
to other black$-$hole geometries, 4$-$dim and higher.

With this motivation we present Hautot's results \cite{Hautot}
emphasizing those points which are more relevant to our analysis and
highlighting the underlying reasoning. This will make more
intelligible
\begin{enumerate}

\item{The necessary modification of Hautot's results, in order to apply them to other black hole geometries.}

\item{A rectification made in subsection
\ref{Laguerre} of an oversight by Hautot.}

\item{More importantly, the
necessary extension of Hautot's results, introduced in subsection \ref{extension},  in order to apply them
to the problem under consideration.}

\end{enumerate}
Most likely a similar extension is needed in order to apply Hautot's results  to other black hole geometries.

\subsection{Necessary and sufficient conditions}
\noindent

By assuming that equation (\ref{hautot}) admits a polynomial solution ${\rm P}_{\rm n}(z)=\sum_{k=0}^{\rm n} \lambda_{k} z^{k}$ of degree $\rm n$
and by substituting  ${\rm P}_{\rm n}(z)$ into equation (\ref{hautot}) we obtain $f=0$, $e=- a \rm n$, and  that the coefficients $\lambda_{k} $, $k=0,1,2,...,$ satisfy
the
three$-$term recurrence relation
 \begin{eqnarray}
 \label{rr}
 a(k-1-\rm n)  \ \lambda_{\it k-\rm 1} + \it (d+k(b+k- \rm 1)) \ \lambda_{\it k} + (\it c- \it k)(\it k+ \rm 1) \ \lambda_{\it k+ \rm 1} &=& 0.
 \end{eqnarray}
If equation (\ref{hautot}) admits polynomial solutions  ${\rm P}_{\rm n}(z)$ then the  determinant of the associated linear homogeneous system ($\rm n+1$) $ \times$ ($\rm n+1$)
is zero.

Let us denote this determinant by $det(\mathcal M)$, where $\mathcal M$ is the ($\rm n+1$) $ \times$ ($\rm n+1$)
matrix of the coefficients of the $\rm n+1$ unknowns $\lambda_{0}, \lambda_{1}, \lambda_{2},...,\lambda_{\rm n+1}$ of this linear homogeneous system.
The
determinant   $det(\mathcal M)$       is tridiagonal and it involves the coefficients
\be
\label{coef}
R_{k}=a(k-1-\rm n), \quad \it S_{\it k} = \it d + \it k(\it b + \it k - \rm 1), \quad \it T_{\it k}=(\it c- \it k)(\it k+ \rm 1),
\ee
$k=0,1,2,...,\rm n$, of the three$-$term recurrence relation (\ref{rr}).

We conclude that the  necessary conditions for the existence of polynomial solutions
to equation (\ref{hautot}) are:
\begin{enumerate}
\item \begin{equation}  \label{con1} \it f \rm = \rm 0  \end{equation}
\item   \begin{equation}  \label{con2} \it e = - a \rm{n} \end{equation}
\item \begin{equation}  \label{con3}  det(\mathcal M)  = \begin{vmatrix}
   S_{\rm 0} & T_{\rm 0} & 0 & \cdot & \cdot & 0 & 0 & 0  \\
R_{\rm 1} & S_{\rm 1} & T_{\rm 1} & \cdot & \cdot & 0 & 0 & 0  \\
\cdot & \cdot & \cdot & \cdot & \cdot  & \cdot & \cdot & \cdot  \\
\cdot & \cdot & \cdot  & \cdot & \cdot & \cdot & \cdot & \cdot  \\
\cdot & \cdot & \cdot & \cdot & \cdot  & \cdot & \cdot & \cdot  \\
\cdot & \cdot & \cdot  & \cdot & \cdot & \cdot & \cdot & \cdot  \\
0 & \cdot & \cdot  & \cdot & \cdot & R_{\rm n - 1} & S_{\rm n -1} & T_{\rm n -1} \\
0 & \cdot & \cdot  & \cdot & \cdot & \cdot & R_{\rm n} & S_{\rm n}
  \end{vmatrix}=0 \end{equation}
 \end{enumerate}

Conditions 1, 2, and 3, are also sufficient for the existence of polynomial solutions to equation (\ref{hautot}).
As pointed out by Hautot \cite{Hautot} the key difficulty in satisfying condition 3   is that the order  of the determinant increases
with the order n of the polynomial solution.


This difficulty, as expected,  appears not only when we try to solve the linear homogeneous system associated
with the differential equation \cite{Hautot} but also when we try to find polynomial solutions to equation (\ref{hautot}) with any of the standard
methods, eg., by studying the structure of the three$-$term recurrence relation (\ref{rr}),
by considering the  continued fraction associated to the three$-$term  recurrence relation (\ref{rr}), by
expanding  the solution of (\ref{hautot}) in terms of confluent hypergeometric functions and the like \cite{Ish,Ish1}, etc..

When $f=0$ equations (\ref{gfkkukku}) and (\ref{hautot}) fall into the confluent Heun class. It is worth noting that
there is a monograph on the Heun's differential equation and its various confluent forms \cite{He1}, and on page 118 of this monograph we read that the necessary and sufficient conditions for the existence of polynomial solutions to equation (\ref{hautot})
are the aforementioned conditions 1, 2, and 3.

Some of the subsequent work on the exact solutions of the Regge$-$Wheeler equation, see
e.g. \cite{r4,r5}, was based on this monograph and on the conditions  1, 2, and 3 which were also stated there. As a result, there is no overlap between that work and this
work which is not based on the conditions 1, 2, and 3, but it is based instead on the sufficient conditions found by Hautot \cite{Hautot} which are stated in the next subsection.



\subsection{Hautot's sufficient conditions  and      $\chi \int \frac {{\rm
d}r_{\!\ast}}{\chi^{2}}$ }
\label{sufficient}
\noindent

Hautot's sufficient conditions are based on the structure of the three$-$term recurrence relation (\ref{rr}) and
on an  elementary fact
about determinants which one can find in any introductory textbook on linear algebra.

If $\mathcal C$ is a square matrix of order $m+q$ and has the form
\begin{equation}
\label{hm}
  \begin{pmatrix}
    \mathcal A & \bf O \\
\mathcal D & \mathcal B
  \end{pmatrix},
\end{equation}
where $\mathcal A$ and $\mathcal B$ are square matrices of order $m$ and $q$ respectively,  $\bf O$ is a $ m \times q$ zero matrix, and $\mathcal D $ is a $ q \times m$  matrix, then
\be
\label{det}
det(\mathcal C)=det(\mathcal A) \cdot det(\mathcal B),
\ee
where, $det(\mathcal C)$, $det(\mathcal A)$, and $det(\mathcal B)$, denote respectively the determinants of the matrices $\mathcal C$, $\mathcal A$, and $\mathcal B$.

The crucial remark made by Hautot is that if $c=j \ \rm( =0, 1,2,...,fixed)$ then the coefficient $T_{j}$ of the three$-$term recurrence relation (\ref{rr}) vanishes, and
the square matrix $\mathcal M$ of order $\rm n +1$, whose determinant appears in condition 3, takes the form of the matrix $\mathcal C$, i.e.,
when $c=j \ \rm( =0, 1,2,...,fixed)$, we have
\begin{equation}
\mathcal M = \begin{pmatrix}
    \mathcal A & \bf O \\
\mathcal D & \mathcal B
  \end{pmatrix}.
\end{equation}
In this case $\mathcal A$, $\mathcal B$ are square matrices of order
$j$+1  and $\rm n- \it j$ respectively,
$\bf O$ is a ($j+1$) $\times$ ($\rm n- \it j$) zero matrix, and
$\mathcal D$ is a ($ \rm n - \it j$) $\times$ ($j+1$) matrix.

The matrices $\mathcal A$, $\mathcal B$ and $\mathcal D$ are given explicitly by
\begin{equation}
\hspace{-0.8cm}
\label{mat}
\mathcal A = \begin{pmatrix}
   S_{\rm 0} & T_{\rm 0} & 0 & \cdot & \cdot & 0 & 0 & 0  \\
R_{\rm 1} & S_{\rm 1} & T_{\rm 1} & \cdot & \cdot & 0 & 0 & 0  \\
\cdot & \cdot & \cdot & \cdot & \cdot  & \cdot & \cdot & \cdot  \\
\cdot & \cdot & \cdot  & \cdot & \cdot & \cdot & \cdot & \cdot  \\
\cdot & \cdot & \cdot & \cdot & \cdot  & \cdot & \cdot & \cdot  \\
\cdot & \cdot & \cdot  & \cdot & \cdot & \cdot & \cdot & \cdot  \\
0 & \cdot & \cdot  & \cdot & \cdot & R_{j-1} & S_{j-1} & T_{j-1} \\
0 & \cdot & \cdot  & \cdot & \cdot & \cdot & R_{j} & S_{j}
  \end{pmatrix},
  \mathcal B = \begin{pmatrix}
   S_{j+1 } & T_{j+1} & 0 & \cdot & \cdot  & 0 & 0 & 0  \\
R_{j+2} & S_{j+2} & T_{j+2} & \cdot & \cdot   & 0 & 0 & 0  \\
\cdot & \cdot & \cdot & \cdot & \cdot   & \cdot & \cdot & \cdot  \\
\cdot & \cdot & \cdot  & \cdot & \cdot   & \cdot & \cdot & \cdot  \\
\cdot & \cdot & \cdot  & \cdot & \cdot   & \cdot & \cdot & \cdot  \\
\cdot & \cdot & \cdot  & \cdot & \cdot   & \cdot & \cdot & \cdot  \\
0 & \cdot & \cdot  & \cdot & \cdot   & R_{\rm n-1} & S_{\rm n-1} & T_{\rm n-1} \\
0 & \cdot & \cdot  & \cdot & \cdot   & \cdot & R_{\rm n} & S_{\rm n}
  \end{pmatrix},
\end{equation}
and
\begin{equation}
\mathcal D = \begin{pmatrix}
   0 & 0 & 0 & \cdot & \cdot & 0 & 0 & R_{j+1}  \\
0 & 0 & 0 & \cdot & \cdot & 0 & 0 & 0  \\
\cdot & \cdot & \cdot & \cdot & \cdot & \cdot & \cdot & \cdot  \\
\cdot & \cdot & \cdot & \cdot & \cdot & \cdot & \cdot & \cdot  \\
0 & 0 & 0 & \cdot & \cdot & 0 & 0 & 0 \\
0 & 0 & 0 & \cdot & \cdot & 0 & 0 & 0
  \end{pmatrix}.
  \end{equation}
In accordance with equation (\ref{det})
\begin{equation}
\label{det1}
det(\mathcal M)=det(\mathcal A) \cdot det(\mathcal B),
\end{equation}
where $det(\mathcal M)$ is given in equation (\ref{con3}), and $(\mathcal A), \  (\mathcal B)$
are given in equation (\ref{mat}). \newline
\noindent

Equation (\ref{det1}), Hautot notices, suggests sufficient conditions for obtaining polynomial
solutions to equation (\ref{hautot}). In fact there are two    possibilities:

\noindent
\it{1st possibility}: \normalfont $det(\mathcal A)=0$, and consequently, $det(\mathcal M)=0$ and as a result equation (\ref{hautot})
has polynomial solutions.

\noindent
\it{2nd possibility}: \normalfont $det(\mathcal B)=0$, and consequently, $det(\mathcal M)=0$ and as a result equation (\ref{hautot})
has polynomial solutions. \newline
\noindent

A few remarks are now in order
\begin{itemize}
\item In the second possibility it is assumed that $det(\mathcal A) \neq 0$, and therefore it is assumed that
$\lambda_{1}=\lambda_{2}=...=\lambda_{j}=0$.

\item It is difficult in general to  pursue  the second possibility further because  $det(\mathcal B)$ shares the same main feature with  $\det(\mathcal M)$: the order  of $det(\mathcal B)$   increases with the order  of the polynomial solution to
    equation (\ref{hautot}); so we restrict our attention to the first possibility.

\item  In the case of the Liouvillian solution $\chi \int \frac {{\rm
d}r_{\!\ast}}{\chi^{2}}$   found by Chandrasekhar,
the first possibility  reduces to the requirement that the frequency of the gravitational
perturbations can only take the values of the frequencies of the algebraically special gravitational perturbations of the Schwarzschild geometry
initially found by Couch and Newman \cite{Couch}!

\item The difficulty in  obtaining the whole set of polynomial
solutions  to equation (\ref{hautot}) is brought again to the fore:
Even in the special case where $c=j$ it is difficult to obtain all the
polynomial solutions to equation (\ref{hautot}) since there are in general
such solutions which arise from the second possibility which cannot in general
be pursued further.


\end{itemize}

Hautot concludes that if we dismiss the second possibility, since it is difficult to be pursued further, and keep only the
first possibility we arrive at the following sufficient conditions for the existence of polynomial
solutions  to equation (\ref{hautot}):
\begin{enumerate}
\item \begin{equation}  \label{con4} \it f \rm = \rm 0  \end{equation}
\item   \begin{equation}  \label{con5} \it e = - a \rm{n} \end{equation}
\item \begin{equation}  \label{con6}  c=j \ \rm( =0, 1,2,...,fixed)  \end{equation}

\item \begin{equation}  \label{con7}  det(\mathcal A)  = \begin{vmatrix}
   S_{\rm 0} & T_{\rm 0} & 0 & \cdot & \cdot & 0 & 0 & 0  \\
R_{\rm 1} & S_{\rm 1} & T_{\rm 1} & \cdot & \cdot & 0 & 0 & 0  \\
\cdot & \cdot & \cdot & \cdot & \cdot  & \cdot & \cdot & \cdot  \\
\cdot & \cdot & \cdot  & \cdot & \cdot & \cdot & \cdot & \cdot  \\
\cdot & \cdot & \cdot & \cdot & \cdot  & \cdot & \cdot & \cdot  \\
\cdot & \cdot & \cdot  & \cdot & \cdot & \cdot & \cdot & \cdot  \\
0 & \cdot & \cdot  & \cdot & \cdot & R_{j-1} & S_{j-1} & T_{j-1} \\
0 & \cdot & \cdot  & \cdot & \cdot & \cdot & R_{j} & S_{j}
  \end{vmatrix}=0 \end{equation}
 \end{enumerate}

\it { Hautot's sufficient conditions, when satisfied, do not
give
the whole set of polynomial solutions to equation (\ref{hautot}), even in the special case $c=j$,
since in general there are polynomial solutions to equation (\ref{hautot}) which arise from the second possibility, $det ( \mathcal B ) =0$,  which is difficult in general to
be pursued further.
}   \normalfont
To put it differently, Hautot's sufficient conditions are not necessary, precisely because
there are in general polynomial solutions  to equation (\ref{hautot}) resulting from the second possibility,
$det ( \mathcal B ) =0$.

In subsection \ref{G7} we proved that the family G7 gives rise to the Liouvillian  solution $\chi \int \frac {{\rm
d}r_{\!\ast}}{\chi^{2}}$, found by Chandrasekhar\cite{Chandr} , which describes algebraically special perturbations of the Schwarzschild geometry
which excite only incoming or only outgoing waves. The frequencies $\sigma$  of these algebraically special perturbations can only take values in a
\it denumerable \normalfont infinite set\cite{Chandr,Couch}  (equation (\ref{nutnuta1}))
$$
{\rm
i}\sigma=
\frac{l(l-1)(l+1)(l+2)}{12}, \quad l=2,3,... \ . $$

In the case of the family G7,
 $\theta(r)$ and $\nu(r)$ are given respectively  by (\ref{gravity8})
and  (\ref{gorgor}), with $\beta=-3$, and equation (\ref{gfkkukku}) reads (equation(\ref{73}))
$$
r(r-2)  {\rm P}^{''}(r)    +   (6-2r-4rs+r^{2}s)  {\rm P}^{'}(r)    +   (2-l(l+1)+6s-r s(1+2s))    {\rm P}(r)=0,
$$
where $s=2{\rm i}\sigma$.
With a change of the independent variable from $r$ to $z=\frac{r}{2}$ the previous equation becomes
\begin{equation}
\label{ch}
z(z-1)  {\rm P}^{''}(z)    +   (2 s z^{2}  - 2 (2 s + 1) z  + 3)  {\rm P}^{'}(z)    +   (- 2 s (2s+1)z + 2 - l(l+1)  + 6s)    {\rm P}(z)=0,
\end{equation}
where now the differentiation is understood with respect to $z$.

This equation, which falls into the confluent Heun class,
has the same form as equation (\ref{hautot}) which is the object of study of
Hautot's paper \cite{Hautot} with
\begin{eqnarray}
f & = & 0,  \\
e & =& - a \rm{n} \normalfont  =   - 2   \it{s}   (\rm{2} \it{s}   + \rm{1}), \ \rm{and}  \normalfont \\
c &=& 3.
\end{eqnarray}

Moreover, the requirement
\begin{equation}  \label{con8}  det(\mathcal A)  = \begin{vmatrix}
S_{\rm 0} & T_{\rm 0} & 0 & 0  \\
R_{\rm 1} & S_{\rm 1} & T_{\rm 1} & 0  \\
0   & R_{2} & S_{2} & T_{2} \\
0  & 0 & R_{3} & S_{3}
  \end{vmatrix}=0, \end{equation}
where the coefficients $R_{k}, S_{\it k}, \ \rm and \  T_{\it k} $, $k=0,1,2,3$,
are the coefficients of  the three$-$term recurrence relation (\ref{rr}) and are given by equation
(\ref{coef}), reads
\begin{equation}  \label{con9}  det(\mathcal A)  = \begin{vmatrix}
2- l(l+1) + 6 s & 3 & 0 & 0  \\
- 2 s (2 s + 1) & 2 s - l (l+1) & 4 & 0  \\
0   & - 4 s^{2} & - 2 s - l (l+1) & 3 \\
0  & 0 & 2 s (1-2s) & 2 - 6 s - l (l+1)
\end{vmatrix}=0.
\end{equation}

Requirement (\ref{con9}) is equivalent to
\begin{equation}
\label{con10}
s=\underline{+} \frac{l(l-1)(l+1)(l+2)}{6} \Leftrightarrow  {\rm
i}\sigma= \underline{+}
\frac{l(l-1)(l+1)(l+2)}{12}, \quad l=2,3,... \ ,
\end{equation}
which are precisely the frequencies\cite{Chandr,Couch} of the algebraically special perturbations
of the Schwarzschild geometry! The frequencies in equation (\ref{con10}) appear with $+$ or $-$ sign.
To restore the ambiguity regarding the sign we note that
so far the frequencies of the algebraically special perturbations appear with the $+$ sign (equation (\ref{nutnuta1}))
because of the form of
the time dependence ${\rm exp} \left[-{\rm i} \sigma t \right]$ we have assumed in the Fourier decomposition
of the perturbing field $\Phi(t,r,\theta,\phi)$  at the beginning of our study, equation (\ref{Fourier}); perturbations
which become infinite as $t \rightarrow +\infty $ are not physically acceptable.

Hautot's sufficient conditions (\ref{con4}), (\ref{con5}), (\ref{con6}), and (\ref{con7}), when satisfied,  lead to a polynomial solution ${\rm P}_{\rm n}(z)=\sum_{k=0}^{\rm n} \lambda_{k} z^{k}$ of degree $\rm n$
to equation (\ref{hautot}).
The  coefficients of this polynomial solution
satisfy the three$-$term recurrence relation (\ref{rr}). Hautot carries on and proves
that when these
sufficient conditions
 are satisfied, the polynomial solution to equation (\ref{hautot}),
resulting from the sufficient conditions, can be expressed in ``closed$-$form''
as a finite sum of
truncated  confluent hypergeometric functions of the first kind. This is the content of  subsection \ref{truncated}.


In subsection \ref{G7},
in the
case of Chandrasekhar's Liouvillian solution $\chi \int \frac {{\rm
d}r_{\!\ast}}{\chi^{2}}$, we found
in ``closed$-$form''
the polynomial solution to the ODE
(\ref{73}), associated to the
family G7. As we said before,
equation (\ref{73}), after
a change of the independent variable from $r$ to $z=\frac{r}{2}$,  becomes equation (\ref{ch})
which falls into the confluent Heun class and
has the same form as equation (\ref{hautot}). 
This suggests that the ``closed$-$form'' polynomial solution, given by equations
(\ref{hutyytuh}), (\ref{hutyytuh1}) and (\ref{nugyygun}),
we
found  to the ODE
(\ref{73})
admits an expansion in terms of truncated confluent hypergeometric functions of the first kind.

In subsection we prove that this is precisely the case. However the proof involves
an unexpected twist: Hautot's results need to be
extended in order to prove that the polynomial solution to the ODE (\ref{hautot}) admits such an expansion. We give this extension
in subsection \ref{extension}.







\subsection{
Polynomial solutions to the confluent Heun equation
}

\label{truncated}


Hautot's sufficient conditions (\ref{con4}), (\ref{con5}), (\ref{con6}), and (\ref{con7}),
when
satisfied, ensure the existence of polynomial solutions to equation (\ref{hautot}).
The coefficients of these solutions satisfy the three$-$term  recurrence relation (\ref{rr}).
In general solutions to three$-$term  recurrence relations cannot be expressed in ``closed$-$form''.
Remarkably, Hautot\cite{Hautot} proved that the polynomial solutions to equation (\ref{hautot})
ensuing from the sufficient conditions he found, can be expressed in ``closed$-$form'' as a finite sum of
truncated  confluent hypergeometric functions.

This is an unexpected result and constitutes the
most  significant part of Hautot's nice paper\cite{Hautot}. We note in passing that several confluent hypergeometric expansions
of  the  solutions  of  the  confluent  Heun  equation have been considered recently in \cite{Ish}.
These expansions  differ from the expansion
given by Hautot  which is more appropriate
for our purposes.

The crucial remark which underlies Hautot's proof is the following.
When Hautot's sufficient conditions (\ref{con4}), (\ref{con5}), (\ref{con6}), and (\ref{con7}) are satisfied
then
equation (\ref{hautot}), where  $f=0$, $e=- a \rm n$, and $c=j$,  admits  polynomial solution ${\rm P}_{\rm n}(z)=\sum_{k=0}^{\rm n} \lambda_{k} z^{k}$ of degree $\rm n$,
and therefore yields
\be
\label{hautot1}
z(z-1) {\rm P}_{\rm n}^{''}(z) + (a z^{2}  + b z + j) {\rm P}_{\rm n}^{'}(z) + (d  - a {\rm n} \normalfont z ){\rm P}_{\rm n}(z)=0.
\ee
This equation, falls into the confluent Heun class and has two regular singular points at $z=0$ and at $z=1$
and one irregular singular point at $z=\infty$.
M\"{o}bius transformations  of the independent variable play a central role in the study
of linear ODEs with rational function coefficients (see e.g. \cite{Ma,El}).
As expected prominent role have the transformations which
interchange and/or shift the locations of the singular points.

Under
the
linear transformation
\be
u=a(1-z)
\ee
of the independent variable $z$ equation (\ref{hautot1}) becomes
\be
\label{con}
 (u-a)(u {\rm P}_{\rm n}^{''}(u) + ((a+b+j)-u){\rm P}_{\rm n}^{'}(u)+{\rm n}{\rm P}_{\rm n}(u))
-u j {\rm P}_{\rm n}^{'}(u) + d {\rm P}_{\rm n}(u)=0
\ee
where for simplicity we use  the same symbol P$_{\rm n}$(u) for the polynomial of degree n,
P$_{\rm n}$($\frac{a-u}{a}$).
\subsubsection{Polynomial solutions
in terms of truncated confluent hypergeometric functions of the first kind
}

\label{hyper}

Regarding equation (\ref{con}) we note that equation
\be
\label{confluent}
u y^{''}(u)   + (q-u) y^{'}(u) - e y(u)=0
\ee
is the confluent hypergeometric equation.
This equation has a regular singular point
at $u=0$ and an irregular singular point at $u=\infty.$ The roots $\rho_{1}, \rho_{2}$ of the  indicial equation associated to the regular singular point  $u=0$, are $\rho_{1}=0$ and $\rho_{2}=1- q.$

For future reference we note the following. According to Fuchs' Theorem
a basis of the space of solutions of equation (\ref{confluent}) in
the neighbourhood $|u|< \infty$  of  $u=0$
depends on $q$, and, when  $q$  is  not  an  integer  is  given by
\begin{eqnarray}
\label{sol}
 y_{1}(u)=\sum_{k=0}^{\infty} \frac{(e)_{k}}{(q)_{k}} \frac{u^{k}}{k!} \equiv F(e,q;u) \quad & \rm{and}  & \quad \normalfont  y_{2}(u) = u^{1-q} F(e+1-q,2-q;u),
\end{eqnarray}
\noindent
when  1-$q$  is  a positive  integer  is  given by
\begin{eqnarray}
\label{sol1a}
 y_{1}(u) & = &
 u^{1-q} F(e+1-q,2-q;u) \quad  \rm{and}   \\
\label{sol1b}
 y_{2}(u) & = & \sum_{\rm{n=0}}^{\infty} \rm {d}_{\rm n} {\it u}^{\rm n}       +  {\rm C} \normalfont \it{y}_{\rm{1}}(\it{u}) \rm {ln}\normalfont |\it{u}|, \
 \rm where \ \rm{C} \ may \ turn \ out \ to \ be\ zero,
\end{eqnarray}
\noindent
and finally, when $q$=1,  is  given by
\begin{eqnarray}
\label{sol2a}
 y_{1}(u) & = &
  F(e,1;u) \quad  \rm{and} \quad {\it y}_{2}(\it u)  =  \sum_{\rm{n=0}}^{\infty} \rm {e}_{\rm n} {\it u}^{\rm n}       +   \normalfont \it{y}_{\rm{1}}(\it{u}) \rm {ln}\normalfont |\it{u}|.
\end{eqnarray}

\noindent
It is the second case, when 1-$q$  is  a positive  integer, which is more relevant in subsection
\ref{extension}.



The symbol
$(w)_{k}$, is the  Pochammer's symbol, and is defined by
\begin{equation}
(w)_{k}=w (w+1) ... (w+k-1).
\end{equation}
The function $F(e,
q;u)$
is called confluent hypergeometric function of the first kind or Kummer's function of the first kind.
When $e=-\rm n$, $\rm n$  being a natural number,
$F(e,
q;u)$ is truncated and it reduces to a polynomial ${\rm P}_{\rm n}(u)=F(-\rm {n}$$,q;u)$  of degree $\rm n$
\be
\label{polsol}
{\rm P}_{\rm n}(u)=F(-\rm {n}, \it q;u)=\rm 1- \frac{\rm n}{\it q} {\it u} + \frac{\rm n (\rm n -1)} {\it q (\it q+\rm 1)} \frac{{\it u}^{\rm 2}}{\rm 2!}+...+\frac{(-1)^{\rm n} \rm n !}{\it q(\it q+\rm 1)...(\it q+\rm n -1)} \frac{{\it u}^{\rm n}}{\rm n!}.
\ee

Obviously equation (\ref{con}) is not satisfied by $\rm P_{\rm n}$$(u)=F(-\rm {n},$$a+b+j;u)$. However the possibility arises
that equation (\ref{con}) is satisfied by a linear combination of truncated confluent hypergeometric functions.
With this motivation Hautot \cite{Hautot} starts with the ansatz
\be
\label{sum}
\rm P_{\rm n}(\it u)= \sum_{\it k=\rm 0}^{j}A_{k}F(k-\rm {n},\it a+b+j;u).
\ee
By following Hautot we substitute $P_{\rm n}(\it u)$ into equation (\ref{con})  with  this sum of truncated confluent hypergeometric
functions of the first kind and we obtain
\begin{eqnarray}
&& (u-a)\left (  \sum_{\it k = \rm 0}^{j}A_{k} (   u F^{''}(k-\rm {n}) + (( \it a+b+j)-u)F^{'}(k-\rm {n})+(\rm n- \it k)F(\it k-\rm {n})) \right ) \nonumber \\
\label{hautot1a}
&& + \sum_{\it k = \rm 0}^{j}A_{k} \left ( -u j F^{'}(k-\rm {n}) + (\it d - \it a \it k) F(k-\rm {n}) + \it u \it k F(\it k-\rm {n}) \right )   =0
\end{eqnarray}
where for simplicity we write $F(k-\rm {n})$ instead of $F(k-\rm {n},\it a+b+j;u)$; the differentiation is with respect to $u$. The polynomial $F(\it k-\rm {n})$ satisfies the confluent hypergeometric equation $u F^{''}(k-\rm {n}) + (( \it a+b+j)-u)F^{'}(k-\rm {n})+(\rm n- \it k)F(\it k-\rm {n})=0$ and therefore the last equation yields
\be
\label{an}
\sum_{\it k = \rm 0}^{j}A_{k} \left ( -u j F^{'}(k-\rm {n}) + (\it d - \it a \it k) F(k-\rm {n}) + \it u \it k F(\it k-\rm {n}) \right )   =0.
\ee

By using the recurrence relations
\begin{eqnarray}
\label{recurrencerel1}
u  F^{'}(k-\rm {n}) &=& (\rm n - \it k) F(\it k-\rm {n}) + (\it k - \rm n  ) F(\it k-\rm {n}+1), \\
\label{recurrencerel2}
u F(\it k-\rm {n}) & = & (\it k - \rm n) F(\it k-\rm {n}+1) - ( 2 (\it k - \rm n) - (\it a+b+j))F(\it k-\rm {n}) \nonumber \\
&-& (\it a+b+j - (\it k - \rm n) )F(\it k-\rm {n}-1)
\end{eqnarray}
satisfied by the confluent hypergeometric functions of the first kind$F(\it k-\rm {n})$ (see e.g. \cite{Abr} p. 506, 507), equation (\ref{an}) yields
\begin{eqnarray}
\label{recrel}
\sum_{\it k = \rm 0}^{j}
&&\left \{ (k-1-j)(k-1-\rm n)  \it A_{\it k-\rm 1} + ( \it d- \it j \rm n + \it k (\it b+\rm 2\it j-\rm 2\it k+\rm 2 \rm n)) \it A_{\it k} \right. \nonumber \\
&& \left. +(\it k+\rm 1) (\it k+ \rm 1-\rm n - (\it a+b+j)) \it A_{\it k+\rm 1} \right \} F(\it k-\rm {n})=0.
\end{eqnarray}

The last condition can be satisfied for non zero values of the coefficients
$A_{k}, \ k=0,1,...,j,$ if and only if
\begin{equation}  \label{conhau}   \begin{vmatrix}
  \mathcal S_{\rm 0} & \mathcal T_{\rm 0} & 0 & \cdot & \cdot & 0 & 0 & 0  \\
\mathcal R_{\rm 1} & \mathcal S_{\rm 1} & \mathcal T_{\rm 1} & \cdot & \cdot & 0 & 0 & 0  \\
\cdot & \cdot & \cdot & \cdot & \cdot  & \cdot & \cdot & \cdot  \\
\cdot & \cdot & \cdot  & \cdot & \cdot & \cdot & \cdot & \cdot  \\
\cdot & \cdot & \cdot & \cdot & \cdot  & \cdot & \cdot & \cdot  \\
\cdot & \cdot & \cdot  & \cdot & \cdot & \cdot & \cdot & \cdot  \\
0 & \cdot & \cdot  & \cdot & \cdot & \mathcal  R_{j-1} & \mathcal S_{j-1} & \mathcal T_{j-1} \\
0 & \cdot & \cdot  & \cdot & \cdot & \cdot & \mathcal R_{j} & \mathcal S_{j}
  \end{vmatrix}=0, \end{equation}
where
\begin{eqnarray}
\label{coefshau}
\mathcal R_{k} & = & (k-1-j)(k-1-\rm n), \\
\mathcal  S_{\it k} & = & \it d - \it j \rm n + \it k(\it b + \rm 2 \it j - \rm 2 \it k + \rm 2 \rm n), \ \rm and, \\
\mathcal  T_{\it k} & = & (\it k+ \rm 1)(\it k+ \rm 1 - \rm n - \it a - \it b - \it j).
\end{eqnarray}

We conclude that when condition (\ref{conhau}) is satisfied then equation (\ref{con}), and therefore equation (\ref{hautot1}),
admits a polynomial solution of degree n. Therefore condition (\ref{conhau}) is equivalent to condition (\ref{con7}).
Indeed, it can be verified that for every value of $j=0,1,2,...,$
the following equation holds
\begin{equation}  \label{equality}   \begin{vmatrix}
   S_{\rm 0} & T_{\rm 0} & 0 & \cdot & \cdot & 0 & 0 & 0  \\
R_{\rm 1} & S_{\rm 1} & T_{\rm 1} & \cdot & \cdot & 0 & 0 & 0  \\
\cdot & \cdot & \cdot & \cdot & \cdot  & \cdot & \cdot & \cdot  \\
\cdot & \cdot & \cdot  & \cdot & \cdot & \cdot & \cdot & \cdot  \\
\cdot & \cdot & \cdot & \cdot & \cdot  & \cdot & \cdot & \cdot  \\
\cdot & \cdot & \cdot  & \cdot & \cdot & \cdot & \cdot & \cdot  \\
0 & \cdot & \cdot  & \cdot & \cdot & R_{j-1} & S_{j-1} & T_{j-1} \\
0 & \cdot & \cdot  & \cdot & \cdot & \cdot & R_{j} & S_{j}
  \end{vmatrix}=
\begin{vmatrix}
  \mathcal S_{\rm 0} & \mathcal T_{\rm 0} & 0 & \cdot & \cdot & 0 & 0 & 0  \\
\mathcal R_{\rm 1} & \mathcal S_{\rm 1} & \mathcal T_{\rm 1} & \cdot & \cdot & 0 & 0 & 0  \\
\cdot & \cdot & \cdot & \cdot & \cdot  & \cdot & \cdot & \cdot  \\
\cdot & \cdot & \cdot  & \cdot & \cdot & \cdot & \cdot & \cdot  \\
\cdot & \cdot & \cdot & \cdot & \cdot  & \cdot & \cdot & \cdot  \\
\cdot & \cdot & \cdot  & \cdot & \cdot & \cdot & \cdot & \cdot  \\
0 & \cdot & \cdot  & \cdot & \cdot & \mathcal  R_{j-1} & \mathcal S_{j-1} & \mathcal T_{j-1} \\
0 & \cdot & \cdot  & \cdot & \cdot & \cdot & \mathcal R_{j} & \mathcal S_{j}
  \end{vmatrix}. \end{equation}

The coefficients $A_{k}, \ k=0,1,...,j,$  in the expansion $
\rm P_{\rm n}(\it u)= \sum_{\it k=\rm 0}^{j}A_{k}F(k-\rm {n},\it a+b+j;u),
$ are determined up to a multiplicative constant,
since the polynomial $\rm P_{\rm n}(\it u)$ satisfies the linear \it homogeneous \normalfont
equation (\ref{con}).

\noindent
Two remarks are here in order
\begin{itemize}

\item The crucial element in Hautot's derivation is the following: The Polynomial  $P_{\rm n}(\it u)$
satisfies equation (\ref{con}), an equation with three singular points. The previous analysis shows that
$P_{\rm n}(\it u)$ admits an expansion
in terms of a finite sum of truncated confluent hypergeometric functions of the first kind which satisfy the confluent hypergeometric equation,
an equation with two singular points. As Hautot points out \cite{Hautot} it seems that the number of singularities has been diminished by one.
This phenomenon
is not accidental.
Hautot  does not give an explanation of it
\cite{Hautot,Hautot1,Hautot2,Hautot3}.
To the best of our knowledge the
first systematic study and explanation of this phenomenon was given by Craster and Shanin in \cite{Shanin}. Interestingly enough, a
recent study\cite{Gr} gives the
prerequisites for a a group theoretical explanation
of this phenomenon. We elaborate more
on this issue in section \ref{D}.


\item We give in some detail Hautot's derivation because in his paper it is incorrectly stated that
expansion of the Polynomial solution $P_{\rm n}(\it u)$ to equation (\ref{con}) into a finite sum
of  Laguerre polynomials  leads to equation (\ref{recrel}) and subsequently to condition (\ref{conhau}).
As the previous analysis shows equation (\ref{recrel}) and subsequently  condition (\ref{conhau})
are derived starting from the ansatz that  the Polynomial solution $P_{\rm n}(\it u)$ admits an expansion
in terms of a finite sum of truncated confluent hypergeometric functions of the first kindand \it not \normalfont in terms
of Laguerre polynomials.
\end{itemize}

\subsubsection{Polynomial solutions
in terms of associated Laguerre polynomials
}

\label{Laguerre}

True the previous analysis can be repeated by expanding $P_{\rm n}(\it u)$ into a finite sum of associated Laguerre
polynomials but then we are not led to equation (\ref{recrel}) and subsequently  to condition (\ref{conhau}) as Hautot
claims \cite{Hautot}.

Indeed starting with the ansatz
\be
\label{sum1}
\rm P_{\rm n}(\it u)= \sum_{\it k=\rm 0}^{j}B_{k}L_{\rm {n}- \it k}^{(\it a + b + j - \rm 1)}(u),
\ee
where $L_{\rm {n}- \it k}^{(\it a + b + j - \rm 1)}(u)$ is the associated Laguerre polynomial which satisfies the equation
\be
u y^{''}(u)   + (a + b + j -u) y^{'}(u) + (\rm {n}- \it k)  y(u)= \rm 0,
\ee
and substituting into  equation (\ref{con}) we obtain
\be
\label{an1}
\sum_{\it k = \rm 0}^{j}B_{k} \left ( -u j L^{'(\it a + b + j - \rm 1)}_{\rm {n}- \it k}(u)
+ (\it d - \it a \it k) L^{^(\it a + b + j - \rm 1)}_{\rm {n}- \it k}(u) + \it u \it k L^{^(\it a + b + j - \rm 1)}_{\rm {n}- \it k}(u) \right )   =0.
\ee

The associated Laguerre polynomials $L^{(\it a + b + j - \rm 1)}_{\rm {n}- \it k}(u)$ satisfy the recurrence relations
\begin{eqnarray}
\label{recrelation1}
u  L^{'(\it a + b + j - \rm 1)}_{\rm {n}- \it k}(u) &=& (\rm {n}- \it k)L^{(\it a + b + j - \rm 1)}_{\rm {n}- \it k}(u)
-(\rm {n}- \it k+ \it a + b + j - \rm 1 ) L^{(\it a + b + j - \rm 1)}_{\rm {n}- \it k}(u),
\\
\label{recrelation2}
u L^{(\it a + b + j - \rm 1)}_{\rm {n}- \it k}(u) & = & -  ( \rm n - \it k + \it a + b + j - \rm 1) L^{(\it a + b + j - \rm 1)}_{\rm {n}- \it k - \rm 1}(u) \nonumber \\ &+& ( 2 ( \rm n - \it k) + \it a+b+j) L^{(\it a + b + j - \rm 1)}_{\rm {n}- \it k}(u) \nonumber \\
&-& (\rm n - \it k + \rm 1 )L^{(\it a + b + j - \rm 1)}_{\rm {n}- \it k + \rm 1}(u).
\end{eqnarray}
We note that  associated Laguerre polynomials $L^{(\it a + b + j - \rm 1)}_{\rm {n}- \it k}(u)$ and  confluent hypergeometric functions
$F(\it k - \rm {n},\it a + b + j,u)$
are
related via
\be
\label{relation}
L^{(\it a + b + j - \rm 1)}_{\rm {n}- \it k}(u)= \left( {\begin{array}{*{20}c} \rm {n}- \it k + \it a + b + j - \rm 1 \\ \rm {n}- \it k \\ \end{array}} \right)
F(\it k - \rm {n},\it a + b + j,u),
\ee
where
\be
\left( {\begin{array}{*{20}c} N \\ K \\ \end{array}} \right)=\frac{N(N-1)(N-2)\cdot\cdot\cdot(N-K+1)}{K!}
\ee
is the binomial coefficient.

The recurrence relations (\ref{recrelation1}) and (\ref{recrelation2}) can be derived from the recurrence relations
(\ref{recurrencerel1}) and (\ref{recurrencerel2}) by using the relation (\ref{relation}).
With the use of the recurrence relations (\ref{recrelation1}) and (\ref{recrelation2}) equation (\ref{an1}) yields
\begin{eqnarray}
\label{recrel1}
\sum_{\it k = \rm 0}^{j}
&&\left \{ (k-1-j)(k-\rm n - \it a - \it b - \it j ) \it B_{\it k-\rm 1} + ( \it d- \it j \rm n + \it k (\it b+\rm 2\it j-\rm 2\it k+\rm 2 \rm n)) \it B_{\it k} \right. \nonumber \\
&& \left. +(\it k+\rm 1) (\it k -\rm n )\it  B_{\it k+\rm 1} \right \} L^{(\it a + b + j - \rm 1)}_{\rm {n}- \it k }(u)=0.
\end{eqnarray}

The last condition can be satisfied for non zero values of the coefficients
$B_{k}, \ k=0,1,...,j,$ if and only if
\begin{equation}  \label{conhau1}   \begin{vmatrix}
  \mathcal S_{\rm 0} & \mathcal W_{\rm 0} & 0 & \cdot & \cdot & 0 & 0 & 0  \\
\mathcal Q_{\rm 1} & \mathcal S_{\rm 1} & \mathcal W_{\rm 1} & \cdot & \cdot & 0 & 0 & 0  \\
\cdot & \cdot & \cdot & \cdot & \cdot  & \cdot & \cdot & \cdot  \\
\cdot & \cdot & \cdot  & \cdot & \cdot & \cdot & \cdot & \cdot  \\
\cdot & \cdot & \cdot & \cdot & \cdot  & \cdot & \cdot & \cdot  \\
\cdot & \cdot & \cdot  & \cdot & \cdot & \cdot & \cdot & \cdot  \\
0 & \cdot & \cdot  & \cdot & \cdot & \mathcal  Q_{j-1} & \mathcal S_{j-1} & \mathcal W_{j-1} \\
0 & \cdot & \cdot  & \cdot & \cdot & \cdot & \mathcal Q_{j} & \mathcal S_{j}
  \end{vmatrix}=0, \end{equation}
where
\begin{eqnarray}
\label{coefshau1}
\mathcal Q_{k} & = & (k-1-j)(k-\rm n - \it a - \it b - \it j), \\
\mathcal  S_{\it k} & = & \it d - \it j \rm n + \it k(\it b + \rm 2 \it j - \rm 2 \it k + \rm 2 \rm n), \ \rm and, \\
\mathcal  W_{\it k} & = & (\it k+ \rm 1)(\it k - \rm n).
\end{eqnarray}

We conclude that when condition (\ref{conhau1}) is satisfied then equation (\ref{con}), and therefore equation (\ref{hautot1}),
admits a polynomial solution of degree n. Therefore condition (\ref{conhau1}) is equivalent to condition (\ref{con7}).
As in the case of confluent hypergeometric functions, it can be verified that for every value of $j=0,1,2,...,$
the following equation holds
\begin{equation}  \label{equality}   \begin{vmatrix}
   S_{\rm 0} & T_{\rm 0} & 0 & \cdot & \cdot & 0 & 0 & 0  \\
R_{\rm 1} & S_{\rm 1} & T_{\rm 1} & \cdot & \cdot & 0 & 0 & 0  \\
\cdot & \cdot & \cdot & \cdot & \cdot  & \cdot & \cdot & \cdot  \\
\cdot & \cdot & \cdot  & \cdot & \cdot & \cdot & \cdot & \cdot  \\
\cdot & \cdot & \cdot & \cdot & \cdot  & \cdot & \cdot & \cdot  \\
\cdot & \cdot & \cdot  & \cdot & \cdot & \cdot & \cdot & \cdot  \\
0 & \cdot & \cdot  & \cdot & \cdot & R_{j-1} & S_{j-1} & T_{j-1} \\
0 & \cdot & \cdot  & \cdot & \cdot & \cdot & R_{j} & S_{j}
  \end{vmatrix}=
\begin{vmatrix}
  \mathcal S_{\rm 0} & \mathcal W_{\rm 0} & 0 & \cdot & \cdot & 0 & 0 & 0  \\
\mathcal Q_{\rm 1} & \mathcal S_{\rm 1} & \mathcal W_{\rm 1} & \cdot & \cdot & 0 & 0 & 0  \\
\cdot & \cdot & \cdot & \cdot & \cdot  & \cdot & \cdot & \cdot  \\
\cdot & \cdot & \cdot  & \cdot & \cdot & \cdot & \cdot & \cdot  \\
\cdot & \cdot & \cdot & \cdot & \cdot  & \cdot & \cdot & \cdot  \\
\cdot & \cdot & \cdot  & \cdot & \cdot & \cdot & \cdot & \cdot  \\
0 & \cdot & \cdot  & \cdot & \cdot & \mathcal  Q_{j-1} & \mathcal S_{j-1} & \mathcal W_{j-1} \\
0 & \cdot & \cdot  & \cdot & \cdot & \cdot & \mathcal Q_{j} & \mathcal S_{j}
  \end{vmatrix}. \end{equation}
The coefficients $B_{k}, \ k=0,1,...,j,$  in the expansion $
{\rm P}_{\rm n}(\it u)= \sum_{\it k=\rm 0}^{j}B_{k}L_{\rm {n}- \it k}^{(\it a + b + j - \rm 1)}(u),
$ are determined up to a multiplicative constant,
since
$\rm P_{\rm n}(\it u)$ satisfies the linear \it homogeneous \normalfont
equation (\ref{con}).

To summarize we conclude that equation (\ref{hautot1}),
$ z(z-1) {\rm P}_{\rm n}^{''}(z) + (a z^{2}  + b z + j) {\rm P}_{\rm n}^{'}(z) + (d  - a {\rm n} \normalfont z ){\rm P}_{\rm n}(z)=0 $,
when Hautot's sufficient condition (\ref{con7}) is satisfied, admits  polynomial solution  ${\rm P}_{\rm n}(z)$ of degree n, which
when it is written as a polynomial in the variable $u=a(1-z)$
admits an
expansion both in terms of a finite sum of truncated confluent hypergeometric functions of the first kindand in terms of a finite sum of associated Laguerre polynomials.


\section{
$\chi \int \frac {{\rm
d}r_{\!\ast}}{\chi^{2}}$ in terms of special
functions
}

\label{summ2}

In subsection \ref{G7} using Kovacic's algorithm we showed that Chandrasekhar's Liouvillian solution $\chi \int \frac {{\rm
d}r_{\!\ast}}{\chi^{2}}$ can be put into ``closed$-$form'' which involves a polynomial solution of
equation  (\ref{73}) which we were also able to give in
``closed$-$form''. Equation  (\ref{73}) after a change of the independent variable from $r$ to $z=\frac{r}{2}$
falls into the confluent Heun class, and in particular it
falls into the  form
(\ref{hautot1}).

This fact provides the
link between the results of subsection \ref{G7} and the results of the subsections \ref{hyper} and \ref{Laguerre}.
As a result we expect to be able to express the polynomial solution of equation  (\ref{73}), we found in
section  \ref{G7}, in terms of a finite sum of truncated confluent hypergeometric functions of the first kindin the form given in subsection \ref{hyper}
or in terms of a finite sum of
associated Laguerre polynomials
in the form given in subsection \ref{Laguerre}.

To our surprise when we try to verify that the ``closed$-$form'' polynomial solution to equation
(\ref{73}) admits such an expansion,
we encounter an obstruction,
since, as
 we show in subsection
\ref{extension},
two of the four
terms of the sum are not defined
in the case of  the Liouvillian solution
$\eta \int \frac { e^{-\int a}}{\eta ^{2}}$
initially found by Chanadrasekhar
\cite{Chandr}.

So it appears, that oddly enough,
although
in the case of the
the Liouvillian solution
$\eta \int \frac { e^{-\int a}}{\eta ^{2}}$,
initially found by Chanadrasekhar
\cite{Chandr},
Hautot's
sufficient conditions are satisfied, Hautot's  polynomial solution
to the associated confluent Heun equation does not arise in this case.

We resolve this apparent paradox
in subsections
\ref{extension}, \ref{truncon} and \ref{Laguerre1}, where
we introduce an extension of
Hautot's results \cite{Hautot,Hautot1,Hautot2,Hautot3}, and
use this extension
in order
to prove that the confluent Heun
equation, associated with Chandrasekhar's
Liouvillian solution
$\eta \int \frac { e^{-\int a}}{\eta ^{2}}$,
does admit polynomial solution which can be expressed both as a sum of truncated confluent hypergeometric functions of the first kind and as a sum of associated Laguerre polynomials, with appropriate function coefficients in each case. Both sums hold for every
value of the angular harmonic index $l=2,3,... \ .$



\subsection{$\chi \int \frac {{\rm
d}r_{\!\ast}}{\chi^{2}}$ and Hautot's results}
\label{hautotresults}

In  subsection \ref{G7} we showed that in the case of the gravitational perturbations of the
Schwarzschild geometry the family G7 gives rise to a liouvillian solution to the
equation
(\ref{adamsmith}) governing  the perturbations of the Schwarzschild geometry, namely to the solution
$\chi \int \frac {{\rm d}r_{\!\ast}}{\chi^{2}}$ initially found by Chandrasekhar\cite{Chandr}.

In particular we showed that this solution is equal to (equation (\ref{grasssop}))
$$
y'=\frac{{\rm
P}(r)e^{\frac{s}{2}r}}{r(r-2)^{s}},$$
where $\rm P(r)$ is a polynomial solution to equation (\ref{73})
$$
r(r-2) {\rm P}^{''}(r)    +  (6-2r-4rs+r^{2}s)    {\rm P}^{'}(r)    +  (2-l(l+1)+6s-r s(1+2s)){\rm P}(r)=0.
$$
With the change of the independent variable $r=w+2$ the last equation reads
\be
\label{cv}
 w(w+2){\rm  P}^{''}(w) + (2-2 w + s (- 4 + w^{2})) {\rm  P}^{'}(w)  + (2-l(l+1)+6s-(w+2) s(1+2s)) {\rm P}(w)=0.
\ee

In subsection \ref{G7} we proved  that the last equation admits
polynomial solutions
${\rm P}(w)=\sum_{{\rm n}=0}^{1+4\sigma_{0}}{\rm P}_{\rm
n}w^{\rm n}$ of degree $1+4\sigma_{0}$ where ${\rm P}_{\rm n}$ are given by
equations (\ref{hutyytuh}), (\ref{hutyytuh1}), and
(\ref{nugyygun})
\begin{eqnarray}
{\rm P}_{\rm n}&=&\frac
{3(-2\sigma_{0})^{{\rm
n}-4\sigma_{0}-1}(4\sigma_{0})!(\mu^{2}-6\sigma_{0}) \left [ ({\rm
n}-4\sigma_{0})\mu^{2}-12\sigma_{0} \right ] } {{\rm
n}!(\mu^{2}+12\sigma_{0})\sigma_{0}\mu^{6}},
\qquad 0\leq{\rm
n}\leq4\sigma_{0}-1,   \nonumber \\
{\rm
P}_{4\sigma_{0}} & = & \frac{\mu^{2}-3}{\sigma_{0}\mu^{4}}, \qquad
{\rm P}_{4\sigma_{0}+1}   =  \frac{1}{2\sigma_{0}\mu^{2}},  \nonumber
\end{eqnarray}
where $\mu^{2}$=($l$$-$1)($l$+2) (equation (\ref{frogfrog})) and
$s =2\sigma_{0}=\frac{l(l-1)(l+1)(l+2)}{6}, \quad l=2,3,... \ .$
\noindent

Equation  (\ref{73}) after a change of the independent variable from $r$ to $z=\frac{r}{2}$
falls into the confluent Heun class,  in particular it
 falls into  the form
(\ref{hautot1}), and it becomes equation (\ref{ch}) which reads
$$
z(z-1)  {\rm P}^{''}(z)    +   (2 s z^{2}  - 2 (2 s + 1) z  + 3)  {\rm P}^{'}(z)    +   (- 2 s (2s+1)z + 2 - l(l+1)  + 6s)    {\rm P}(z)=0.
$$

This fact provides the
link between the results of subsection \ref{G7} and the results of the subsections \ref{hyper} and \ref{Laguerre}.
With a new change of the independent variable from $z$ to $u=2 s$(1$-$$z$) equation (\ref{ch}) falls into the
form (\ref{con}) and in particular it becomes
\be
\label{cons}
u(u-2s) {\rm P}^{''}(u) +(-u^{2}- 2 u - 2 s(-2 s + 1)  ){\rm P}^{'}(u)+ ((2s+1)u  + 2 - l(l+1)  + 4s(1-s))    {\rm P}(u)=0.
\ee
The change of the independent variable $w$ to
\be
\label{change}
u = - s w \ee
transforms equation (\ref{cv}) into equation (\ref{cons}).

We conclude that we can restate what we proved in subsection \ref{G7} as follows: Equation (\ref{cons}) has
polynomial solution
${\rm P}(u)=\sum_{{\rm n}=0}^{1+2 s}{
P}_{\rm
n}u^{\rm n}$ of degree $1+2s$
with coefficients ${
P}_{\rm n}$ given by
\begin{eqnarray}
\label{coef1}
{
P}_{\rm n}&= - &\frac
{6 s^{- 2 s -1}(2 s)!((l-1)(l+2)- 3 s) \left [ ({\rm
n}- 2 s )(l-1)(l+2)- 6 s \right ] } {{\rm
n}!((l-1)(l+2)+ 6 s) s (l-1)^{3}(l+2)^{3}},
\\
\label{coef2}
{
P}_{2 s} & = & \frac{2 (l-1) (l+2) - 6}{ s^{2s+1} (l-1)^{2} (l+2)^{2}    }, \qquad
{
P}_{2 s + 1}   = -  \frac{1}{ s^{2s+2} (l-1)(l+2)},
\end{eqnarray}
where $0\leq{\rm n}\leq 2 s -1$ and $s=\frac{l(l-1)(l+1)(l+2)}{6}, \quad l=2,3,... \ .$

We note that the coefficients $P_{\rm n}$, $0\leq{\rm n}\leq 2 s +1$, are related to the
coefficients ${\rm P}_{\rm n}$, $0\leq{\rm n}\leq 2 s +1$,
given by
equations (\ref{hutyytuh}), (\ref{hutyytuh1}), and
(\ref{nugyygun}),
of the
polynomial solution
${\rm P}(w)=\sum_{{\rm n}=0}^{1+4\sigma_{0}}{\rm P}_{\rm
n}w^{\rm n}$, by
\be
\label{change1}
{ P}_{\rm n}=\frac{{\rm P}_{\rm n}}{(-s)^{\rm n}},
\quad 0\leq{\rm n}\leq 2 s +1.
\ee


As explained at the end of  subsection \ref{sufficient} and in
subsection  \ref{hyper}
Hautot proved that when
$s=\frac{l(l-1)(l+1)(l+2)}{6}, \quad l=2,3,...,$ equation (\ref{cons}) has
polynomial solution
${\rm P_{2s+1}}(u)$ of degree $2s+1$ which admits an expansion in terms of a finite sum of truncated  confluent hypergeometric functions of the first kind.
By comparing equations
(\ref{ch}) and (\ref{hautot1})  we obtain the values
of the parameters $a, b, j, \rm{n}$, \normalfont and   $d,$
\be
\label{par}
a=2s, \ b= -2(1+2s), \ j=3, \ \rm{n}=2\it{s}+\rm{1}, \  \it{d=\rm{2}-\it{l}(\it{l}+\rm{1})+\rm{6}\it{s}}.
\ee
This expansion is given
in the general case by equation (\ref{sum})
which now (equation (\ref{par}))   reads
\be
\label{expansion}
\rm P_{2 \it{s}+\rm{1}}(\it u)= \sum_{\it k=\rm 0}^{\rm {3}}A_{k}F(k-(\rm{2} \it{s} + \rm {1}), \rm {1} - \rm{2} \it{s} ; \it{u})
\ee
The constants $A_{k}$ are
determined, up to an arbitrary multiplicative constant, as solutions to
the 4 $\times$ 4 linear homogeneous system
resulting from equation (\ref{recrel}).

Thus we expect that the polynomial solution
${\rm P}(u)$ of degree $2 s + 1$ we found in subsection  \ref{G7} to equation (\ref{cons}),
given by equations
(\ref{coef1}) and (\ref{coef2}), admits expansion
of the form (\ref{expansion}) in terms of truncated
confluent hypergeometric functions of the first kind.
This brings us to the next subsection.



\subsection{Extension of Hautot's results}

\label{extension}

As soon as we try to apply Hautot's results
and form the sum
$\sum_{\it k=\rm 0}^{\rm {3}}A_{k}F(k-(\rm{2} \it{s} + \rm {1}), \rm {1} - \rm{2} \it{s} ; \it{u})$
we encounter an obstruction since
$F(k-(\rm{2} \it{s} + \rm {1}), \rm {1} - \rm{2} \it{s} ; \it{u})$ is not defined when $k=0$ and when $k=1.$
So it appears, that oddly enough,
although
in the case of the
the Liouvillian solution
$\eta \int \frac { e^{-\int a}}{\eta ^{2}}$,
initially found by Chanadrasekhar
\cite{Chandr},
Hautot's
sufficient conditions are satisfied, Hautot's  polynomial solution
to the associated confluent Heun equation does not arise in this case.

In order to resolve  this apparent paradox
we need to appropriately extend
Hautot's results \cite{Hautot,Hautot1,Hautot2,Hautot3}
and
use this extension
in order
to prove that the confluent Heun
equation, associated with Chandrasekhar's
Liouvillian solution
$\eta \int \frac { e^{-\int a}}{\eta ^{2}}$,
does admit polynomial solution which can be expressed  as a sum of truncated confluent hypergeometric functions of the first kind. This is the content of the
following Theorem.
\begin{thrm}
\label{Th1}
Equation (\ref{cons})
$$
u(u-2s) {\rm P}^{''}(u) +(-u^{2}- 2 u - 2 s(-2 s + 1)  ){\rm P}^{'}(u)+ ((2s+1)u  + 2 - l(l+1)  + 4s(1-s))    {\rm P}(u)=0
$$
admits a polynomial solution
$$
\rm P_{2 \it{s}+\rm{1}}(\it u)
= \sum_{\it k=\rm 0}^{\rm 1}A_{k} \phi((\rm 2 \it  s + \rm 1) - \it  k;u)+
\sum_{\it k=\rm 2}^{\rm 3}A_{k}F(k-(\rm 2 \it s + \rm 1),\rm 1 - \rm 2 \it s; \it u)
$$
of degree $2 s + 1$, if and only if
$$
s=\underline{+} \frac{l(l-1)(l+1)(l+2)}{6},
$$
where
\begin{eqnarray}
\phi(2 s + 1;u) & = & u^{2 s} F(-1, 2 s+1;u)=u^{2 s} \left (1-\frac{1}{2 s + 1} u  \right ), \nonumber  \\
\phi(2 s;u) & = & u^{2 s} F(0, 2 s+1;u)=u^{2 s}, \nonumber  \\
F(-(\rm 2 \it s - \rm 1),\rm 1 - \rm 2 \it s; \it u)&=& 1 + u +  \frac{{\it u }^{\rm 2}}{\rm 2!} +\frac{{\it u }^{\rm 3}}{\rm 3!}+ ... +  \frac{{\it u }^{\rm 2 \it s - \rm 1}}{(\rm 2 \it s - \rm 1)!}, \nonumber \\
F(-(\rm 2 \it s - \rm 2),\rm 1 - \rm 2 \it s; u)&=& 1 +    \frac{\rm 2 -\rm 2 \it s }{\rm 1 - \rm 2 \it s}   u     + \frac{\rm 3 - \rm 2 \it s }{\rm 1 - \rm 2 \it s}     \frac{{\it u }^{\rm 2}}{\rm 2!} +...+
\frac{\rm (- 1) }{\rm 1 - \rm 2 \it s}     \frac{{\it u}^{\rm 2 \it s - \rm 2}}{(\rm 2 \it s - \rm 2) !}, \nonumber
\end{eqnarray}
and  where,
\begin{eqnarray}
A_{1}= - \frac{l^{2}+l+1}{2 s +1}A_{0}, \ \
A_{2}= - \frac{ 3   (2s)!}{2 s + 1} A_{0}, \ \
A_{3}= - \frac{(l^{2}+l-3)  (2s)!}{2 s + 1} A_{0}, \nonumber
\end{eqnarray}
 $A_{0}$ is arbitrary.
\end{thrm}
\noindent \textbf{Proof} \hspace{0.1cm}

The confluent Heun
equation
associated with the family G7,
and consequently
associated with Chandrasekhar's
Liouvillian solution
$\eta \int \frac { e^{-\int a}}{\eta ^{2}}$,
is equation (\ref{ch}) which reads
$$
z(z-1)  {\rm P}^{''}(z)    +   (2 s z^{2}  - 2 (2 s + 1) z  + 3)  {\rm P}^{'}(z)    +   (- 2 s (2s+1)z + 2 - l(l+1)  + 6s)    {\rm P}(z)=0.
$$

As we stated in subsection \ref{hautotresults}
with a new change of the independent variable from $z$ to $u=2 s$(1$-$$z$) equation (\ref{ch})
becomes equation (\ref{cons}) which reads
$$
u(u-2s) {\rm P}^{''}(u) +(-u^{2}- 2 u - 2 s(-2 s + 1)  ){\rm P}^{'}(u)+ ((2s+1)u  + 2 - l(l+1)  + 4s(1-s))    {\rm P}(u)=0.
$$
We rewrite the last equation as follows (see equations (\ref{hautot1})     and (\ref{con}))
\be
\label{conaa}
 (u-2 s)(u {\rm P}^{''}(u) + (1-2 s-u){\rm P}^{'}(u)+(2 s + 1){\rm P}(u))
-3 u  {\rm P}^{'}(u) + (2 - l(l+1)  + 6 s) {\rm P}(u)=0.
\ee

Of crucial importance is the solution space of the
 confluent hypergeometric equation (equation (\ref{confluent}))
  \be
  \label{coneq}
  u y^{''}(u)   + (1-2 s-u) y^{'}(u) - e y(u)=0,
  \ee
$ e=k - (2 s + 1), \ \it{k}=\rm{0,1,2,3},$
which appears in equation (\ref{conaa}). Strictly speaking
in  (\ref{conaa}) appears only the confluent hypergeometric
equation (\ref{coneq}) with $\it{k}=\rm{0}$ but the confluent hypergeometric
equations with $\it{k}=\rm{1,2,3}$ are also going to be useful
in our proof of this Theorem (compare equations (\ref{confluent}), (\ref{sum}), and   (\ref{hautot1a})), so we include
them here.

The key observation is that the confluent hypergeometric equation  (\ref{coneq})
 admits a polynomial solution of
degree $2 s + 1$ when $k=0$ and a polynomial solution of
degree $2 s $ when $k=1$. We denote these solutions
by $\phi(2 s + 1;u)$ and $\phi(2 s;u)$ respectively.



This suggests, as a possible way out to the aforementioned obstruction, to replace
$F(-(\rm{2} \it{s} + \rm {1}), \rm {1} - \rm{2} \it{s} ; \it{u})$ and $F(-\rm{2} \it{s}, \rm {1} - \rm{2} \it{s} ; \it{u})$ in the sum
$\sum_{\it k=\rm 0}^{\rm {3}}A_{k}F(k-(\rm{2} \it{s} + \rm {1}), \rm {1} - \rm{2} \it{s} ; \it{u})$
by $\phi(2 s + 1;u)$ and $\phi(2 s;u)$ respectively.
The polynomial solutions $\phi(2 s;u)$ and $\phi(2 s + 1;u)$ are not truncated
confluent hypergeometric functions of the first
kind but they involve in their expressions such functions.

On the other hand the polynomials
$ \it{F}(-(\rm{2} \it{s} - \rm{1}), \rm{1} - \rm{2} \it{s}; u)$ and $ \it{F}(-(\rm{2} \it{s} - \rm{2}), \rm{1} - \rm{2} \it{s}; u)$
are truncated confluent hypergeometric functions of the first
kind defined by equation (\ref{polsol}), and are given explicitly by
\begin{eqnarray}
F(-(\rm 2 \it s - \rm 1),\rm 1 - \rm 2 \it s; \it u)&=& 1 + u +  \frac{{\it u }^{\rm 2}}{\rm 2!} +\frac{{\it u }^{\rm 3}}{\rm 3!}+ ... +  \frac{{\it u }^{\rm 2 \it s - \rm 1}}{(\rm 2 \it s - \rm 1)!},  \\
F(-(\rm 2 \it s - \rm 2),\rm 1 - \rm 2 \it s; u)&=& 1 +    \frac{\rm 2 -\rm 2 \it s }{\rm 1 - \rm 2 \it s}   u     + \frac{\rm 3 - \rm 2 \it s }{\rm 1 - \rm 2 \it s}     \frac{{\it u }^{\rm 2}}{\rm 2!} +...+
\frac{\rm (- 1) }{\rm 1 - \rm 2 \it s}     \frac{{\it u}^{\rm 2 \it s - \rm 2}}{(\rm 2 \it s - \rm 2) !}.
\end{eqnarray}

It is useful to examine how the four polynomial solutions $\phi(2 s + 1;u), \phi(2 s;u), \it{F}(-(\rm{2} \it{s} - \rm{1}), \rm{1} - \rm{2} \it{s}; u), \it{F}(-(\rm{2} \it{s} - \rm{2}), \rm{1} - \rm{2} \it{s}; u)$ to the confluent hypergeometric equation
$u y^{''}(u)   + (1 - 2 s -u) y^{'}(u) - e y(u)=0$, $ e=k - (2 s + 1), \ \it{k}=\rm{0,1,2,3},$
result from its two solutions $y_{1}(u)$, and $y_{2}(u)$, given by
equations (\ref{sol1a}) and (\ref{sol1b})
respectively.

It will
thus
be made apparent that the
confluent hypergeometric equation does not have
any other polynomial solutions in the corresponding cases  $e=-(2 s + 1), \  e=-2 s, \ e=-(2 s - 1)$, and
$e=-(2 s - 2)$, and as a result,
the uniqueness of the sum
$\sum_{\it k=\rm 0}^{\rm {3}}A_{k}F(k-(\rm{2} \it{s} + \rm {1}), \rm {1} - \rm{2} \it{s} ; \it{u})$ will
become evident.


We start with the polynomial solutions
$\phi(2 s + 1;u)$ and $\phi(2 s;u)$.
Since $1-q=2s$ is a positive integer,
according to equation  (\ref{sol1a}),
the confluent hypergeometric equation
 admits polynomial solutions
when $e=-(2 s + 1)$ ($k=0$) and
when $e=-2 s$ ($k=1$) which are given respectively by
\begin{eqnarray}
\label{soltrun1}
\phi(2 s + 1;u) & = & u^{2 s} F(-1, 2 s+1;u)=u^{2 s} \left (1-\frac{1}{2 s + 1} u  \right ), \ \rm{and} \normalfont \\
\label{soltrun2}
\phi(2 s;u) & = & u^{2 s} F(0, 2 s+1;u)=u^{2 s}.
\end{eqnarray}
It is worth mentioning that both polynomial
solutions are obtained from the first  solution $y_{1}(u)$, given by
equation (\ref{sol1a}), which in both cases
terminates
to give solutions  (\ref{soltrun1}) and
(\ref{soltrun2}).

In none of the two cases
the corresponding confluent hypergeometric equation 
admits another polynomial solution
since it is easy to check that in
neither of the two cases
the second linearly independent solution $y_{2}(u)$,
given by equation (\ref{sol1b}),
terminates
to a polynomial solution; in particular, in neither of the two cases the constant C which appears in
equation (\ref{sol1b}) is equal to zero.

Thus we have: \newline
\noindent
When  $e=-(2 s + 1)$ ($k=0$),
\begin{eqnarray}
y_{1}(u) &=& \phi(2 s + 1;u)  =  u^{2 s} F(-1, 2 s+1;u)=u^{2 s}    \left (1-\frac{1}{2 s + 1} u \right ), \ \rm{and}  \\
y_{2}(u) & = & \sum_{\rm{n=0}}^{\infty} \rm {f}_{\rm n} {\it u}^{\rm n}       +  {\rm C} \normalfont \it{y}_{\rm{1}}(\it{u}) \rm {ln} \normalfont |\it{u}|, \
 \rm where \ \rm{C}\neq 0.
\end{eqnarray}

\noindent
When  $e=-2 s$ ($k=1$),
\begin{eqnarray}
y_{1}(u) &=& \phi(2 s;u)  =  u^{2 s} F(0, 2 s+1;u)=u^{2 s}, \ \rm{and}  \\
y_{2}(u) & = & \sum_{\rm{n=0}}^{\infty} \rm {g}_{\rm n} {\it u}^{\rm n}       +  {\rm C} \normalfont \it{y}_{\rm{1}}(\it{u}) \rm {ln} \normalfont |\it{u}|, \
 \rm where \ \rm{C}\neq 0.
\end{eqnarray}

Moreover, the confluent hypergeometric equation
admits polynomial solutions also
in the two remaining cases, i.e., when
$e=-(2 s - 1)$ ($k=2$), and
when $e=-(2 s - 2)$ ($k=3$), which are given respectively by
\begin{eqnarray}
F(-(2 s - 1), 1 - 2 s; u) \ \rm{and} \normalfont \
\it{F}\normalfont (-(\rm{2} \it{s}\normalfont - \rm{2}), 1 - 2 \it{s}; u).
\end{eqnarray}
Both of these polynomial solutions result now from
the second linearly independent solution $y_{2}(u)$
(equation (\ref{sol1b})).

In particular we have: \newline
\noindent
When  $e=-(2 s - 1)$ ($k=2$),
\begin{eqnarray}
y_{1}(u) &=&
u^{2 s} F(1, 2 s+1;u) \ \rm{which} \
\rm{does} \ \rm{not} \ \rm{terminate},
\normalfont
\ \rm{and},  \\
y_{2}(u) & = & \sum_{\rm{n=0}}^{\infty} \rm {h}_{\rm n} {\it u}^{\rm n} = \rm {h}_{\rm 0}
\it{F}(-(\rm{2} \it{s} - \rm{1}), \rm{1} - \rm{2} \it{s}; u) + \rm {h}_{2 \it{s}} \it{y}_{\rm{1}}(\it{u}),
 \\
 && \rm {where} \ \rm {h}_{\rm 0}, \ \rm {h}_{2 \it{s}} \
   \rm{arbitrary} \ \rm{and} \ \rm{C}= 0. \nonumber
\end{eqnarray}
Finally when $e=-(2 s - 2)$ ($k=3$),
\begin{eqnarray}
y_{1}(u) &=&
u^{2 s} F(2, 2 s+1;u) \ \rm{which} \
\rm{does} \ \rm{not} \ \rm{terminate},
\normalfont
\ \rm{and},  \\
y_{2}(u) & = & \sum_{\rm{n=0}}^{\infty} \rm {j}_{\rm n} {\it u}^{\rm n} = \rm {j}_{\rm 0}
\it{F}(-(\rm{2} \it{s} - \rm{2}), \rm{1} - \rm{2} \it{s}; u) + \rm {j}_{2 \it{s}} \it{y}_{\rm{1}}(\it{u}),
 \\
 && \rm {where} \ \rm {j}_{\rm 0}, \ \rm {j}_{2 \it{s}} \
   \rm{arbitrary} \ \rm{and} \ \rm{C}= 0. \nonumber
\end{eqnarray}


We showed how the four polynomial solutions $\phi(2 s + 1;u), \phi(2 s;u), \it{F}(-  (\rm{2} \it{s} - \rm{1}), \rm{1} - \rm{2} \it{s}; u), \it{F}(-
(\rm{2} \it{s} - \rm{2}), \rm{1} - \rm{2} \it{s}; u)$ to the confluent hypergeometric equation
result from its two solutions $y_{1}(u)$ and $y_{2}(u)$, given by
equations (\ref{sol1a}) and (\ref{sol1b})
respectively. We follow through now Hautot's
analysis by substituting
$F(-(\rm{2} \it{s} + \rm {1}), \rm {1} - \rm{2} \it{s} ; \it{u})$ and $F(-\rm{2} \it{s}, \rm {1} - \rm{2} \it{s} ; \it{u})$ in the sum
$\sum_{\it k=\rm 0}^{\rm {3}}A_{k}F(k-(\rm{2} \it{s} + \rm {1}), \rm {1} - \rm{2} \it{s} ; \it{u})$
with $\phi(2 s + 1;u)$ and $\phi(2 s;u)$ respectively.


We thus substitute
\be
\label{exp}
\rm P_{2 \it{s}+\rm{1}}(\it u)
= \sum_{\it k=\rm 0}^{\rm 1}A_{k} \phi((\rm 2 \it  s + \rm 1) - \it  k;u)+
\sum_{\it k=\rm 2}^{\rm 3}A_{k}F(k-(\rm 2 \it s + \rm 1),\rm 1 - \rm 2 \it s; \it u)
\ee
into equation (\ref{conaa}) and we obtain
\begin{eqnarray}
&& \hspace{-0.6cm} \sum_{\it k = \rm 0}^{1}A_{k} \left ( - 3 u  
\phi^{'}((\rm 2 \it  s + \rm 1) - \it  k )
+ (\it d - \rm 2 \normalfont  \it s \it k) \phi((\rm 2 \it  s + \rm 1) - \it  k )  + \it u \it k \phi((\rm 2 \it  s + \rm 1) - \it  k ) \right )   + \nonumber
\\
\label{34}
&& \hspace{-0.6cm} \sum_{\it k = \rm 2}^{3}A_{k} \left ( - 3 u  F^{'}(k-(\rm 2 \it  s + \rm 1)) + (\it d - \rm 2 \it s \it k) F(k-(\rm 2 \it  s + \rm 1)) + \it u \it k F(\it k-(\rm 2 \it  s + \rm 1)) \right )=0,
\end{eqnarray}
where $d=2 - l(l+1) + 6 s$, and    where for simplicity, in the rest of
this subsection, we write
$\phi((\rm 2 \it  s + \rm 1) - \it  k)$
instead of $\phi((\rm 2 \it  s + \rm 1) - \it  k;u),$
and $F(k-(\rm 2 \it  s + \rm 1))$ instead of $F(k-(\rm 2 \it  s + \rm 1), 1 - 2 \it s; \it u)$.

By using appropriate
recurrence relations for
the four polynomial solutions
$\phi(2 s + 1), \phi(2 s), \it{F}(-  (\rm{2} \it{s} - \rm{1})), \it{F}(-(\rm{2} \it{s} - \rm{2}))$ the last equation yields a $4 \times 4$ homogeneous system for the coefficients $A_{k}, \ k=0,1,2,3.$
We give
in turn for the various values of $k$ the appropriate recurrence relations. These are the same with the recurrence relations
 (\ref{recurrencerel1}) and (\ref{recurrencerel2})
only when
$k=3,$ which, for completeness,
we also give explicitly.

We
have,  when $k=0$
\begin{eqnarray}
\label{recrel2}
u  \phi^{'}(2 s + 1) &=& (2 s + 1) \phi(2 s + 1) - \phi(2 s), \\
\label{recrel3}
u  \phi(2 s + 1) & = & - \phi(2 s) + ( 2 s + 3)  \phi(2 s + 1) - (2 s + 2) \phi(2 s + 2),
\end{eqnarray}
when $k=1$
\begin{eqnarray}
\label{recrel4}
u  \phi^{'}(2 s) &=& 2 s  \phi(2 s), \\
\label{recrel5}
u  \phi(2 s) & = & (2 s + 1) \phi(2 s) - ( 2 s + 1)  \phi(2 s + 1),
\end{eqnarray}
when $k=2$
\begin{eqnarray}
\label{recrel6}
u  F^{'}(-(2 s - 1)) &=& (2 s - 1) F(-(2 s - 1)) - (2 s - 1 ) F(-(2 s - 2)), \\
\label{recrel7}
u  F(-(2 s - 1)) & = & -(2 s - 1) F(-(2 s - 2)) - ( 1 - 2 s)F(-(2 s - 1)) + \nonumber \\
&+&  \frac{1}{(2 s - 1)!} \phi(2 s), \ \rm and
\end{eqnarray}
when $k=3$,
\begin{eqnarray}
\label{recrel8}
u  F^{'}(-(2 s - 2)) &=& (2 s - 2) F(-(2 s - 2)) - (2 s - 2 ) F(-(2 s - 3)), \\
\label{recrel9}
u  F(-(2 s - 2)) & = & -(2 s - 2) F(-(2 s - 3)) + ( 2 s- 3)F(-(2 s - 2)) + \nonumber \\
&+& F(-(2 s - 1)).
\end{eqnarray}

By substituting  recurrence relations
(\ref{recrel2})$-$(\ref{recrel9})
into equation (\ref{34}) we obtain the following
$4 \times 4$ homogeneous system for the coefficients $A_{k}, \ k=0,1,2,3,$
\begin{eqnarray}
\label{sys1}
-(1+l+l^{2})A_{0}-(2 s + 1)A_{1}&=&0  \\
\label{sys2}
3 A_{0} + (3 - l(l+1)) A_{1} + \frac{2}{(2 s - 1)!}A_{2} &=& 0  \\
\label{sys3}
(3 - l(l+1))A_{2} + 3 A_{3} &=& 0  \\
\label{sys4}
(2 s - 1)A_{2} + (-1-l(l+1)) A_{3} &=& 0.
\end{eqnarray}
This system has non zero solutions
$A_{k}, \ k=0,1,2,3,$
if and only if
\begin{equation}
\begin{vmatrix}
-(1+l+l^{2}) & -(2 s + 1) & 0 & 0  \\
3 & 3 - l(l+1) & \frac{2}{(2 s - 1)!} & 0  \\
0   & 0 & 3 - l(l+1) & 3 \\
0  & 0 & 2 s - 1 & -1-l(l+1)
\end{vmatrix}=0.
\end{equation}

The last condition gives
\begin{equation}
\label{tersol}
s=\underline{+} \frac{l(l-1)(l+1)(l+2)}{6}, 
\quad l=2,3,...  ,
\end{equation}
which are precisely the frequencies\cite{Chandr,Couch} of the algebraically special perturbations
of the Schwarzschild geometry!; with the $+$
 sign, in accord  with the convention we follow
here for the time dependence of the perturbing field
in its Fourier decomposition (equation (\ref{Fourier})).

When condition  (\ref{tersol}) is satisfied
the non zero solutions
 $A_{k}, \ k=0,1,2,3,$ of the system of
 equations  (\ref{sys1})$-$(\ref{sys4}) is given by
\begin{eqnarray}
\label{coefs}
A_{1}= - \frac{l^{2}+l+1}{2 s +1}A_{0}, \ \
A_{2}= - \frac{ 3   (2s)!}{2 s + 1} A_{0}, \ \
A_{3}= - \frac{(l^{2}+l-3)  (2s)!}{2 s + 1} A_{0},
\end{eqnarray}
where $A_{0}$ is arbitrary and
where it is understood that $s= \frac{l(l-1)(l+1)(l+2)}{6}$.
This completes the proof.


\subsection{$\chi \int \frac {{\rm
d}r_{\!\ast}}{\chi^{2}}$ in terms of truncated confluent hypergeometric functions of the first kind}

\label{truncon}

Theorem \ref{Th1}
very strongly suggests that the
polynomial solution
${\rm P}(u)$ of degree $2 s + 1$ we found in subsection  \ref{G7} to equation (\ref{cons}),
given by equations
(\ref{coef1}) and (\ref{coef2}), does admit
expansion
of the form (\ref{exp}),
in terms of truncated
confluent hypergeometric functions of the first kind. This in turn implies that the polynomial
$\rm P (\it w)$  which appears in the second
Liouvillian solution
$\chi \int \frac {{\rm
d}r_{\!\ast}}{\chi^{2}},$
given by the family
G7,
can be expressed as
a sum of truncated confluent hypergeometric functions of the first kind.
It turns out that this is indeed the case, and it is the content of the following Theorem.

\begin{thrm}
\label{Th2}

The second Liouvillian solution $\chi \int \frac {{\rm d}r_{\!\ast}}{\chi^{2}}$  to the master
equation (\ref{adamsmith}), given by the family G7, in the case of the gravitational perturbations of the Schwarzschild geometry,
is a product of elementary functions, one
of them being a polynomial $\rm P(\it w)$
which admits an expansion in terms of
truncated confluent hypergeometric functions
of the first kind
with appropriate
function coefficients. In particular the
following holds
\begin{eqnarray}
 \chi \int \frac {{\rm
d}r_{\!\ast}}{\chi^{2}} & = & \frac{{\rm
P}(w)e^{\frac{s}{2}r}}{r(r-2)^{s}} = \nonumber \\
  & = &
 \frac{e^{\frac{s}{2}r}}{r(r-2)^{s}} \left (
\sum_{\it k=\rm 0}^{\rm 1}A_{k} \phi((\rm 2 \it  s + \rm 1) - \it  k; - s w)+
\sum_{\it k=\rm 2}^{\rm 3}A_{k}F(k-(\rm 2 \it s + \rm 1),\rm 1 - \rm 2 \it s; \it - s w) \right ), \nonumber
\end{eqnarray}
where $w=r-2,$
\begin{eqnarray}
\phi(\rm 2 \it  s + \rm 1; - \it s w) &=&    (- s w)^{2 s} F(-1, 2 s+1; - s w)=(s w)^{2 s}    \left (1+\frac{s}{2 s + 1}  w \right ),  \nonumber \\
\phi(\rm 2 \it  s; - \it s w) &=& (- s w)^{2 s}  F(0, 2 s+1; - s w)=(s w)^{2 s},  \nonumber \\
F(-(\rm 2 \it s - \rm 1),\rm 1 - \rm 2 \it s; \it - s w)&=&
1 - s w +  \frac{{\it ( s w)}^{\rm 2}}{\rm 2!} - \frac{{\it ( s w)}^{\rm 3}}{\rm 3!}+ ... -  \frac{{\it ( s w)}^{\rm 2 \it s - \rm 1}}{(\rm 2 \it s - \rm 1)!}, \nonumber \\
F(-(\rm 2 \it s - \rm 2),\rm 1 - \rm 2 \it s; \it - s w)&=&
1 -    \frac{\rm 2 -\rm 2 \it s }{\rm 1 - \rm 2 \it s}   s w + \frac{\rm 3 - \rm 2 \it s }{\rm 1 - \rm 2 \it s}     \frac{{\it ( s w)}^{\rm 2}}{\rm 2!} +...+
 \frac{{\it ( s w)}^{\rm 2 \it s - \rm 2}}{(\rm 2 \it s - \rm 1) !}, \nonumber
\end{eqnarray}
\begin{eqnarray}
\label{coefs}
A_{1}= - \frac{l^{2}+l+1}{2 s +1}A_{0}, \
A_{2}= - \frac{ 3   (2s)!}{2 s + 1} A_{0}, \
A_{3}= - \frac{(l^{2}+l-3)  (2s)!}{2 s + 1} A_{0}, \ A_{0}=\frac{s^{-2-2 s}(1+ 2 s)}{(l-1)(l+2)}, \nonumber
\end{eqnarray}
where the coefficients ${\rm P}_{\rm
n}$ of the polynomial
${\rm P}(w)=\sum_{{\rm n}=0}^{1+
2 s
}{\rm P}_{\rm
n}w^{\rm n}$,
of degree $2 s + 1$,
are given by
equations (\ref{hutyytuh}), (\ref{hutyytuh1}), and
(\ref{nugyygun}), and where $s=\frac{l(l-1)(l+1)(l+2)}{6}, \ l=2,3,...  \ .$
\end{thrm}

\noindent \textbf{Proof} \hspace{0.1cm}

We can easily verify that if we choose \be \label{coefs1} A_{0}=\frac{s^{-2-2 s}(1+ 2 s)}{(l-1)(l+2)} \ee
then the expansion (equation (\ref{exp}))
$$
\rm P_{2 \it{s}+\rm{1}}(\it u)
= \sum_{\it k=\rm 0}^{\rm 1}A_{k} \phi((\rm 2 \it  s + \rm 1) - \it  k;u)+
\sum_{\it k=\rm 2}^{\rm 3}A_{k}F(k-(\rm 2 \it s + \rm 1),\rm 1 - \rm 2 \it s; \it u)
$$
indeed holds, where $A_{1},A_{2},A_{3}$ are given
by equation (\ref{coefs}), and
\be
\rm P_{\rm 2 \it{s}+\rm{1}}(\it u)=\sum_{\rm n=\rm 0}^{\rm 2 \it{s}+\rm{1}} {
P}_{\rm n}   u^{\rm n},
\ee
where the coefficients ${
P}_{\rm n}$ are given by equations
(\ref{coef1}) and (\ref{coef2}), and where it is
understood that $s= \frac{l(l-1)(l+1)(l+2)}{6}$.

$\rm P_{\rm 2 \it{s}+\rm{1}}(\it u)$ is a polynomial
solution to equation (\ref{cons}).  There is an
associated polynomial solution
${\rm P}(w)=\sum_{{\rm n}=0}^{1+
2 s
}{\rm P}_{\rm
n}w^{\rm n}$,
of degree $2 s + 1$,
to equation
(\ref{cv}), with coefficients
${\rm P}_{\rm n}$ given by
equations (\ref{hutyytuh}), (\ref{hutyytuh1}), and
(\ref{nugyygun}).
From
$u = - s w$ and
$
{ P}_{\rm n}=\frac{{\rm P}_{\rm n}}{(-s)^{\rm n}}, \
 0\leq{\rm n}\leq 2 s +1,
$
(equations (\ref{change}) and (\ref{change1})
correspondingly),
we conclude that $\rm P_{\rm 2 \it{s}+\rm{1}}(\it u)$ and ${\rm P}(w)$ are equal when they are
evaluated at $u=-s w$ and $ w$ respectively.
Indeed,
\be
\label{tel}
\rm P_{\rm 2 \it{s}+\rm{1}}(\it u)=\sum_{\rm n=\rm 0}^{\rm 2 \it{s}+\rm{1}} {
P}_{\rm n}   u^{\rm n}=\sum_{\rm n=\rm 0}^{\rm 2 \it{s}+\rm{1}} \frac{{\rm P}_{\rm n}}{(-s)^{\rm n}}
(-s)^{\rm n}
w^{\rm n}= \sum_{\rm n=\rm 0}^{\rm 2 \it{s}+\rm{1}}
{\rm P}_{\rm n}
w^{\rm n}={\rm P}(w).
\ee

Equations (\ref{exp}) and (\ref{tel})
yield
\begin{eqnarray}
{\rm P}(w) & = & \sum_{\rm n=\rm 0}^{\rm 2 \it{s}+\rm{1}}
{\rm P}_{\rm n}
w^{\rm n} = \nonumber \\
\label{tel1}
& = &  \sum_{\it k=\rm 0}^{\rm 1}A_{k} \phi((\rm 2 \it  s + \rm 1) - \it  k; - s w)+
\sum_{\it k=\rm 2}^{\rm 3}A_{k}F(k-(\rm 2 \it s + \rm 1),\rm 1 - \rm 2 \it s; \it - s w).
\end{eqnarray}

${\rm P}(w)$ is precisely the polynomial which appears in the second Liouvillian solution   $\chi \int \frac {{\rm
d}r_{\!\ast}}{\chi^{2}},$
initially found by Chandrasekhar\cite{Chandr},
to the RWE (RWE is the master equation
(\ref{adamsmith}) when $\beta=-3$),
which describes the first order
gravitational
perturbations of the
Schwarzschild geometry.
Equations (\ref{grasssop}), (\ref{derossored}) and
(\ref{tel1}) give
\begin{eqnarray}
 \chi \int \frac {{\rm
d}r_{\!\ast}}{\chi^{2}} & = & \frac{{\rm
P}(w)e^{\frac{s}{2}r}}{r(r-2)^{s}} = \nonumber \\
\label{interms}
  & = &
 \frac{e^{\frac{s}{2}r}}{r(r-2)^{s}} \left (
\sum_{\it k=\rm 0}^{\rm 1}A_{k} \phi((\rm 2 \it  s + \rm 1) - \it  k; - s w)+
\sum_{\it k=\rm 2}^{\rm 3}A_{k}F(k-(\rm 2 \it s + \rm 1),\rm 1 - \rm 2 \it s; \it - s w) \right ), \nonumber \\
&&
\end{eqnarray}
where $w=r-2.$ From equations (\ref{polsol}), (\ref{soltrun1}), and (\ref{soltrun2}), we have
\begin{eqnarray}
\label{pol1}
\phi(\rm 2 \it  s + \rm 1; - \it s w) &=&    (- s w)^{2 s} F(-1, 2 s+1; - s w)=(-s w)^{2 s}    \left (1-\frac{1}{2 s + 1} (- s w) \right ) = \nonumber \\
&=& (s w)^{2 s}    \left (1+\frac{s}{2 s + 1}  w \right ), \\
\label{pol2}
\phi(\rm 2 \it  s; - \it s w) &=& (- s w)^{2 s}  F(0, 2 s+1; - s w)=(s w)^{2 s},  \\
\label{pol3}
F(-(\rm 2 \it s - \rm 1),\rm 1 - \rm 2 \it s; \it - s w)&=& 1 + (- s w) +  \frac{{\it (- s w)}^{\rm 2}}{\rm 2!} +\frac{{\it (- s w)}^{\rm 3}}{\rm 3!}+ ... +  \frac{{\it (- s w)}^{\rm 2 \it s - \rm 1}}{(\rm 2 \it s - \rm 1)!}=
\nonumber \\
&=& 1 - s w +  \frac{{\it ( s w)}^{\rm 2}}{\rm 2!} - \frac{{\it ( s w)}^{\rm 3}}{\rm 3!}+ ... -  \frac{{\it ( s w)}^{\rm 2 \it s - \rm 1}}{(\rm 2 \it s - \rm 1)!}, \\
\label{pol4}
F(-(\rm 2 \it s - \rm 2),\rm 1 - \rm 2 \it s; \it - s w)&=& 1 +    \frac{\rm 2 -\rm 2 \it s }{\rm 1 - \rm 2 \it s}   (- s w) + \frac{\rm 3 - \rm 2 \it s }{\rm 1 - \rm 2 \it s}     \frac{{\it (- s w)}^{\rm 2}}{\rm 2!} +...+
\frac{\rm (- 1) }{\rm 1 - \rm 2 \it s}     \frac{{\it (- s w)}^{\rm 2 \it s - \rm 2}}{(\rm 2 \it s - \rm 2) !}=
\nonumber \\
&=& 1 -    \frac{\rm 2 -\rm 2 \it s }{\rm 1 - \rm 2 \it s}   s w + \frac{\rm 3 - \rm 2 \it s }{\rm 1 - \rm 2 \it s}     \frac{{\it ( s w)}^{\rm 2}}{\rm 2!} +...+
 \frac{{\it ( s w)}^{\rm 2 \it s - \rm 2}}{(\rm 2 \it s - \rm 1) !},
\end{eqnarray}
and where,  $A_{0}$, and   $A_{1}, \ A_{2}, \ A_{3}$ are given respectively by equations (\ref{coefs1}) and (\ref{coefs}). Moreover, it is understood that
$s= \frac{l(l-1)(l+1)(l+2)}{6}, \ l=2,3,... \ .$
This completes the proof.

In subsections
 \ref{G7} and
 \ref{elementary}, by using Kovacic's algorithm, we showed that the
second Liouvillian solution $\chi \int \frac {{\rm
d}r_{\!\ast}}{\chi^{2}}$ has an elementary functions answer for \it every \normalfont $l=2,3,... \ .$
In this subsection we go one step further and
combine the results, derived in subsections \ref{G7} and \ref{elementary}, from applying Kovacic's algorithm
to the RWE,
with the extension of Hautot's results we obtained in subsection
\ref{extension}, and
conclude, in equation (\ref{interms}),
that  the polynomial ${\rm P}(w)$
appearing in the
elementary functions answer of $\chi \int \frac {{\rm
d}r_{\!\ast}}{\chi^{2}}$, is in fact
a linear combination of
truncated confluent hypergeometric functions of
the first kind, with appropriate function coefficients.

\subsubsection{The simplest case $l=2$}
To give an example we consider the
case where $l$ takes its lowest value
$l=2.$
Equation  (\ref{tersol}),
with the $+$ sign, gives $s=4$ when $l=2$,
and, as a result, equation (\ref{interms}) gives

\begin{eqnarray}
 \chi \int \frac {{\rm
d}r_{\!\ast}}{\chi^{2}} & = & \frac{{\rm
P}(w)e^{2 r}}{r(r-2)^{4}} = \nonumber \\
  & = &
  \label{expan}
 \frac{e^{ 2 r}}{r(r-2)^{4}} \left (
A_{\rm 0} \ \phi(9; -  4 \it w) +
A_{\rm 1} \ \phi(\rm 8; - 4 \it w)+
A_{\rm 2} \ F(- \rm 7,  - \rm 7      ; \it - \rm 4 \it w)+ A_{\rm 3} \ F(- \rm 6, - \rm 7 ; \it - \rm 4 \it w) \right ), \nonumber \\
&&
\end{eqnarray}
where $w=r-2.$

The coefficients ${\rm P}_{\rm
n}$ of the polynomial
${\rm P}(w)=\sum_{{\rm n}=0}^{1+
2 s
}{\rm P}_{\rm
n}w^{\rm n}$,
appearing in the
elementary functions answer
(\ref{expan})
of $\chi \int \frac {{\rm
d}r_{\!\ast}}{\chi^{2}}$, are
given by
equations (\ref{hutyytuh}), (\ref{hutyytuh1}), and
(\ref{nugyygun}), and so we obtain
\be
\label{pol9}
{\rm P}_{\rm 9}(w)= - \frac{945}{16384}+ \frac{1755}{8192}w - \frac{405}{1024} w^{2} +
\frac{495}{1024}w^{3}-\frac{225}{512}w^{4}+
\frac{81}{256}w^{5}-\frac{3}{16}w^{6}+\frac{3}{32}w^{7}
+ \frac{1}{32}w^{8}+\frac{1}{16}w^{9},
\ee
where the subscript 9 in ${\rm P}_{\rm 9}(w)$
indicates the degree of the polynomial
${\rm P}_{\rm 9}(w)$.
When $l=2$ the coefficients
$A_{0}$, and   $A_{1}, \ A_{2}, \ A_{3}$,
given respectively by equations (\ref{coefs1}) and (\ref{coefs}), are equal to
\be \label{coefs2} A_{0}=\frac{9}{4194304}, \ A_{1}= - \frac{7}{4194304}, \   A_{2} = - \frac{945}{32768}, \   A_{3} = - \frac{945}{32768}. \ee

When $s=4$ the polynomials
$\phi(\rm 2 \it  s + \rm 1; - \it s w)$,
$\phi(\rm 2 \it  s; - \it s w)$,
$F(-(\rm 2 \it s - \rm 1),\rm 1 - \rm 2 \it s; \it - s w)$, and
$F(-(\rm 2 \it s - \rm 2),\rm 1 - \rm 2 \it s; \it - s w)$,
given respectively by equations
(\ref{pol1}), (\ref{pol2}),
(\ref{pol3}), and (\ref{pol4}),
are equal to
\begin{eqnarray}
\label{pol1a}
\phi(\rm 2 \it  s + \rm 1; - \it s w) &=&
\phi(9; -  4 \it w) = (\rm 4 \it w)^{\rm 8}
 \left (\rm 1+\frac{\rm 4}{\rm 9} \it  w \right ), \\
\label{pol2a}
\phi(\rm 2 \it  s; - \it s w) &=&
\phi(\rm 8; - 4 \it w)=(\rm 4 \it w)^{\rm 8}, \\
F(-(\rm 2 \it s - \rm 1),\rm 1 - \rm 2 \it s; \it - s w)&=&
F(- \rm 7,  - \rm 7      ; \it - \rm 4 \it w)=
\nonumber \\
\label{pol3a}
&=& 1 - 4 w +  \frac{{\it ( \rm 4 \it w)}^{\rm 2}}{\rm 2!} - \frac{{\it ( \rm 4 \it w)}^{\rm 3}}{\rm 3!}+
\frac{{\it ( \rm 4 \it w)}^{\rm 4}}{\rm 4!}-
\frac{{\it ( \rm 4 \it w)}^{\rm 5}}{\rm 5!}+
\frac{{\it ( \rm 4 \it w)}^{\rm 6}}{\rm 6!}-
\frac{{\it ( \rm 4 \it w)}^{\rm 7}}{\rm 7!}
, \nonumber \\
&& \\
F(-(\rm 2 \it s - \rm 2),\rm 1 - \rm 2 \it s; \it - s w)&=&
F(- \rm 6, - \rm 7 ; \it - \rm 4 \it w) =
\nonumber \\
\label{pol4a}
&=& 1 -    \frac{ 6  }{  7} \  4 w + \frac{ 5 }{ 7 }     \frac{{\it ( \rm 4 \it w)}^{\rm 2}}{\rm 2!} -  \frac{ 4 }{ 7 }     \frac{{\it ( \rm 4 \it w)}^{\rm 3}}{\rm 3!}
+ \frac{ 3 }{ 7 }     \frac{{\it ( \rm 4 \it w)}^{\rm 4}}{\rm 4!}
- \frac{ 2 }{ 7 }     \frac{{\it ( \rm 4 \it w)}^{\rm 5}}{\rm 5!}+
\frac{ 1 }{ 7 }     \frac{{\it ( \rm 4 \it w)}^{\rm 6}}{\rm 6!}, \nonumber \\
&&
\end{eqnarray}

We easily check that (\ref{tel1}) is indeed valid
when $s=4$, i.e., we easily check that
\begin{eqnarray}
{\rm P}_{\rm 9}(w)&=&
A_{\rm 0} \ \phi(9; -  4 \it w) +
A_{\rm 1} \ \phi(\rm 8; - 4 \it w)+
A_{\rm 2} \ F(- \rm 7,  - \rm 7      ; \it - \rm 4 \it w)+ A_{\rm 3} \ F(- \rm 6, - \rm 7 ; \it - \rm 4 \it w), \nonumber \\
&&,
\end{eqnarray}
where ${\rm P}_{\rm 9}(w)$ is given by
equation (\ref{pol9}),  $A_{0}, \ A_{1}, \ A_{2}, \ A_{3}$ are given by equation (\ref{coefs2}), and
$ \phi(\rm 9; -  \rm 4 \it w), \ \phi(\rm 8; -  \rm 4 \it w), \
F(- \rm 7,  - \rm 7      ; \it - \rm 4 \it w),$
and $F(- \rm 6, - \rm 7 ; \it - \rm 4 \it w)$,
are given respectively by equations (\ref{pol1a}), (\ref{pol2a}), (\ref{pol3a}), and (\ref{pol4a}).

Thus, by combining the results of subsections
 \ref{G7},
 \ref{elementary} and  \ref{extension},
with the results of this subsection,
we conclude that
when $l=2$ the second Liouvillian solution is
\begin{eqnarray}
 \chi \int \frac {{\rm
d}r_{\!\ast}}{\chi^{2}} & = &
 \frac{e^{ 2 r}}{r(r-2)^{4}} \left(
\frac{9}{4194304} \ (\rm 4 \it w       )^{\rm 8}
 \left (\rm 1+\frac{\rm 4}{\rm 9} \it  w \right ) - \frac{\rm 7}{\rm 4194304} \ (\rm 4 \it w)^{\rm 8} - \right. \nonumber \\
 &-& \left.  \frac{945}{32768} \
 \left( 1 - 4 w +  \frac{{\it ( \rm 4 \it w     )}^{\rm 2}}{\rm 2!} - \frac{{\it ( \rm 4 \it w     )}^{\rm 3}}{\rm 3!}+
\frac{{\it ( \rm 4 \it w )}^{\rm 4}}{\rm 4!}-
\frac{{\it ( \rm 4 \it w     )}^{\rm 5}}{\rm 5!}+
\frac{{\it ( \rm 4 \it w    )}^{\rm 6}}{\rm 6!}-
\frac{{\it ( \rm 4 \it w      )}^{\rm 7}}{\rm 7!} \right)
\right. - \nonumber \\
&-& \left.
\frac{945}{32768} \
\left(
1 -    \frac{ 6  }{  7} \  4 w + \frac{ 5 }{ 7 }     \frac{{\it ( \rm 4 \it w)}^{\rm 2}}{\rm 2!} -  \frac{ 4 }{ 7 }     \frac{{\it ( \rm 4 \it w)}^{\rm 3}}{\rm 3!}
+ \frac{ 3 }{ 7 }     \frac{{\it ( \rm 4 \it w)}^{\rm 4}}{\rm 4!}
- \frac{ 2 }{ 7 }     \frac{{\it ( \rm 4 \it w)}^{\rm 5}}{\rm 5!}+
\frac{ 1 }{ 7 }     \frac{{\it ( \rm 4 \it w)}^{\rm 6}}{\rm 6!}
\right)
\right), \nonumber \\
&&
\end{eqnarray}
where $w=r- \rm 2.$

\subsection{$\chi \int \frac {{\rm
d}r_{\!\ast}}{\chi^{2}}$ in terms of associated Laguerre polynomials}
\label{Laguerre1}

As we noted in subsection
\ref{Laguerre},
associated Laguerre polynomials
$L^{(\it a) }_{\rm {n}}(u)$
and confluent hypergeometric functions
of the first kind
$F( \rm {n},\it a + \rm 1, \it u)$
are
related via
\be
\label{relation1}
L^{(\it a) }_{\rm {n}}(u)= \left( {\begin{array}{*{20}c} \rm {n} + \it a  \\ \rm {n} \\ \end{array}} \right)
F( \rm {n},\it a + \rm 1, \it u),
\ee
where
\be
\left( {\begin{array}{*{20}c} A \\ K \\ \end{array}} \right)=\frac{A(A-1)(A-2)\cdot\cdot\cdot(A-K+1)}{K!}
\ee
is the generalized binomial coefficient,
i.e.,
$A \in R$ and $K \in N$,  $R$ and $N$ are the set of real numbers and the set of natural numbers respectively.

Thus the associated Laguerre polynomials $L^{(\it a) }_{\rm {n}}(u)$ satisfy the differential equation
\be
u L^{(\it a)''}_{\rm {n}}(u)   + (\it a + \rm 1 - \it u) L^{(\it a)' }_{\rm {n}}(\it u) + \rm {n} \it L^{(\it a) }_{\rm {n}}(\it u)= \rm 0.
\ee
Equation (\ref{relation1}) is in fact equivalent
to equation (\ref{relation}), given in subsection \ref{Laguerre}, which expresses the relation between
associated Laguerre polynomials
and confluent hypergeometric functions of the first kind.

Relation (\ref{relation1}) suggests
that the analysis of the previous subsection can be repeated by expanding
${\rm P}(w)$ not as a sum of
confluent hypergeometric functions of the first kind, with appropriate function
coefficients, as it is done in
equation (\ref{tel1}), but instead,
by expanding
${\rm P}(w)$
as a sum of
associated Laguerre polynomialls, with appropriate function
coefficients, in the form
\begin{eqnarray}
{\rm P}(w) & = &
\sum_{\rm n=\rm 0}^{\rm 2 \it{s}+\rm{1}}
{\rm P}_{\rm n}
w^{\rm n} = \nonumber \\
\label{tel2}
& = &  \sum_{\it k=\rm 0}^{\rm 1} \mathcal A_{k} \varphi((\rm 2 \it  s + \rm 1) - \it  k; - s w)+
\sum_{\it k=\rm 2}^{\rm 3} \mathcal A_{k}L^{ (- \rm 2 \it s )}_{\rm { (\rm 2 \it s + \rm 1) - \it k }}(- \it s w),
\end{eqnarray}
where,
from equation (\ref{relation1}) we have
\begin{eqnarray}
\label{polyn1}
L^{(\rm 2 \it s )}_{\rm {1}}(- \it s w) &=& (\rm 2 \it s + \rm 1) \it F(- \rm 1, \rm 2 \it s + \rm 1; - \it s w ), \\
\label{polyn2}
L^{(\rm 2 \it s )}_{\rm {0}}(- \it s w) &=& \it F( \rm 0, \rm 2 \it s + \rm 1; - \it s w ), \\
\label{polyn3}
L^{ (- \rm 2 \it s )}_{\rm 2 \it s - \rm 1}(- \it s w) &=& - \it F( - (\rm 2 \it s - \rm 1),  \rm 1 - \rm 2 \it s; - \it s w ), \\
\label{polyn4}
L^{ (- \rm 2 \it s )}_{\rm 2 \it s - \rm 2}(- \it s w) &=& (2 s - 1) \it F( - (\rm 2 \it s - \rm 2),  \rm 1 - \rm 2 \it s; - \it s w ),
\end{eqnarray}
and subsequently,     by
taking also into account
equations (\ref{polyn1}), (\ref{polyn2}),
(\ref{polyn3}), and (\ref{polyn4}),
we have
\begin{eqnarray}
\label{pol11}
\varphi(\rm 2 \it  s + \rm 1; - \it s w) &=&    (- s w)^{2 s}
L^{(\rm 2 \it s )}_{\rm {1}}(- \it s w)=
( s w)^{\rm 2 \it s}
\left (\it s w + \rm 2 \it s + \rm 1 \right ),
\\
\label{pol22}
\varphi(\rm 2 \it  s; - \it s w) &=& (- s w)^{2 s}  L^{(\rm 2 \it s )}_{\rm {0}}(- \it s w)=(s w)^{\rm 2 \it s},  \\
\label{pol33}
L^{ (- \rm 2 \it s )}_{\rm 2 \it s - \rm 1}(- \it s w)&=& - \left (
\it F( - (\rm 2 \it s - \rm 1),  \rm 1 - \rm 2 \it s; - \it s w )
\right )=
\nonumber \\
&=&
- \left (
1 - s w +  \frac{{\it ( s w)}^{\rm 2}}{\rm 2!} - \frac{{\it ( s w)}^{\rm 3}}{\rm 3!}+ ... -  \frac{{\it ( s w)}^{\rm 2 \it s - \rm 1}}{(\rm 2 \it s - \rm 1)!} \right ), \\
\label{pol44}
L^{ (- \rm 2 \it s )}_{\rm 2 \it s - \rm 2}(- \it s w)&=&
(2 s - 1) \it F( - (\rm 2 \it s - \rm 2),  \rm 1 - \rm 2 \it s; - \it s w )=
\nonumber \\
&=& (2 s - 1) \left ( 1 -    \frac{\rm 2 -\rm 2 \it s }{\rm 1 - \rm 2 \it s}   s w + \frac{\rm 3 - \rm 2 \it s }{\rm 1 - \rm 2 \it s}     \frac{{\it ( s w)}^{\rm 2}}{\rm 2!} +...+
 \frac{{\it ( s w)}^{\rm 2 \it s - \rm 2}}{(\rm 2 \it s - \rm 1) !} \right ).
\end{eqnarray}

\noindent
 ${\rm P}(w)$ is the
polynomial solution
of degree $2 s + 1$,
to equation
(\ref{cv}), with coefficients
${\rm P}_{\rm n}$ given by
equations (\ref{hutyytuh}), (\ref{hutyytuh1}), and
(\ref{nugyygun}).

We follow through the analysis carried out in subsections \ref{extension} and \ref{truncon}: By taking into account equation (\ref{tel}), by substituting   the ansatz (\ref{tel2}) into
equation (\ref{con}) and
by using appropriate
recurrence relations for
the four polynomial solutions
$\varphi(\rm 2 \it  s + \rm 1; - \it s w)$, $\varphi(\rm 2 \it  s; - \it s w)$, $L^{ (- \rm 2 \it s )}_{\rm 2 \it s - \rm 1}(- \it s w)$, and $L^{ (- \rm 2 \it s )}_{\rm 2 \it s - \rm 2}(- \it s w)$, we obtain
a $4 \times 4$ homogeneous system for the coefficients $\mathcal A_{k}, \ k=0,1,2,3.$

It is straightforward to verify that the necessary and sufficient condition
for obtaining non zero solutions to
this system is $s=\frac{l(l-1)(l+1)(l+2)}{6}, \ l=2,3,...  \ .$ When this condition
is satisfied
we obtain
\begin{eqnarray}
\label{coefs3}
\mathcal A_{1}= -(l^{2}+l+1) \mathcal A_{0}, \ \
\mathcal A_{2}=   3   (2s)!   \mathcal A_{0}, \ \
\mathcal A_{3}= - \frac{(l^{2}+l-3)  (2s)!}{2 s - 1} \mathcal A_{0},
\end{eqnarray}

We  easily find that
if we choose \be \label{coefs4} \mathcal A_{0}=\frac{s^{-2-2 s}}{(l-1)(l+2)} \ee
then the expansion (equation (\ref{tel2}))
\begin{eqnarray}
{\rm P}(w)
& = &  \sum_{\it k=\rm 0}^{\rm 1} \mathcal A_{k} \varphi((\rm 2 \it  s + \rm 1) - \it  k; - s w)+
\sum_{\it k=\rm 2}^{\rm 3} \mathcal A_{k}L^{ (- \rm 2 \it s )}_{\rm { (\rm 2 \it s + \rm 1) - \it k }}(- \it s w), \nonumber
\end{eqnarray}
indeed holds, where $\mathcal A_{1}, \mathcal A_{2}, \mathcal A_{3}$ are given
by equation (\ref{coefs3}), and
\be
{\rm P}(w)  =
\sum_{\rm n=\rm 0}^{\rm 2 \it{s}+\rm{1}}
{\rm P}_{\rm n}
w^{\rm n}
\ee
where the coefficients ${\rm P}_{\rm n}$ are given by
equations (\ref{hutyytuh}), (\ref{hutyytuh1}), and
(\ref{nugyygun}), and where it is
understood that $s= \frac{l(l-1)(l+1)(l+2)}{6}$.

${\rm P}(w)$ is
the polynomial which appears in the second Liouvillian solution   $\chi \int \frac {{\rm
d}r_{\!\ast}}{\chi^{2}},$
initially found by Chandrasekhar\cite{Chandr},
to the RWE.
Equations (\ref{grasssop}), (\ref{derossored}) and
(\ref{tel2}) give
\begin{eqnarray}
 \chi \int \frac {{\rm
d}r_{\!\ast}}{\chi^{2}} & = & \frac{{\rm
P}(w)e^{\frac{s}{2}r}}{r(r-2)^{s}} = \nonumber \\
\label{interms1}
  & = &
 \frac{e^{\frac{s}{2}r}}{r(r-2)^{s}} \left (
\sum_{\it k=\rm 0}^{\rm 1} \mathcal A_{k} \varphi((\rm 2 \it  s + \rm 1) - \it  k; - s w)+
\sum_{\it k=\rm 2}^{\rm 3} \mathcal A_{k}L^{ (- \rm 2 \it s )}_{\rm { (\rm 2 \it s + \rm 1) - \it k }}(- \it s w)
\right ),
\nonumber \\
&&
\end{eqnarray}
where $w=r-\rm 2.$  The four polynomials $\varphi(\rm 2 \it  s + \rm 1; - \it s w)$, $\varphi(\rm 2 \it  s; - \it s w)$, $L^{ (- \rm 2 \it s )}_{\rm 2 \it s - \rm 1}(- \it s w)$, and $L^{ (- \rm 2 \it s )}_{\rm 2 \it s - \rm 2}(- \it s w)$, are given respectively by equations (\ref{pol11}), (\ref{pol22}), (\ref{pol33}), and
(\ref{pol44}).


We conclude, in equation (\ref{interms1}),
that  the polynomial ${\rm P}(w)$
appearing in the
elementary functions answer of $\chi \int \frac {{\rm
d}r_{\!\ast}}{\chi^{2}}$,
admits an expansion in terms of
associated Laguerre polynomials,
with appropriate function coefficients.
We thus have the following Theorem.
\begin{thrm}
\label{Th3}

The second Liouvillian solution $\chi \int \frac {{\rm d}r_{\!\ast}}{\chi^{2}}$  to the master
equation (\ref{adamsmith}), given by the family G7, in the case of the gravitational perturbations of the Schwarzschild geometry,
is a product of elementary functions, one
of them being a polynomial $\rm P(\it w)$
which admits an expansion in terms of
associated Laguerre polynomials with appropriate
function coefficients. In particular the
following holds
\begin{eqnarray}
 \chi \int \frac {{\rm
d}r_{\!\ast}}{\chi^{2}} & = & \frac{{\rm
P}(w)e^{\frac{s}{2}r}}{r(r-2)^{s}} = \nonumber \\
  & = &
 \frac{e^{\frac{s}{2}r}}{r(r-2)^{s}} \left (
\sum_{\it k=\rm 0}^{\rm 1} \mathcal A_{k} \varphi((\rm 2 \it  s + \rm 1) - \it  k; - s w)+
\sum_{\it k=\rm 2}^{\rm 3} \mathcal A_{k}L^{ (- \rm 2 \it s )}_{\rm { (\rm 2 \it s + \rm 1) - \it k }}(- \it s w)
\right ),
\nonumber
\end{eqnarray}
where $w=r-2,$
\begin{eqnarray}
\varphi(\rm 2 \it  s + \rm 1; - \it s w) &=&    (- s w)^{2 s}
L^{(\rm 2 \it s )}_{\rm {1}}(- \it s w)=
( s w)^{\rm 2 \it s}
\left (\it s w + \rm 2 \it s + \rm 1 \right ),
 \nonumber
\\
\varphi(\rm 2 \it  s; - \it s w) &=& (- s w)^{2 s}  L^{(\rm 2 \it s )}_{\rm {0}}(- \it s w)=(s w)^{\rm 2 \it s},   \nonumber \\
L^{ (- \rm 2 \it s )}_{\rm 2 \it s - \rm 1}(- \it s w)
&=&
- \left (
1 - s w +  \frac{{\it ( s w)}^{\rm 2}}{\rm 2!} - \frac{{\it ( s w)}^{\rm 3}}{\rm 3!}+ ... -  \frac{{\it ( s w)}^{\rm 2 \it s - \rm 1}}{(\rm 2 \it s - \rm 1)!} \right ), \nonumber \\
L^{ (- \rm 2 \it s )}_{\rm 2 \it s - \rm 2}(- \it s w)
&=& (2 s - 1) \left ( 1 -    \frac{\rm 2 -\rm 2 \it s }{\rm 1 - \rm 2 \it s}   s w + \frac{\rm 3 - \rm 2 \it s }{\rm 1 - \rm 2 \it s}     \frac{{\it ( s w)}^{\rm 2}}{\rm 2!} +...+
 \frac{{\it ( s w)}^{\rm 2 \it s - \rm 2}}{(\rm 2 \it s - \rm 1) !} \right ), \nonumber
\end{eqnarray}
\begin{eqnarray}
\mathcal A_{1}= -(l^{2}+l+1) \mathcal A_{0}, \
\mathcal A_{2}=   3   (2s)!   \mathcal A_{0}, \
\mathcal A_{3}= - \frac{(l^{2}+l-3)  (2s)!}{2 s - 1} \mathcal A_{0}, \ \mathcal A_{0}=\frac{s^{-2-2 s}}{(l-1)(l+2)}, \nonumber
\end{eqnarray}
where the coefficients ${\rm P}_{\rm
n}$ of the polynomial
${\rm P}(w)=\sum_{{\rm n}=0}^{1+
2 s
}{\rm P}_{\rm
n}w^{\rm n}$,
of degree $2 s + 1$,
are given by
equations (\ref{hutyytuh}), (\ref{hutyytuh1}), and
(\ref{nugyygun}), and where $s=\frac{l(l-1)(l+1)(l+2)}{6}, \ l=2,3,...  \ .$
\end{thrm}

In subsection \ref{truncon}
we also  proved that ${\rm P}(w)$
admits also an expansion in terms of
truncated confluent hypergeometric functions of
the first kind
with appropriate function coefficients.


\subsubsection{The simplest case $l=2$}

To give an example we consider the
case where $l$ takes its lowest value
$l=2.$
Equation  (\ref{tersol}),
gives $s=4$ when $l=2$,
and, as a result, equation (\ref{interms1}) gives

\begin{eqnarray}
 \chi \int \frac {{\rm
d}r_{\!\ast}}{\chi^{2}} & = & \frac{{\rm
P}_{\rm 9}(w)e^{2 r}}{r(r-2)^{4}} = \nonumber \\
  & = &
  \label{expan1}
 \frac{e^{ 2 r}}{r(r-2)^{4}} \left (
\mathcal A_{\rm 0} \ \varphi(9; -  4 \it w) +
\mathcal A_{\rm 1} \ \varphi(\rm 8; - 4 \it w)+
\mathcal A_{\rm 2} \ L^{ (- \rm 8 )}_{\rm 7}(- \rm 4 \it w)+ \mathcal A_{\rm 3} \ L^{ (- \rm 8 )}_{\rm 6}(- \rm 4 \it w) \right ), \nonumber \\
&&
\end{eqnarray}
where $w=r-\rm 2.$  The polynomial ${\rm P}_{\rm 9}(w)$ of degree 9  is given by equation (\ref{pol9}).

We evaluate the coefficients
$\mathcal A_{\rm 0},$ $\mathcal A_{\rm 1},$
$\mathcal A_{\rm 2},$ $\mathcal A_{\rm 3}$
 from equations (\ref{coefs3})
and (\ref{coefs4}), and we evaluate the polynomials
$\varphi(9; -  4 \it w)$, $\varphi(\rm 8; - 4 \it w)$, $L^{ (- \rm 8 )}_{\rm 7}(- \rm 4 \it w)$, and $L^{ (- \rm 8 )}_{\rm 6}(- \rm 4 \it w) $  from equations (\ref{pol11}), (\ref{pol22}), (\ref{pol33}), and (\ref{pol44}) respectively,  and we obtain
\begin{eqnarray}
 \chi \int \frac {{\rm
d}r_{\!\ast}}{\chi^{2}} & = &
 \frac{e^{ 2 r}}{r(r-2)^{4}} \left(
\frac{1}{4194304} \ (\rm 4 \it w       )^{\rm 8}
 \left ( \rm 4   \it  w + \rm 9  \right ) - \frac{\rm 7}{\rm 4194304} \ (\rm 4 \it w)^{\rm 8} - \right. \nonumber \\
 &-& \left.  \frac{945}{32768} \
 \left( 1 - 4 w +  \frac{{\it ( \rm 4 \it w     )}^{\rm 2}}{\rm 2!} - \frac{{\it ( \rm 4 \it w     )}^{\rm 3}}{\rm 3!}+
\frac{{\it ( \rm 4 \it w )}^{\rm 4}}{\rm 4!}-
\frac{{\it ( \rm 4 \it w     )}^{\rm 5}}{\rm 5!}+
\frac{{\it ( \rm 4 \it w    )}^{\rm 6}}{\rm 6!}-
\frac{{\it ( \rm 4 \it w      )}^{\rm 7}}{\rm 7!} \right)
\right. - \nonumber \\
&-& \left.
\frac{945}{229376} \
\left(
7 -     6     (4 w) +  5      \frac{{\it ( \rm 4 \it w)}^{\rm 2}}{\rm 2!} -   4     \frac{{\it ( \rm 4 \it w)}^{\rm 3}}{\rm 3!}
+  3     \frac{{\it ( \rm 4 \it w)}^{\rm 4}}{\rm 4!}
-  2     \frac{{\it ( \rm 4 \it w)}^{\rm 5}}{\rm 5!}+
\frac{{\it ( \rm 4 \it w)}^{\rm 6}}{\rm 6!}
\right)
\right), \nonumber \\
&&
\end{eqnarray}
where $w=r- \rm 2.$

\section{The remaining families G3, E3, E7 and S3}
\label{rem}
\noindent

We examine now
the remaining families G3, E3, E7 and S3.
In subsection  \ref{confluent} we show that
the ODEs associated with all the remaining families
G3, E3, E7 and S3 fall into the confluent Heun class after a trivial change of the independent variable $r.$  In subsection  \ref{elementarysym}
we use the elementary symmetries of the confluent Heun equation to show that the families G7 and G3
are related with such an elementary symmetry,
and also to show that the families E7 and E3
are related with such an elementary symmetry.

In subsection  \ref{sections3} we prove that
the family S3 does not give rise to a
Liouvillian solution to the master equation
(\ref{adamsmith}). In subsection  \ref{evidence}
we give strong evidence which supports the thesis
that the families G3, E7 and E3 do not give rise to Liouvillian solutions to the master equation
(\ref{adamsmith}). In subsection  \ref{reduction}
we conclude that the four remaining families
G3, E3, E7 and S3 are reduced into the two families
G3 and E7 and we put forward the conjecture,
by taking into account the evidence given
in subsection  \ref{evidence}, that the families
G3 and E7 do not give rise to Liouvillian solutions to the master equation
(\ref{adamsmith}).

Thus in this section we consider
problem 3
in the list of most difficult problems, mentioned in
subsection \ref{kovalg}, we have to solve when we   apply Kovacic's algorithm.
In fact
problem 3
 in
subsection \ref{kovalg},
in the problem under consideration,
amounts to proving that the families
G3, E7, E3, and S3 do not give rise
to Liouvillian solutions to the master
equation (\ref{adamsmith}).

Couch and Holder\cite{C} address also
problem 3.
We improve the results derived by
C.H.\cite{C} on the solution of
problem 3. Namely, the results
derived in subsections
\ref{confluent}, \ref{elementarysym},
\ref{sections3},
\ref{evidence}, and \ref{reduction},
are new. There are some
discrepancies  between the results
we derive  on the solution of problem 3
and the results derived by C.H.\cite{C}
on the solution of problem 3.
We compare our results on the solution of problem 3 with the results of C.H. on the solution of problem 3 in subsection \ref{comparison}.
In the last subsection \ref{comments}
we make some comments on the proper
implementation of Kovacic's algorithm
for any second order ODE with rational
function coefficients.

\subsection{The remaining families fall into the confluent Heun class}

\label{confluent}

The first key observation is that the
ODEs
$$
{\rm P}^{''}+2\theta{\rm P}^{'}+ ( \theta^{2}+\theta^{'}-\nu  )
{\rm P} =0
$$
(equation (\ref{gfkkukku})), where
$$
\nu(r)= \frac { \frac{s^{2}}{4}r^{4} + l(l+1)r^{2} +
2[\beta-l(l+1)-1]r+3-4\beta}{r^{2}(r-2)^{2}}
$$
(equation (\ref{ygflygflygflty})),
and
$\beta=-3,0,1$  distinguishes the various
types of external perturbations, gravitational, electromagnetic or
massless scalar fields respectively,
associated with the families
G3, E3, E7 and S3 fall into the confluent Heun class after a trivial change of the independent variable. It is to be noted that the ODE
associated with the
family G7,
which gives rise to the Liouvillian solution
$\chi \int \frac {{\rm
d}r_{\!\ast}}{\chi^{2}}$,
falls also into the confluent Heun class after the same trivial change of the independent variable. We also include the family G7
because it is going to be useful in subsection
\ref{elementary}.

\hspace{6.3 cm}  {\bf The family G7}

In the case of the family G7
we have (equation (\ref{gravity7}))
$$
\theta(r)= \frac{-3/2}{r} +
\frac{\frac{1}{2}-s}{r-2}+\frac{s}{2},
$$
and therefore equation (\ref{gfkkukku}) reads
\be
\label{rem0}
r(r-2) {\rm P}^{''}(r)    +
(6-2r(2 s +  1)+r^{2}s)
{\rm P}^{'}(r)    +
( 2-l(l+1)+6s-r s(2s+1))
 {\rm P}(r)=0.
\ee

Equation (\ref{rem0})
after the change of the independent variable
$r=2 z$ becomes
\be
\label{ch0}
z(z-1) {\rm p}^{''}(z)    +
(3-2z(2 s + 1)+ 2 z^{2}s)
{\rm p}^{'}(z)    +
( 2-l(l+1)+6s-2 z s(2s+1))
 {\rm p}(z)=0
\ee
which falls into the confluent Heun class.
It is trivially verified that equation
(\ref{rem0}) admits a polynomial solution
$\rm P ( \it r)$ if and only if equation
(\ref{ch0}) admits a polynomial solution
$\rm p ( \it z)$. If such polynomial solutions
exist then $\rm p ( \it z)=\rm P ( 2 \it z)$.
$\vspace{0.1cm}$

\noindent
We have in turn now for the remaining families
G3, E3, E7 and S3:

\hspace{6.3 cm}  {\bf The family G3}

In the case of the family G3
we have (equation (\ref{ygflygflygflty}))
$$
\theta(r)= \frac{5/2}{r} +
\frac{\frac{1}{2}-s}{r-2}+\frac{s}{2},
$$
and therefore equation (\ref{gfkkukku}) reads
\be
\label{rem1}
r(r-2) {\rm P}^{''}(r)    +
(-10-2r(2 s - 3)+r^{2}s)
{\rm P}^{'}(r)    +
( 6-l(l+1)-10s-r s(2s-3))
 {\rm P}(r)=0.
\ee

Equation (\ref{rem1})
after the change of the independent variable
$r=2 z$ becomes
\be
\label{ch1}
z(z-1) {\rm p}^{''}(z)    +
(-5-2z(2 s - 3)+ 2 z^{2}s)
{\rm p}^{'}(z)    +
( 6-l(l+1)-10s-2 z s(2s-3))
 {\rm p}(z)=0
\ee
which falls into the confluent Heun class.
It is trivially verified that equation
(\ref{rem1}) admits a polynomial solution
$\rm P ( \it r)$ if and only if equation
(\ref{ch1}) admits a polynomial solution
$\rm p ( \it z)$. If such polynomial solutions
exist then $\rm p ( \it z)=\rm P ( 2 \it z)$.



\hspace{6.3 cm}  {\bf The family E7}

In the case of the family E7
we have (equation (\ref{yumkfcmufmkfcym}))
$$
\theta(r)=\frac{-1/2}{r} +
\frac{\frac{1}{2}-s}{r-2}+\frac{s}{2},
$$
and therefore equation (\ref{gfkkukku}) reads
\be
\label{rem2}
r(r-2) {\rm P}^{''}(r)    +
(2-4r s +r^{2}s)
{\rm P}^{'}(r)    +
( -l(l+1)+ 2 s- 2  s^{2} r)
 {\rm P}(r)=0.
\ee

Equation (\ref{rem2})
after the change of the independent variable
$r=2 z$ becomes
\be
\label{ch2}
z(z-1) {\rm P}^{''}(z)    +
(1-4 z s+ 2 z^{2}s)
{\rm P}^{'}(z)    +
( -l(l+1)+ 2 s-4 s^{2}z )
 {\rm P}(z)=0
\ee
which falls into the confluent Heun class.
It is trivially verified that equation
(\ref{rem2}) admits a polynomial solution
$\rm P ( \it r)$ if and only if equation
(\ref{ch2}) admits a polynomial solution
$\rm p ( \it z)$. If such polynomial solutions
exist then $\rm p ( \it z)=\rm P ( 2 \it z)$.



\hspace{6.3 cm}  {\bf The family E3}

In the case of the family E3
we have (equation (\ref{ugh.ujhb.b.b,b,b.}))
$$
\theta(r)=\frac{3/2}{r} +
\frac{\frac{1}{2}-s}{r-2}+\frac{s}{2},
$$
and therefore equation (\ref{gfkkukku}) reads
\be
\label{rem3}
r(r-2) {\rm P}^{''}(r)    +
(-6 - 2r(2 s - 2 )+r^{2}s)
{\rm P}^{'}(r)    +
( -l(l+1)- 6 s + 2 - 2 r s( 2 s - 2))
 {\rm P}(r)=0.
\ee

Equation (\ref{rem3})
after the change of the independent variable
$r=2 z$ becomes
\be
\label{ch3}
z(z-1) {\rm P}^{''}(z)    +
(-3 - 2 z( 2 s- 2 )+ 2 z^{2}s)
{\rm P}^{'}(z)    +
( -l(l+1)-  6 s + 2 - 4 z s( 2 s - 2))
 {\rm P}(z)=0
\ee
which falls into the confluent Heun class.
It is trivially verified that equation
(\ref{rem3}) admits a polynomial solution
$\rm P ( \it r)$ if and only if equation
(\ref{ch3}) admits a polynomial solution
$\rm p ( \it z)$. If such polynomial solutions
exist then $\rm p ( \it z)=\rm P ( 2 \it z)$.


\hspace{6.3 cm}  {\bf The family S3}

In the case of the family S3
we have (equation (\ref{scalar11}))
$$
\theta(r)=\frac{1/2}{r} +
\frac{\frac{1}{2}-s}{r-2}+\frac{s}{2},
$$
and therefore equation (\ref{gfkkukku}) reads
\be
\label{rem4}
r(r-2) {\rm P}^{''}(r)    +
(-2   -   2r(   2 s - 1 )+r^{2}s)
{\rm P}^{'}(r)    +
( -l(l+1)- 2 s  - r s( 2 s - 1))
 {\rm P}(r)=0.
\ee

Equation (\ref{rem4})
after the change of the independent variable
$r=2 z$ becomes
\be
\label{ch4}
z(z-1) {\rm P}^{''}(z)    +
(-1 - 2 z( 2 s- 1 )+ 2 z^{2}s)
{\rm P}^{'}(z)    +
( -l(l+1)-  2 s - 2 z s( 2 s - 1))
 {\rm P}(z)=0
\ee
which falls into the confluent Heun class.
It is trivially verified that equation
(\ref{rem4}) admits a polynomial solution
$\rm P ( \it r)$ if and only if equation
(\ref{ch4}) admits a polynomial solution
$\rm p ( \it z)$. If such polynomial solutions
exist then $\rm p ( \it z)=\rm P ( 2 \it z)$.



\subsection{Elementary symmetries of the
confluent Heun equation
and the
remaining families}

\label{elementarysym}

It is well known (see e.g. \cite{El}, \cite{Kaz})
that there are 16 variable substitutions which preserve the form of the confluent Heun equation
\be
\label{conheun}
z(z-1) {\rm p}^{''}(z) + (a z^{2}  + b z + c) {\rm p}^{'}(z) + (d + e z ){\rm p}(z)=0.
\ee

For convenience, we rewrite equation (\ref{conheun}) in the form
\be
\mathcal R_{(a,b,c,d,e)} {\rm p}(z)=0,
\ee
where $\mathcal R_{(a,b,c,d,e)}$ is the linear operator
\be
\mathcal R_{(a,b,c,d,e)}= z(z-1)    \frac{{\rm d}^{2}}{{\rm d}z^{2}}+ + (a z^{2}  + b z + c)        \frac{{\rm d}}{{\rm d}z} + + (d + e z ).
\ee
The subscript $(a,b,c,d,e)$ in $\mathcal R_{(a,b,c,d,e)}$
emphasizes the dependence of $\mathcal R_{(a,b,c,d,e)}$
on the parameters $a,b,c,d$ and $e$.

Out of these 16 variable substitutions the one
which is the most relevant here is the following:
The substitution
\be
\label{sub}
{\rm p}(z)=z^{1+c} {\rm p}_{1}(z)
\ee
in equation (\ref{conheun}) yields
\be
R_{(a,b+2+2c,-2-c,d + (b+c)(c+1),
e + a c + a)}{\rm p}_{1}(z)=0.
\ee

Therefore we have the equivalence
\be
\label{equiv}
\mathcal R_{(a,b,c,d,e)} {\rm p}(z)=0 \Leftrightarrow R_{(a,b+2+2c,-2-c,
  d + (b+c)(c+1), e + a c + a)}{\rm p}_{1}(z)=0.
\ee
We note that in transformation (\ref{sub}) there is no change of the independent variable $z$. Consequently the positions of the singular points
remain fixed under this transformation. For this
reason, transformation (\ref{sub}), and more generally transformations which do not involve
change of the independent variable are called
homotopic \cite{He1} (from the greek words
``homo''=``same'' and ``topos''=``place''.)

We prove now that the confluent Heun equations
associated with the families G7 and G3 are
related with the equivalence relation
(\ref{equiv}). The same applies to the confluent Heun equations associated with the families E7 and E3. This has implications for the solution
of the problem we are considering in this section; in particular it gives useful equivalent
formulation of the problem we need to solve
for the family G3, and it
 reduces the consideration of the  families E7 and E3 into the consideration of the family
 E7 only. The details are as follows:

\vspace{0.3cm}

\hspace{5 cm}  {\bf The families G7 and G3}

The confluent Heun equation associated with the family G7 is (equation (\ref{ch0}))
$$
z(z-1) {\rm p}^{''}(z)    +
(3-2z(2 s + 1)+ 2 z^{2}s)
{\rm p}^{'}(z)    +
( 2-l(l+1)+6s-2 z s(2s+1))
 {\rm p}(z)=0.
$$
Comparison with the general form of the
confluent Heun equation (\ref{conheun}) gives
\be
a=2 s, \ b= - 2 (2 s + 1), \ c=3, \ d=2-l(l+1)+6s, \  \rm and  \ \it e=- \rm 2 \it s  \rm (2 \it s \rm + 1).
\ee

According to the equivalence relation
(\ref{equiv}), the substitution \be
\label{subs}
 {\rm p}(z)=z^{4} {\rm p}_{1}(z) \ee into equation (\ref{ch0}) yields
$$
z(z-1) {\rm p}_{1}^{''}(z)    +
(-5-2z(2 s - 3)+ 2 z^{2}s)
{\rm p}_{1}^{'}(z)    +
( 6-l(l+1)-10s-2 z s(2s-3))
 {\rm p}_{1}(z)=0,
$$
which is nothing but the confluent Heun equation
associated with the family G3 (equation (\ref{ch1})).

In subsections \ref{G7} and \ref{hautotresults}
 we proved that the confluent Heun equation
 associated with the family G7  (equation (\ref{ch0}))
admits
a polynomial solution
\be
\label{polsol}
{\rm P}(u)=\sum_{{\rm n}=0}^{1+2 s}{
P}_{\rm
n}u^{\rm n}=
\sum_{{\rm n}=0}^{1+2 s}{
P}_{\rm
n}(2 s (1-z))^{\rm n}= {\rm p}(z)=
{\rm p}_{0} + {\rm p}_{1} z + {\rm p}_{2} z^{2} + {\rm p}_{3} z^{3} + ... + {\rm p}_{1+2 s} z^{1+2 s} \ ,
\ee
of degree $1+2s$
with coefficients ${
P}_{\rm n}
$ given by (equations (\ref{coef1}) and (\ref{coef2}) respectively)
\begin{eqnarray}
{
P}_{\rm n}&= - &\frac
{6 s^{- 2 s -1}(2 s)!((l-1)(l+2)- 3 s) \left [ ({\rm
n}- 2 s )(l-1)(l+2)- 6 s \right ] } {{\rm
n}!((l-1)(l+2)+ 6 s) s (l-1)^{3}(l+2)^{3}},
\nonumber
\\
{
P}_{2 s} & = & \frac{2 (l-1) (l+2) - 6}{ s^{2s+1} (l-1)^{2} (l+2)^{2}    }, \qquad
{
P}_{2 s + 1}   = -  \frac{1}{ s^{2s+2} (l-1)(l+2)}, \nonumber
\end{eqnarray}
where $0\leq{\rm n}\leq 2 s -1$ and $s=\frac{l(l-1)(l+1)(l+2)}{6}, \quad l=2,3,... \ .$

The key fact here is that the coefficients ${\rm p}_{0},{\rm p}_{1},{\rm p}_{2},{\rm p}_{3}$ of the polynomial
solution (\ref{polsol}) to the confluent Heun equation (\ref{ch0}) associated with the family G7 are different from zero for \it
every
value of the angular harmonic index \normalfont $l=2,3,... \ .$ In fact  we can easily show by using equations (\ref{coef1}) and (\ref{coef2}) that
\be
\label{unequal}
sgn(\rm p_{\rm n})=(-1)^{\rm n + 1}, \
\rm n=0,1,2,...,2 \it s \rm + 1, \ \  \forall l=2,3,... \ .
\ee
So in particular for the first four coefficients we have
\be
\label{unequa2}
{\rm p}_{0}<0, \ {\rm p}_{1}>0, \ {\rm p}_{2} <0, \ {\rm p}_{3} >0, \  \forall l=2,3,... \ .
\ee

In subsection \ref{evidence} we give strong evidence that the family G3 does not give rise to a Liouvillian solution to the master equation
(\ref{adamsmith}).
We can reformulate the problem of proving
that the family G3 does not give rise to
a
Liouvillian solution
to the master equation
(\ref{adamsmith})
by using
the equivalence of families G7 and G3
which was shown to exist by setting $\rm p (\it z)$ equal to $z^{4}\rm p_{1}(\it z)$ into equation (\ref{ch0}). In the following two
Corollaries we give two such reformulations
relating the families G7 and G3.

\begin{cor}
\label{cor1}
In order to prove that the
the family G3 does not give rise to a
Liouvillian solution to the master equation (\ref{adamsmith}) it suffices to prove that
the confluent Heun equation (\ref{ch0})
 associated with the family G7  admits
 only one polynomial solution, namely the
 polynomial solution we found in subsections \ref{G7} and \ref{hautotresults}, given by equations
(\ref{coef1}), (\ref{coef2}), and (\ref{polsol}).
\end{cor}
\noindent \textbf{Proof} \hspace{0.1cm}
Let us assume that the confluent Heun equation (\ref{ch0})
 associated with the family G7  admits
 only one polynomial solution, namely the
 polynomial solution
 \be
 \label{polsol1}
{\rm p}(z)= \sum_{{\rm n}=0}^{1+2 s}{ \rm
p}_{\rm
n} z^{\rm n} =
{\rm p}_{0} + {\rm p}_{1} z + {\rm p}_{2} z^{2} + {\rm p}_{3} z^{3} + ... + {\rm p}_{1+2 s} z^{1+2 s}\ ,
 \ee
  we found in subsections \ref{G7} and \ref{hautotresults}, given by equations
 (\ref{coef1}), (\ref{coef2}), and (\ref{polsol}). The coefficients
 ${\rm p}_{0}, {\rm p}_{1}, {\rm p}_{2}, {\rm p}_{3}$ of this polynomial solution are different from zero, in particular we have $
{\rm p}_{0}<0, \ {\rm p}_{1}>0, \ {\rm p}_{2} <0, \ {\rm p}_{3} >0, \  \forall l=2,3,...,$
(equation (\ref{unequa2})).

This implies that the confluent Heun equation (\ref{ch1})
 associated with the family G3 does not admit
 any polynomial solution, because if it did,
 according to equation (\ref{subs}),
 the confluent Heun equation (\ref{ch0})
 associated with the family G3 would  admit
  polynomial solution whose smallest power
  would be 4. But this cannot happen since,
  according to equation  (\ref{subs}), the
  coefficients ${\rm p}_{0}, {\rm p}_{1}, {\rm p}_{2}, {\rm p}_{3}$ are different from zero
and we have assumed that the only polynomial solution to the confluent Heun equation (\ref{ch0})
 associated with the family G7 is the one given by equation  (\ref{polsol1}).
 We conclude that the confluent Heun equation (\ref{ch1})
 associated with the family G3 does not admit
 any polynomial solution and therefore the
family G3 does not give rise to any Liouvillian
solution to the master equation  (\ref{adamsmith}). This completes the proof.

\vspace{0.2cm}

We note that, strictly speaking, for the proof of
Corollary \ref{cor1} we do not need equations (\ref{unequal})
and (\ref{unequa2}), we
only need to know that at least one of the coefficients ${\rm p}_{0}, {\rm p}_{1}, {\rm p}_{2}, {\rm p}_{3}$  is different from zero.
This remark suggests
a second reformulation of
the problem of proving that the family G3
does not give rise to a Liouvillian solution to
the master equation (\ref{adamsmith}).
This is the content of the next Corollary.

\begin{cor}
\label{cor2}
In order to prove that the
the family G3 does not give rise to a
Liouvillian solution to the master equation (\ref{adamsmith}) it suffices to prove that
 if the confluent Heun equation (\ref{ch0})
 associated with the family G7  admits
 a polynomial solution
$$
{\rm p}(z)= \sum_{{\rm n}=0}^{1+2 s}{ \rm
p}_{\rm
n} z^{\rm n} =
{\rm p}_{0} + {\rm p}_{1} z + {\rm p}_{2} z^{2} + {\rm p}_{3} z^{3} + ... + {\rm p}_{1 + 2 s} z^{1+ 2 s} \ ,
 $$
then at least one of the coefficients
${\rm p}_{0} , {\rm p}_{1}, {\rm p}_{2}, {\rm p}_{3}$ is different from zero.
\end{cor}

\noindent \textbf{Proof} \hspace{0.1cm}
The proof is similar to the proof of
Corollary \ref{cor1} and as such it is
omitted.

\vspace{0.3cm}

\hspace{5 cm}  {\bf The families E7 and E3}

The confluent Heun equation associated with the family E7 is (equation (\ref{ch2}))
$$
z(z-1) {\rm p}^{''}(z)    +
(1-4 z s+ 2 z^{2}s)
{\rm p}^{'}(z)    +
( -l(l+1)+ 2 s-4 s^{2}z )
 {\rm p}(z)=0.
$$

\noindent
Comparison with the general form of the
confluent Heun equation (\ref{conheun}) gives
\be
a=2 s, \ b= - 4 s, \ c=1, \ d=-l(l+1)+ 2 s, \  \rm and  \ \it e=- \rm 4 \it s^{2}.
\ee

According to the equivalence relation
(\ref{equiv}), the substitution \be
\label{subs}
 {\rm p}(z)=z^{2} {\rm p}_{1}(z) \ee into equation (\ref{ch2}) yields
$$
z(z-1) {\rm p_{1}}^{''}(z)    +
(-3 - 2 z( 2 s- 2 )+ 2 z^{2}s)
{\rm p_{1}}^{'}(z)    +
( -l(l+1)-  6 s + 2 - 4 z s( 2 s - 2))
 {\rm p_{1}}(z)=0
$$
which is nothing but the confluent Heun equation
associated with the family E3 (equation (\ref{ch3})).

In subsection \ref{evidence} we give strong evidence which supports the thesis that the families E7 and  E3 do not give rise to
Liouvillian solutions to the master equation
(\ref{adamsmith}).
Remarkably  the equivalence of the
families E7 and  E3, shown by the substitution
(\ref{subs}) into equation (\ref{ch2}),
enables us to reduce the consideration of the two cases  E7 and  E3  into the consideration of the
case E7 only. This is precisely the content of the next  Corollary.


\begin{cor}
\label{cor3}
In order to prove that the
the family E3 does not give rise to
a Liouvillian solution to the master equation (\ref{adamsmith}) it suffices to prove that
the family E7 does not give rise to
a Liouvillian solution to the master equation (\ref{adamsmith}).
\end{cor}
\noindent \textbf{Proof} \hspace{0.1cm}
This is an immediate consequence of the equivalence of the families E7 and E3.
Equation (\ref{subs}) implies that if
the confluent Heun equation (\ref{ch3}),
associated with the family E3, admits
a polynomial solution $\rm p_{1}(z)$,
then the confluent Heun equation (\ref{ch2}),
associated with the family E7, admits
also  a polynomial solution $ {\rm p}(z)=z^{2} {\rm p}_{1}(z) $. Thus if equation
(\ref{ch2}) does not have
 a polynomial solution, and therefore, the family
E7 does not give rise to a
Liouvillian solution to the master equation
(\ref{adamsmith}),
then
equation
(\ref{ch3}) does not have a polynomial
solution either, and therefore, the family
E3 does not give rise to a
Liouvillian solution to the master equation
(\ref{adamsmith}). This completes the proof.




\subsection{The family S3 does not give rise to a Liouvillian solution}

\label{sections3}

For the family
S3 we prove the following
\begin{thrm}
\label{s3}
The family S3 does not give
rise to a Liouvillian solution
to the master equation (\ref{adamsmith}).
\end{thrm}

\noindent \textbf{Proof} \hspace{0.1cm}
A Liouvillian solution arises
from the family S3
if there is  polynomial solution  P to equation
(\ref{gfkkukku})
$$
{\rm P}^{''}+2\theta{\rm P}^{'}+ ( \theta^{2}+\theta^{'}-\nu  )
{\rm P} =0$$
which in this case (equations (\ref{gorgor}) and (\ref{scalar11}) with $\beta=1$) yields
\be
\label{73aaa}
  {\rm P}^{''}(r)    + \frac {   s  r^{2} -4rs + 2r - 2}  {r(r-2)} {\rm P}^{'}(r)    +  \frac    {  -r s (2s-1) -l(l+1) - 2 s } {r(r-2)}   {\rm P}(r)=0.
\ee
The last equation has two regular singular points at $r=0$,
$r=2$, and,  one irregular singular point of Poincare rank 1 at $r=
\infty .$ The
roots of the associated indicial equations are the following
$$
\begin{array}{ccccc}
Singular \; points & & & &  \! \!     \! \!      \! \!   \! \! \!
\!    \! \!    \! \!    \! \!   \! \!   Roots
\\
\! \! \! r=0 & & & & \! \! \! \! \! \!  \! \! \! \! \!   \! \! \!
\!  \! \! \!  \! \!  \!
\rho =0 \ (double \ root),
\\
\! \! r=2 & & & & \! \! \! \! \! \!  \! \! \!  \! \! \! \! \! \!  \! \! \! \rho_{1}=0 \ , \ \rho_{2}=
 2s,
\\
\;  \;     r=
\infty & & & &
     \rho= 2s-1.
\end{array}
$$
One can easily check that
if   equation  (\ref{73aaa}) admits a polynomial solution at all then
this solution will have degree $2 s - 1$, which is precisely the degree
determined by  Kovacic's algorithm.
One can easily check that equation  (\ref{73aaa})
cannot admit polynomial solution of zero
degree. Thus in this proof we will assume hereafter that $2 s$
is an integer which takes values in the set
$ \{2,3,4,...    \}$.

According to Fuchs' Theorem
a basis of the space of solutions of equation (\ref{73aaa})
 in
the interval (0,2),
when the indicial equation associated to
the singular point $r=0$ has a double root
$\rho=0$,
is
given by
\begin{eqnarray}
\label{iuhij;uhnaaab} \hspace{1.5cm} p_{1}(r) & = &
\sum_{{\rm n}=0}^{\infty}{\rm P}_{\rm n}r^{\rm n},  \quad \rm P_{0} \neq 0,
\quad {\rm and,}
\\
\label{iuhij;uhn1aaab} \hspace{1.5cm} p_{2}(r) & = &
\sum_{{\rm n}=0}^{\infty}{\rm k}_{\rm n}r^{\rm n} +
p_{1}{\rm ln}r,  \quad \rm k_{0} \neq 0.
\end{eqnarray}

By substituting ${\rm P}(r)= \sum_{\rm n = 0}^{+ \infty} \rm  P_{\rm n} r^{\rm n}$
into equation (\ref{73aaa}),
we find that the coefficients
$\rm P_{\rm n}$ satisfy the following
three$-$term recurrence relation
\begin{eqnarray}
\label{rrr10aaa}
&  & s ({\rm n}    - 2 s ) \rm P_{{\rm n}-1}
+
(   {\rm n}^{2}  +  {\rm n}   - 4 \it s   {\rm n}
- \it l(l+ \rm 1) - 2 \it s)
\rm  P_{{\rm n}} \nonumber \\
&   &  - 2 ({\rm n}  +1)^{2} \rm  P_{{\rm n}+1}=
 0,
\quad \rm{where}, \ \    {\rm P}_{-1}=0.
\end{eqnarray}
Equation
(\ref{rrr10aaa}) gives the three$-$term recurrence relation satisfied by the coefficients $ \rm P_{\rm n}$
of the power series solution
$p_{1}(r)  =
\sum_{{\rm n}=0}^{\infty}{\rm P}_{\rm n}r^{\rm n},  \rm P_{0} \neq 0,$ given by equation (\ref{iuhij;uhnaaab}).

If    equation  (\ref{73aaa}) admits a polynomial solution $\mathcal P(r)=\sum_{\rm n = 0}^{ 2 s - 1} \rm  P_{\rm n} r^{\rm n}$ then this solution
in the interval (0,2) will admit an expansion
in terms of $p_{1}(r)$ and $p_{2}(r)$ which span
the solution space in this interval. Therefore
\be
\mathcal P(r)= \alpha p_{1}(r) +  \beta p_{2}(r),
\ee
for some $\alpha, \beta \in R.$  The functional forms (\ref{iuhij;uhnaaab})and  (\ref{iuhij;uhn1aaab}) of $p_{1}(r)$ and
$p_{2}(r)$ respectively imply that $\beta=0$ and that
$p_{1}(r)$ has to truncate so that to give a polynomial solution of degree 2$s$-1.
Therefore
\be
\label{prop}
\mathcal P(r)= \alpha \sum_{{\rm n}=0}^{\rm 2 \it s - \rm 1}{\rm P}_{\rm n}r^{\rm n},  \quad \rm P_{0} \neq 0,
\ee
for some $\alpha \in R.$

The coefficients
$\rm P_{\rm n}$ of the power series solution $p_{1}(r)$ (equation (\ref{iuhij;uhnaaab}))  satisfy the three$-$term recurrence relation
(\ref{rrr10aaa}). In particular
the three$-$term recurrence relation
(\ref{rrr10aaa})
for $\rm n = 2 \it s - \rm 1$ and  for $\rm n = 2 \it s  $
 gives respectively
\begin{eqnarray}
\label{rec1}
- s \rm P_{2 \it s - \rm 2} - ( \it l(l+ \rm 1) + 4 \it s^{\rm 2}) \rm P_{2 \it s - \rm 1} - 8 \it s^{\rm 2} \rm P_{2 \it s }&=&0,  \quad \rm and, \\
\label{rec2}
- ( \it l(l+ \rm 1) + 4 \it s^{\rm 2}) \rm P_{2 \it s} - 2 (2 \it s+ \rm 1)^{2} \rm P_{2 \it s + \rm 1}&=&0.
\end{eqnarray}

From equations (\ref{rec1}), (\ref{rec2})
and the fact that the coefficient $ - \rm 2 (\rm n + \rm 1)^{2}$   of $\rm P_{\rm n + 1}$
in the three$-$term recurrence relation
(\ref{rrr10aaa}) is different from zero
for every $\rm n$, and in particular
for every $\rm n= \rm 2 \it s + \rm 1, \rm 2 \it s + \rm 2,... \ ,$
we infer that the power series
solution $p_{1}(r)$, given by equation
(\ref{iuhij;uhnaaab}), does indeed terminate and gives a polynomial solution of degree 2$s$-1 if and only if
\be
\label{tercon}
- s \rm P_{2 \it s - \rm 2} - ( \it l(l+ \rm 1) + 4 \it s^{\rm 2}) \rm P_{2 \it s - \rm 1}=0.
\ee
In this case the coefficients of the polynomials $p_{1}(r)$ and of $\mathcal P(r)$ are proportional
as equation (\ref{prop}) implies.
In any case the coefficients of $\mathcal P(r)$
can only be determined up to a multiplicative constant. Equation (\ref{tercon}) gives
\be
\label{tercona}
\frac{\rm P_{\rm 2 \it s - \rm 1}}{\rm P_{\rm 2 \it s - \rm 2}} = - \frac{s}{\it l(l+ \rm 1) + 4 \it s^{\rm 2}}.
\ee

Moreover, according to Fuchs' Theorem, another basis of the space of solutions of equation (\ref{73aaa})
in
the punctured neighbourhood $(0,2)\cup (2,4)$  of 2, when the difference
$\rho_{2}-\rho_{1}=2 s - 0= 2 s \in \{2,3,4,...    \}$
of the roots $\rho_{1}, \rho_{2}$ of the indicial
equation associated to the singular point $r=2$
is a positive integer,
is
given by
\begin{eqnarray}
\label{iuhij;uhnaaa} \hspace{1.5cm} p_{3}(r) & = &
(r-2)^{2 s} \sum_{{\rm n}=0}^{\infty}  P_{\rm n}(r-2)^{\rm n},  \quad  P_{0} \neq 0,
\quad {\rm and,}
\\
\label{iuhij;uhn1aaa} \hspace{1.5cm} p_{4}(r) & = &
\sum_{{\rm n}=0}^{\infty}{\rm m}_{\rm n}(r-2)^{\rm n} +{\rm C}
p_{1}{\rm ln}|r-2|,  \quad \rm m_{0} \neq 0.
\end{eqnarray}
The constant $\rm C$ may turn out to be zero.

By substituting ${\rm P}(r)=(r-2)^{\rho} \sum_{\rm n = 0}^{+ \infty}  P_{\rm n} (r-2)^{\rm n}$
into equation (\ref{73aaa}), where $\rho=0$ or
$\rho=2 s$, we find that the coefficients
$P_{\rm n}$ satisfy the following
three$-$term recurrence relation
\begin{eqnarray}
\label{rrr10aaab}
&  & s ({\rm n}  + \rho  - 2 s )  P_{{\rm n}-1}
+
(({\rm n} + \rho) ({\rm n} + \rho +1 - 2 s)
- l(l+1) - 4 s^{2}) P_{{\rm n}} \nonumber \\
&   & + ({\rm n} + \rho +1)( 2 ({\rm n} + \rho +1)      - 4 s)  P_{{\rm n}+1}=
 0,
\quad \rm{where}, \ \    {\it P}_{-1}=0.
\end{eqnarray}

When $\rho=2 s$ equation
(\ref{rrr10aaab}) gives the three$-$term recurrence relation satisfied by the coefficients $ P_{\rm n}$
of the power series solution
$p_{3}(r)  =
(r-2)^{2 s} \sum_{{\rm n}=0}^{\infty}  P_{\rm n}(r-2)^{\rm n},   P_{0} \neq 0,$ given by equation (\ref{iuhij;uhnaaa}).
When $\rho=0$ equation
(\ref{rrr10aaab}) gives the three$-$term recurrence relation satisfied by the coefficients ${\rm m}_{\rm n}$
of the
solution
$p_{4}(r)  =
\sum_{{\rm n}=0}^{\infty}{\rm m}_{\rm n}(r-2)^{\rm n} +{\rm C}
p_{1}{\rm ln}|r-2|,  {\rm m}_{\rm 0} \neq 0,$ given by equation (\ref{iuhij;uhn1aaa}), \it
only when $\rm C = 0.$ \normalfont When
$\rm C \neq 0$ the coefficients
${\rm m}_{\rm n}$
of the power series
$\sum_{{\rm n}=0}^{\infty}{\rm m}_{\rm n}(r-2)^{\rm n}$
appearing in the solution
$p_{4}(r)$,
given by equation (\ref{iuhij;uhn1aaa}),
are determined by substituting  $p_{4}(r)$
into equation (\ref{73aaa}).

The  polynomial solution $\mathcal P(r)=\sum_{\rm n = 0}^{ 2 s - 1} \rm  P_{\rm n} r^{\rm n}$
will also admit an expansion
in terms of $p_{3}(r)$ and $p_{4}(r)$
in the interval (0,2), since
$p_{3}(r)$ and $p_{4}(r)$
also span
the solution space in this interval. Therefore
\be
\mathcal P(r)= \gamma p_{3}(r) + \delta  p_{4}(r),
\ee
for some $\gamma, \delta \in R.$  The functional forms (\ref{iuhij;uhnaaa}) and (\ref{iuhij;uhn1aaa}) of
$p_{3}(r)$ and $p_{4}(r)$
imply that $\gamma=0$, $\rm C=0,$
and that the power series
$\sum_{{\rm n}=0}^{\infty}{\rm m}_{\rm n}(r-2)^{\rm n}, \ \rm m_{0} \neq 0$,
which appears in
$p_{4}(r)$ has to truncate so that to give a polynomial solution of degree 2$s$-1.
Therefore
\be
\label{prop1}
\mathcal P(r)= \delta \sum_{{\rm n}=0}^{\rm 2 \it s - \rm 1 }{\rm m}_{\rm n}(r-2)^{\rm n},
\ \rm m_{0} \neq 0,
\ee
for some $\delta \in R.$
The coefficients
$ P_{\rm n}$ of the power series solution $p_{3}(r)$ (equation (\ref{iuhij;uhnaaa}))  satisfy the three$-$term recurrence relation
(\ref{rrr10aaab}) when $\rho=2 s$.
When $\rm C=0$ the coefficients ${\rm m}_{\rm n}$ of the power series $\sum_{{\rm n}=0}^{\infty}{\rm m}_{\rm n}(r-2)^{\rm n}, \ \rm m_{0} \neq 0$,
satisfy also the three$-$term recurrence relation
(\ref{rrr10aaab}) with $\rho=0$.

In particular
the three$-$term recurrence relation
(\ref{rrr10aaab})
for $\rm n = 2 \it s - \rm 1$ and  for $\rm n = 2 \it s  $,
when
$\rho=0$,  gives respectively
\begin{eqnarray}
\label{rec1a}
- s \rm m_{2 \it s - \rm 2} - ( \it l(l+ \rm 1) + 4 \it s^{\rm 2}) \rm m_{2 \it s - \rm 1} + 0 \rm m_{\rm 2 \it s} &=&0, \quad \rm and,  \\
\label{rec2a}
0 \rm m_{\rm 2 \it s- \rm 1} - ( \it l(l+ \rm 1) + 2    \it s  ( \rm 2 \it s - \rm 1)   ) \rm m_{2 \it s} + 2 (2 \it s+ \rm 1) \rm m_{2 \it s + \rm 1}&=&0.
\end{eqnarray}

The following two remarks are now in order
\begin{enumerate}
\item{The coefficients of $\rm m_{\rm n - \rm 1}$, $\rm m_{\rm n}$ and of $\rm m_{\rm n + 1}$ of the three$-$term recurrence relation
(\ref{rrr10aaab}) contain $\rm n$ only
when $\rm n$  is added to $\rho.$
Consequently the coefficients of the relations
resulting from (\ref{rrr10aaab}) when $\rho=0$
and $\rm n = 2 \it s, \rm n = 2 \it s+ \rm 1, \rm n = 2 \it s+ \rm 2$, and $ \rm n = 2 \it s+ \rm 3, ...,$ are the same with the coefficients of the relations resulting from (\ref{rrr10aaab})
when $\rho=2 \it s$
and $\rm n = 0, \rm n =  \rm 1, \rm n =  \rm 2$, \rm and $ \rm n =  \rm 3, ... \ .$}
\item{Both the coefficient of $\rm m_{\rm 2 \it s}$ in (\ref{rec1a})
    and the coefficient of $\rm m_{\rm 2 \it s- \rm 1}$
    in (\ref{rec2a}) are zero.}
\end{enumerate}

From remarks 1 and 2 and equations
(\ref{rec1a}) and (\ref{rec2a})
we conclude that when $\rm C = 0$
we have
\be
p_{4}(r)  =
\sum_{{\rm n}=0}^{\infty}{\rm m}_{\rm n}(r-2)^{\rm n}=\sum_{{\rm n}=0}^{\rm 2 \it s - \rm 1}{\rm m}_{\rm n}(r-2)^{\rm n} + p_{3}(r), \ \rm m_{0} \neq 0.
\ee
Since $p_{3}(r)$ is a solution to equation (\ref{73aaa}), and equation (\ref{73aaa}) is linear,
the polynomial $\sum_{{\rm n}=0}^{\rm 2 \it s - \rm 1}{\rm m}_{\rm n}(r-2)^{\rm n}$ of degree
$ 2 \it s - \rm 1$ is also a solution to
equation (\ref{73aaa}). This explains how the
polynomial solution $$\mathcal P(r)= \delta \sum_{{\rm n}=0}^{\rm 2 \it s - \rm 1 }{\rm m}_{\rm n}(r-2)^{\rm n},
\ \rm m_{0} \neq 0,$$  (equation (\ref{prop1})), if it exists, is derived from
the basis $p_{3}(r), \ p_{4}(r)$, given by equations (\ref{iuhij;uhnaaa}) and (\ref{iuhij;uhn1aaa}), spanning  the solution space of equation (\ref{73aaa}) in the
the punctured neighbourhood $(0,2)\cup (2,4)$  of 2.
We note that equation   (\ref{rec1a}) gives
\be
\label{cond}
\frac{\rm m_{\rm 2 \it s - \rm 1}}{\rm m_{\rm 2 \it s - \rm 2}} = - \frac{s}{\it l(l+ \rm 1) + 4 \it s^{\rm 2}}.
\ee

The key observation is that,
according to Fuchs' Theorem,
in the interval (0,2),
both $p_{1}(r), \ p_{2}(r)$,
given by equations (\ref{iuhij;uhnaaab}), (\ref{iuhij;uhn1aaab})
respectively, and
$p_{3}(r), \ p_{4}(r)$,
given by equations (\ref{iuhij;uhnaaa}), (\ref{iuhij;uhn1aaa})
correspondingly,
constitute a basis of
the space of solutions of equation (\ref{73aaa}).
Thus in the interval (0,2) both equations
(\ref{prop}) and (\ref{prop1}) hold simultaneously. Therefore in the interval (0,2) we obtain
\be
\label{prop2}
\mathcal P(r)= \alpha \sum_{{\rm n}=0}^{\rm 2 \it s - \rm 1}{\rm P}_{\rm n}r^{\rm n}=
\delta \sum_{{\rm n}=0}^{\rm 2 \it s - \rm 1 }{\rm m}_{\rm n}(r-2)^{\rm n},
\quad \rm P_{0} \neq 0,
\ \rm m_{0} \neq 0,
\ee
where $\alpha, \ \delta$ are   non zero   real numbers.

From equation (\ref{prop2}), by equating the coefficients of $r^{\rm 2 \it s - \rm 1}$ and $r^{\rm 2 \it s - \rm 2}$, we obtain
\begin{eqnarray}
\label{terminal1}
\alpha {\rm P}_{\rm 2 \it s - \rm 1} & = & \delta {\rm m}_{\rm 2 \it s - \rm 1}, \\
\label{terminal2}
\alpha {\rm P}_{\rm 2 \it s - \rm 2} & = & \delta \left (  (- \rm 2)   \left( {\begin{array}{*{20}c} \rm 2 \it s - \rm 1 \\ \rm 2 \it s - \rm 2 \\ \end{array}} \right)
{\rm m}_{\rm 2 \it s - \rm 1}  + {\rm m}_{\rm 2 \it s - \rm 2} \right ).
\end{eqnarray}
 Division of equations (\ref{terminal1}) and (\ref{terminal2}) by parts  yields
\be
\label{comb}
\frac{{\rm P}_{\rm 2 \it s - \rm 1}}{{\rm P}_{\rm 2 \it s - \rm 2}}= \frac{1}{  2 (\rm 1 - \rm 2 \it s ) +  \frac{{\rm m}_{\rm 2 \it s - \rm 2}}{{\rm m}_{\rm 2 \it s - \rm 1}}
}.
\ee

By combining  equations  (\ref{tercona}), (\ref{cond}) and    (\ref{comb}) we have
\be
s=0 \ \rm or \ \it s=\frac{\rm 1}{\rm 2}.
\ee

Therefore the compatibility condition (\ref{prop2}),
which must be satisfied if a polynomial solution
to equation (\ref{73aaa}) exists, can be satisfied if and only if $s=0 \ \rm or \ \it s=\frac{\rm 1}{\rm 2}.$ However, when $s=0$ the degree
$\rm d_{\rm s}=\rm 2 \it s - \rm 1$
of the sought polynomial becomes negative,
and when $s=\frac{\rm 1}{\rm 2}$
the degree
$\rm d_{\rm s}$
of the sought polynomial becomes zero.
One can easily check
that equation (\ref{73aaa}) does not admit
polynomial solution of zero degree when
$\it s=\frac{\rm 1}{\rm 2}.$
Consequently the requirement that the compatibility condition (\ref{prop2})
 must be satisfied leads to a contradiction.
We conclude that there is no polynomial solution
to equation (\ref{73aaa}). This completes the proof.

\vspace{0.5cm}

 The method employed in the proof of Theorem \ref{s3}
 applied to the
other remaining cases G3, E3 and E7 does not
lead to contradictions, as it happens in the
case of the family S3,
but leads instead to trivial identities,
and thus no conclusions can be drawn for the
families G3, E3 and E7 by applying the method employed in  the proof
of Theorem \ref{s3}. For the families G3, E3 and E7, in the next subsection, we give strong evidence which supports the thesis that the
families G3, E3 and E7 do not give rise to
Liouvillian solutions to the master equation
(\ref{adamsmith}).


\subsection{Evidence for the  families G3, E3 and E7}
\label{evidence}

In this subsection we give strong evidence which supports the thesis that the
families G3, E3 and E7 do not give rise to
Liouvillian solutions to the master equation
(\ref{adamsmith}).

We start with the family G3.
The family G3 will give rise to a
Liouvillian solution to the master equation
(\ref{adamsmith}) if there is
polynomial solution P to
equation
(\ref{gfkkukku})
$$
{\rm P}^{''}+2\theta{\rm P}^{'}+ ( \theta^{2}+\theta^{'}-\nu  )
{\rm P} =0$$ where, (equation (\ref{ygflygflygflty})),
$$
\theta(r)=
 \frac{5/2}{r} +
\frac{\frac{1}{2}-s}{r-2}+\frac{s}{2}
$$
and, (equation (\ref{gorgor})), $$
\nu(r)= \frac { \frac{s^{2}}{4}r^{4} + l(l+1)r^{2} +
2[\beta-l(l+1)-1]r+3-4\beta}{r^{2}(r-2)^{2}} \ ,
$$
where, $\; \beta=-3.$
One easily finds that if a a polynomial solution P to equation
(\ref{gfkkukku}) exists at all, then this solution will have
degree d=$\;2s-3 ;$ this is precisely the degree of the sought
polynomial dictated by Kovacic's algorithm.

By substituting $
 {\rm P}(r)=\sum_{{\rm n}=0}^{2s-3}{\rm P}_{{\rm n}}r^{{\rm n}} $
into equation (\ref{gfkkukku}) we obtain
that the following 3$-$term recurrence relation holds for the
coefficients $\; {\rm P}_{{\rm n}}\;$ \be \label{liygv,yhvyfvktfc}
-s(2s-2-{\rm n}){\rm P}_{{\rm n}-1}+ ({\rm n}^{2}+5{\rm n}-4s{\rm
n} + 6 - l(l+1)-10s){\rm P}_{{\rm n}} -2({\rm n}+1)({\rm n}+5)
{\rm P}_{{\rm n}+1}=0,  \ee where, ${\rm n} \in
\{0,1,2,...,2s-3 \}$, and, ${\rm P}_{-1}={\rm P}_{2s-2}=0.$

We find that for polynomials of degree d=0,...,500 the determinants
${\rm D}_{{\rm d}+1}$ of the coefficients of the associated $\;
({\rm d}+1) \times ({\rm d}+1) \;$ linear homogeneous systems are
either strictly positive or strictly negative.
In particular, \be \label{ggfgfggfgfgfjk} sgn({\rm D}_{{\rm
d}+1})=(-1)^{{\rm d}+1} \ , \ \ \ {\rm d}=0,1,2,...,500.\ee
We stopped at ${\rm d}=500$ because the required time for the calculation of determinants of matrices of order  $\rm n \times \rm n$, with computers of medium performance, tends
to become large when $\rm n$ approaches and
exceeds the value 500.

In general,
from equation (\ref{liygv,yhvyfvktfc}) we obtain that the
determinant of the associated $\;(2s-2) \times (2s-2) \;$ linear
homogeneous system is tridiagonal and has the form
\noindent

\begin{equation}  \label{conhau1}   \begin{vmatrix}

$\tiny $-$ \rm(\it l\rm(\it l\rm+\rm 1)+10\it s $-$\rm 6$)
& $\tiny $-$10$  & $\tiny 0$ & \cdot & \cdot & $\tiny 0$ & $\tiny 0$ & $\tiny 0$  \\
$\tiny $-$\it s\rm(2\it s $-$\rm3) $& $ \tiny $-$(\it l \rm(\it l \rm + \rm 1)+ 14\it s $-$\rm 12)$ & $\tiny$-$ 24$ & \cdot & \cdot & $\tiny 0 $  & $ \tiny  0 $ & $ \tiny 0  $\\
\cdot & \cdot & \cdot & \cdot & \cdot  & \cdot & \cdot & \cdot  \\
\cdot & \cdot & \cdot  & \cdot & \cdot & \cdot & \cdot & \cdot  \\
\cdot & \cdot & \cdot & \cdot & \cdot  & \cdot & \cdot & \cdot  \\
\cdot & \cdot & \cdot  & \cdot & \cdot & \cdot & \cdot & \cdot  \\
$ \tiny 0 $ & \cdot & \cdot  & \cdot & \cdot &
$   \tiny $-$ 2\it s $
& $ \tiny $-$
(\it l \rm(\it l $+$\rm1)$+$4\it
$s^{\rm 2}$$-$\rm 2)
$
& $ \tiny $-$ \rm(8(\it s$-$\rm2)(\it s$+$\rm1) $+$ \rm10\rm) $ \\
$ \tiny 0 $& \cdot & \cdot  & \cdot & \cdot & \cdot & $\tiny $-$ \it s$ & $ \tiny $-$ \rm(\it l \rm(\it l$+$\rm 1\rm) $+$\rm4$s^{2}$)$
  \end{vmatrix}.
  \end{equation}


Equation (\ref{ggfgfggfgfgfjk}) suggests that the determinant of
the aforementioned tridiagonal matrix is different from zero for
any value of d. By expanding along the last column we find
that this determinant satisfies
the following 3$-$term
recurrence relation
\be \label{.uig.lgl,ygk,hv,lyug}
{\rm D}_{{\rm n}+1}=({\rm
n}^{2}+5{\rm n}-4s{\rm n}+6-l(l+1)-10s){\rm D}_{{\rm n}}-2{\rm
n}s({\rm n}+4)(2s-2-{\rm n}){\rm D}_{{\rm n}-1}, \ee where it is
understood that ${\rm
n}=2s-3,$ ${\rm n} \in \{0,1,2,... \}$, ${\rm D}_{0}=1$, and ${\rm D}_{-1}=0$.
\vspace{0.1cm} \noindent

We examine the similarly
the remaining families
cases
E3 and E7.
By substituting  $
 {\rm P}(r)=\sum_{{\rm n}=0}^{2s-2}{\rm P}_{{\rm n}}r^{{\rm n}}$, and, $
 {\rm P}(r)=\sum_{{\rm n}=0}^{2s}{\rm P}_{{\rm n}}r^{{\rm n}}$
into equation (\ref{gfkkukku}) we obtain that the coefficients $\; {\rm P}_{{\rm n}}\;$ satisfy respectively the
following 3$-$term recurrence relations
\be \label{R1}
-s(2s-1-{\rm n}){\rm P}_{{\rm n}-1}+ ({\rm n}^{2}+3{\rm n}-4s{\rm
n} + 2 - l(l+1)-6s){\rm P}_{{\rm n}} -2({\rm n}+1)({\rm n}+3)
{\rm P}_{{\rm n}+1}=0,  \ee where, ${\rm n} \in
\{0,1,2,...,2s-2 \}$, and, ${\rm P}_{-1}={\rm P}_{2s-1}=0,$
\be \label{R2}
-s(2s+1-{\rm n}){\rm P}_{{\rm n}-1}+ ({\rm n}^{2}-{\rm n}-4s{\rm
n} - l(l+1)+2s){\rm P}_{{\rm n}} -2({\rm n}+1)({\rm n}-1)
{\rm P}_{{\rm n}+1}=0,  \ee where, ${\rm n} \in
\{0,1,2,...,2s\}$, and, ${\rm P}_{-1}={\rm P}_{2s+1}=0.$

We find again that for polynomials of degree d=0,...,500 the determinants
${\rm D}_{{\rm d}+1}$ of the coefficients of the associated $\;
({\rm d}+1) \times ({\rm d}+1) \;$ linear homogeneous systems are
either strictly positive or strictly negative, i.e., we find that
\it{equation} \normalfont   (\ref{ggfgfggfgfgfjk}) \it {holds in all the remaining cases} \normalfont G3, E3 and E7.}

From equations (\ref{R1}) and (\ref{R2})
we obtain that the
determinants ${\rm D}_{{\rm n}}$ of the associated
linear homogeneous systems are tridiagonal and have respectively the form
\begin{eqnarray} \label{D1}
{\rm d}_{({\rm i}+1)({\rm i})}&=&-s(2s-1-{\rm i}), \nonumber \\
{\rm d}_{({\rm i}+1)({\rm i}+1)}&=&({\rm i}^{2}+3{\rm i}-4s{\rm i} + 2 - l(l+1)-6s), \\
{\rm d}_{({\rm i}+1)({\rm i}+2)}&=& -2({\rm i}+1)({\rm i}+3), \nonumber
\end{eqnarray} where, ${\rm i} \in \{0,1,2,...,2s-2 \}$,
\begin{eqnarray} \label{D2}
{\rm d}_{({\rm i}+1)({\rm i})}&=&-s(2s+1-{\rm i}), \nonumber \\
{\rm d}_{({\rm i}+1)({\rm i}+1)}&=&({\rm i}^{2}-{\rm i}-4s{\rm i} - l(l+1)+2s), \\
{\rm d}_{({\rm i}+1)({\rm i}+2)}&=&-2({\rm i}+1)({\rm i}-1), \nonumber
\end{eqnarray} where, ${\rm i} \in \{0,1,2,...,2s \}$.
In both
cases  ${\rm d}_{({\rm k})({\rm j})}$
denotes the element of ${\rm D}_{{\rm n}}$ which belongs to the row ${\rm k}$ and the column ${\rm j}$.

By expanding the previous determinants  along their  last columns  we find
that they satisfy respectively
the following 3$-$term
recurrence relations
\be \label{RD1}
{\rm D}_{{\rm n}+1}=({\rm
n}^{2}+3{\rm n}-4s{\rm n}+2-l(l+1)-6s){\rm D}_{{\rm n}}-2{\rm
n}s({\rm n}+2)(2s-1-{\rm n}){\rm D}_{{\rm n}-1}, \ee where it is
understood that ${\rm
n}=2s-2,$ ${\rm n} \in \{0,1,2,... \}$, ${\rm D}_{0}=1$, and ${\rm D}_{-1}=0$, and,
\be \label{RD2}
{\rm D}_{{\rm n}+1}=({\rm
n}^{2}-{\rm n}-4s{\rm n}-l(l+1)+2s){\rm D}_{{\rm n}}-2{\rm
n}s({\rm n}-2)(2s+1-{\rm n}){\rm D}_{{\rm n}-1}, \ee where it is
understood that ${\rm
n}=2s,$ ${\rm n} \in \{0,1,2,... \}$, ${\rm D}_{0}=1$, ${\rm D}_{-1}=0$.

\indent Equation (\ref{ggfgfggfgfgfjk}) suggests
that the remaining families G3, E3, E7  do not give any new
Liouvillian solutions. Proving this amounts to proving that  equations (\ref{gfkkukku}) in the remaining families
do not have any polynomial solutions. This in fact is an intriguing and difficult problem and
can be cast into different forms. To mention a few:
\begin{enumerate}
\item{Employing (\ref{.uig.lgl,ygk,hv,lyug}), (\ref{RD1}), (\ref{RD2}), and the properties of the tridiagonal
determinants, to prove that equation (\ref{ggfgfggfgfgfjk}) holds for any value of d.}
\item{Using
(\ref{.uig.lgl,ygk,hv,lyug}), (\ref{RD1}), (\ref{RD2}),
and the properties of continued fractions, to prove that equation (\ref{ggfgfggfgfgfjk}) holds for any value of d.}
\item{ Proving that the 3$-$term recurrence
relations (\ref{liygv,yhvyfvktfc}), (\ref{R1}), and (\ref{R2})
do not terminate.}
\item{Extending (\ref{gfkkukku}) to the complex plane and studying its monodromy group in order to prove that
the existence of polynomial solutions is not compatible with the structure of its singularities.}
\end{enumerate}

The main difficulty in approaches  1, 2 and to a lesser extent in 3, which complicates the analysis
considerably, is that in all the remaining families
G3, E3, and E7,
the order of the associated tridiagonal determinants increases with the degree of the corresponding polynomial, e.g.,
in the family
E7 the degree of the polynomial ${\rm P}(r)=\sum_{{\rm n}=0}^{2s}{\rm P}_{{\rm n}}r^{{\rm n}}$ increases with the order of the
determinant (\ref{D2}).

All approaches 1, 2, 3, 4, and variations thereof, were attempted with various degrees of success providing very useful insight
into the nature of the problem but without any definite result.

\subsection{Reduction of the four remaining families to two families and a Conjecture}

\label{reduction}




Kovacic's algorithm (section \ref{KA}, Theorem \ref{kovacic}) constructs all the Liouvillian solutions of equation $y''(x)=\nu(x) y(x)$
  (equation (\ref{katrouba})), $\nu(x)$ is
a rational function over the field of complex numbers, in the form
$\eta(x)=e^{\int\omega(x){\rm d}x}$, where
$\omega(x)$ is an algebraic function
of $x$ of degree $\rm n$ which in general takes the values 1,2,4,6 or 12. The algebraic
function $\omega(x)$
solves an irreducible
algebraic equation $  \Pi(\omega,x)=\sum_{{\rm
i}=0}^{{\rm n}} \frac{{\rm P}_{{\rm i}}(x)} {({\rm n}-{\rm i})!}
\omega^{{\rm i}}=0 $, where ${\rm P}_{{\rm i}}(x)$ are rational
functions of $x$,   (equation (\ref{doladola})).

The search of Kovacic's algorithm for $\Pi(\omega,x)$ is based on
the knowledge of the poles of $\nu$ and consists in constructing
and testing a finite number of possible candidates for
$\Pi(\omega,x)$. If no $\Pi(\omega,x)$ is found then the
differential equation (\ref{katrouba}) has no Liouvillian
solutions.
For each $\rm n$ Kovacic's algorithm constructs
a different finite set of possible candidates for
$\Pi(\omega,x)$.

Each candidate $\Pi(\omega,x)$,
in particular the coefficients $\frac{{\rm P}_{{\rm i}}(x)} {({\rm n}-{\rm i})!}$ of the
algebraic equation $\Pi(\omega,x)=0$,
depend on a rational function $\theta(x)$ and an
associated polynomial $\rm P$ of degree $\rm d$ which solves an ODE with rational function coefficients of order $\rm n + 1.$
The coefficients of this ODE depend on
$\theta$ and on $\nu$. For convenience we
call a pair (d,$\theta$),  a family.

To conclude  the degree $\rm n$ of the
algebraic function $\omega(x)$  depends on
the structure of poles of $\nu(x)$ and can take
the values 1,2,4,6 and 12. For each value of $\rm n$ Kovacic's algorithm
constructs a finite set of families (d,$\theta$).
For each family (d,$\theta$) Kovacic's algorithm
constructs an ODE of order $\rm n + 1.$
If this ODE has a polynomial solution P
of degree d then Kovacic's algorithm constructs, from this polynomial P, an algebraic function
$\omega(x)$, of degree n, which solves the
associated algebraic equation $\Pi(\omega,x)=0.$

Therefore, with running the risk of oversimplification, Kovacic's algorithm reduces the problem of
finding Liouvillian solutions to   equation (\ref{katrouba}), to the problem of finding
polynomial solutions to  an associated set $\mathcal D$  of linear ODEs
whose coefficients
are rational functions over the field of complex numbers.

In section \ref{AKA} we showed, that in the
problem we are examining in this paper, the
structure of poles of the master equation (\ref{adamsmith})
dictates that the possible values of the degree
n of the sought algebraic functions $\omega(x)$
are 1 and 2.
In section \ref{AKA} we gave explicitly, in detail, the families (d,$\theta$) constructed by Kovacic's algorithm, when $\rm n=1,$
and we showed that the only  families remaining for consideration, when $\rm n=1,$
are the families named
G3, G7, G8, E3, E7, and S3.

In subsection \ref{1l} we considered the
family G8. In subsection \ref{G7} we considered the family G7. Therefore the results of the
section \ref{AKA} and the results of the subsections \ref{1l} and  \ref{G7} show
that,
the only  families remaining for consideration, when $\rm n=1,$
are the families
G3, E3, E7, and S3.
In this section we show that the four families G3, E3, E7, and S3 are reduced into two families G3 and E7.
The case $\rm n=2$ is only treated in appendix \ref{app} where we show that, when $\rm n=2$,
there are no families to consider.

We now explain how
 the four families G3, E3, E7, and S3 are reduced into two families G3 and E7.
In subsection \ref{elementarysym}, in Corollary \ref{cor3}, we showed that the examination of the two cases E3 and E7 can be reduced into the
examination of the case E7 only:   In order to prove that the
the family E3 does not give rise to
a Liouvillian solution to the master equation (\ref{adamsmith}) it suffices to prove that
the family E7 does not give rise to
a Liouvillian solution to the master equation (\ref{adamsmith}).

In subsection \ref{sections3} we showed that
the family S3 does not give
rise to a Liouvillian solution
to the master equation (\ref{adamsmith}).
Therefore the four remaining families G3, E3, E7, and S3 are reduced into two
families, the families G3 and E7.
In subsection \ref{evidence} we gave strong evidence which supports the thesis that the
families G3, E7 (and E3) do not give rise to
Liouvillian solutions to the master equation
(\ref{adamsmith}).




Therefore it is tempting to put forward the following

\vspace{0.1cm}

\noindent {\bf Conjecture} { \it
The families G3 and E7 do not give rise
to Liouvillian solutions to the master
equation (\ref{adamsmith}). Consequently
the only ``closed$-$form'' solutions
to the equations describing scalar, electromagnetic and
gravitational radiative perturbations on the static part of the
Schwarzschild space$-$time are those found in the gravitational case, in subsections \ref{1l} and \ref{G7}, and which were
initially found by Chandrasekhar\cite{Chandr}.}

\vspace{0.2cm}

We can examine if the family G3 gives rise to
Liouvillian solutions to the master equation
(\ref{adamsmith})  either by using the methods
1, 2, 3, and 4 given in subsection \ref{evidence}
or/and by using the Corollaries \ref{cor1} and
\ref{cor2} given in subsection \ref{elementary}.

\subsection{Comparison of our results with the results of Couch and Holder}
\label{comparison}

Couch and Holder\cite{C} also
consider
problem 3,
in the list of most difficult problems,  mentioned in
subsection \ref{kovalg}, which we have to solve when we   apply Kovacic's algorithm:
They give a partial proof that there are no
families which give rise to Liouvillian solutions to the equation
which governs the evolution of first order
perturbations, scalar, electromagnetic, or gravitational, to the Schwarzschild geometry
apart from those we found in subsections
\ref{1l} and \ref{G7}, and which were  found initially by
Chandrasekhar\cite{Chandr}.

The main differences between our results on the solution of problem 3 and of the results derived by C.H.\cite{C} on the solution of the same problem are the following:

\begin{enumerate}
\item{In subsection \ref{reduction} we showed that when n=1 the only families remaining for consideration are the families G3 and E7. In subsection \ref{evidence} we gave strong evidence which supports the thesis
    that the families G3 and E7 do not give rise to Liouvillian solutions to the master equation (\ref{adamsmith}). Thus, we conjectured in subsection \ref{reduction},
    that the only Liouvillian solutions
    to the master equation (\ref{adamsmith}) are the Liouvillian solutions found in subsections
\ref{1l} and \ref{G7}. In order to prove that the Conjecture is true we need to prove
that
the families G3 and E7 do not give rise to Liouvillian solutions to the master equation (\ref{adamsmith}).
On the other hand C.H. prove\cite{C}, in their consideration of the problem, that when n=1 no
families give rise to Liouvillian solutions to the master equation which governs the evolution of the first order perturbations of the Schwarzschild geometry  apart from the family which gives rise to the Liouvillian solution we found in subsection
\ref{1l}.    }

\item{
In our consideration of the problem we
are examining in this paper, we proved,
in subsection \ref{1l}, that, in the case of
the gravitational perturbations of the Schwarzschild geometry,     the family G8
gives rise to the Liouvillian solution
$\chi$ to the master equation (\ref{adamsmith}), and in subsection
 \ref{G7}, we proved that,      the family G7
gives rise to the Liouvillian solution
$\chi \int \frac {{\rm
d}r_{\!\ast}}{\chi^{2}}$ to the master equation (\ref{adamsmith}).
On the other hand in the consideration of the
problem by C.H.\cite{C} there exists a family
which gives rise to the Liouvillian solution
$\chi$ but it appears, curiously enough, that there exists no family which gives rise to the Liouvillian solution $\chi \int \frac {{\rm
d}r_{\!\ast}}{\chi^{2}}$. Consequently, needless to say, the proof
we gave in subsection \ref{G7}
is totally absent in the consideration
of the problem by C.H.\cite{C}.
}

\item{In appendix \ref{app} we show that when $\rm n=2$ there are no remaining families
    which could give rise to Liouvillian solutions to the master equation (\ref{adamsmith}).
On the other hand C.H. show\cite{C}, in their consideration of the problem, that when n=2
there are remaining families which could give rise to Liouvillian solutions to the master
equation which governs the evolution of first
order perturbations of the Schwarzschild geometry. As we already have pointed out
in section \ref{FF},
to a great extent C.H. present their results in \cite{C}
in a succinct form. Due to the succinct form
of the presentation of the results in \cite{C}
it is not easy by studying \cite{C} to figure out how many families could give rise to Liouvillian solutions in the study of the
problem by C.H.\cite{C} when $\rm n=2.$
In any case, irrespective of the number of remaining families which could give rise to Liouvillian
solutions in \cite{C} when $\rm n=2$ it is not
easy to prove or disprove that these remaining families do not give rise to Liouvillian solutions to the master
equation which governs the evolution of first
order perturbations of the Schwarzschild geometry. To quote C.H.\cite{C} ``... apparently, it is not straightforward to see  how to settle the question of existence of type 2 Liouvillian solutions in the Schwarzschild case''. In the terminology advocated by C.H. ``type 2 Liouvillian solutions''
are the Liouvillian solutions obtained when $\rm n=2.$
}
\end{enumerate}


In relation with the differences 1, 2, and 3, we note the following:
C.H. \it do not start from the master equation
\normalfont (\ref{adamsmith}).
There is an alternative treatment of
the first order perturbations of the
Schwarzschild geometry, via the Newman$-$Penrose formalism (chapter 29 in \cite{Chandr2}), which leads to a 
differential equation (equation (1) in \cite{C}) different from the master
equation (\ref{adamsmith}). For convenience, let us call $\mathcal P \mathcal E$
the differential equation derived
via
the
Newman$-$Penrose formalism.
The exact form of the differential equation $\mathcal P \mathcal E$ is not
needed.


The following statement, made by C.H.\cite{C}, on the relation between the
master equation (\ref{adamsmith}), which we use as our point of departure, and the
differential equation $\mathcal P \mathcal E$, which is used by C.H.\cite{C} as their point of departure,
is relevant here.

``...The equations resulting from different approaches are related, essentially by linear differential operators. Several of these relations are given in Ref. \cite{Chandr2} and from them it is easy to show that every Liouvillian solution  to $\mathcal P \mathcal E$ generates a Liouvillian solution to the related differential equation (\ref{adamsmith}), and, conversely. We omit the details of the essentially bookkeeping argument which proves this. Hence by finding all Liouvillian solutions of $\mathcal P \mathcal E$  one finds the Liouvillian solutions in other characterizations of the perturbations.''

This statement is correct but it is not evident which are its implications for
the differences 1, 2, and 3, unless one
repeats all over again the analysis of C.H.\cite{C}. The main reasons are: 1) There is no bijective correspondence between the families which arise when we apply Kovacic's algorithm to the master equation (\ref{adamsmith}) and the families which arise when  we apply Kovacic's algorithm to the differential
equation
$\mathcal P \mathcal E$, neither when n=1, nor when n=2. 2)
When we apply Kovacic's algorithm
to the differential equation
$   y''+ay'+by=0, $
(equation \ref{ant}), where $a \, {\rm and} \, b$ are rational functions of a complex variable $x$,
it is not generally true \cite{Duval1} that if one of the
retained families gives rise to a Liouvillian solution  $\eta$ to
equation (\ref{ant}) then another retained family will give rise
to $\eta \int \frac { e^{-\int a}}{\eta ^{2}}$.

However,  the previous statement quoted by C.H. suggests that in the study of the
problem by C.H.\cite{C} there ought to exist a retained family which gives rise
to the Liouvillian solution $\chi \int \frac {{\rm
d}r_{\!\ast}}{\chi^{2}}$. But such a family does not appear in \cite{C}. So either this can be attributed to the
remarks 1) and 2) of the previous paragraph or there is a flaw in the
consideration of the problem by C.H.\cite{C}. To decide between the two
alternatives we need to repeat
all over again the analysis of C.H.\cite{C}, since the succinct
 form
of the presentation of the results in \cite{C} does not allow us to reach a conclusion on the cause of this discrepancy.

Moreover,  the previous statement quoted by C.H. suggests that in our consideration of the problem  the families G3 and E7 do not give rise to Liouvillian solutions to the master equation (\ref{adamsmith}). Furthermore, the same statement,
in the consideration of the problem
by C.H.\cite{C}, suggests that  when n=2 
the remaining
families which could give rise to Liouvillian solutions to the differential
equation $\mathcal P \mathcal E$ do not
in fact do so.

This is in accord with the
evidence given in subsection \ref{evidence} which supports the thesis
that the families families G3 and E7 do not give rise to Liouvillian solutions to the master equation (\ref{adamsmith}), and also, it is in accord with the evidence given by C.H.\cite{C} that the remaining families when n=2 do not in fact give rise to Liouvillian solutions to the
differential
equation $\mathcal P \mathcal E$.

What is even more interesting is that we could in
principle use the bijective correspondence between the Liouvillian solutions of the master equation (\ref{adamsmith}) and the Liouvillian solutions of the differential
equation $\mathcal P \mathcal E$ in order to prove that, in our consideration of the problem, the families G3 and E7 do not give rise to Liouvillian solutions
to the master equation (\ref{adamsmith}), and also, in order to prove that, in the consideration of the problem by C.H.\cite{C}, the remaining families when
n=2 do not give rise to Liouvillian solutions to the  differential
equation
$\mathcal P \mathcal E$.
We expect that the proof will be in line with the proof of the Corollaries
\ref{cor1}, \ref{cor2} and \ref{cor3}.
In order to examine if this is indeed the
case we need again to repeat all over the analysis of Couch and Holder in \cite{C}. 
We leave
for future investigation the problem
of repeating all over the analysis of Couch and Holder in \cite{C} and of clarifying with this analysis the issues raised before.


\subsubsection{The proof of Couch and Holder}

The nicest thing in \cite{C}
is the proof given by Couch and Holder
that when n=1
no
families give rise to Liouvillian solutions to the
differential equation
$\mathcal P \mathcal E$
apart from the family
which gives rise to the Liouvillian solution we found in subsection
\ref{1l}. The proof starts with a
clever rearrangement of the  three$-$term recurrence relations associated with the
remaining families and is done with
induction. Interestingly enough, in our consideration of the problem,
when we try to prove, by following the
method of the aforementioned proof  given by C.H.\cite{C},
that the remaining families
G3, E3, E7 and S3 do not give rise to Liouvillian solutions, we fail bluntly.

The failure is caused by the structure
of the three$-$term recurrence relations associated with the
remaining families G3, E3, E7 and S3.
For this reason we need to employ different methods in order to prove that the remaining families do not give rise to Liouvillian solutions to the master equation (\ref{adamsmith}). One such method is
used in subsection \ref{sections3} where we proved that the family S3 does not give rise to Liouvillian solution to the master equation (\ref{adamsmith}).

Proving that a retained family does not give rise to a Liouvillian solution is
one of the most difficult problems we have to solve when we apply Kovacic's algorithm. This is
problem 3,
in the list of most difficult problems,  mentioned in
subsection \ref{kovalg}, which we have to solve when we   apply Kovacic's algorithm.

This in fact is an intriguing and difficult
problem, it
can be cast into different forms, and
it can be approached with a variety of
solution methods. Four such methods are the methods 1, 2, 3, and 4, given in subsection \ref{evidence}.
Another
method is the one we employ in subsection
\ref{sections3}.
A concrete realization of the method 3  given in subsection \ref{evidence} is the method
of proof given by C.H.\cite{C}.

A key observation regarding the work
of C.H.
is that at the very beginning
of their study
they  use the transformation (\ref{tran}),
which leaves equation
$\mathcal P \mathcal E$
quasi$-$invariant.
C.H. apply Kovacic's algorithm
to the differential equation
$\mathcal P \mathcal E$
by taking into account
its quasi$-$invariance
under the transformation (\ref{tran}).

The quasi$-$invariance of
$\mathcal P \mathcal E$
implies in particular
(we quote C.H.\cite{C})
``So, without loss of generality, we on
occasion take $M$=1 or $\frac{1}{2}$
in order to simplify expressions.''
Throughout their long paper\cite{C} C.H. take $M=1$ with one single exception,
their proof
that
no
families
, when n=1,  give rise to Liouvillian solutions to the
differential equation
$\mathcal P \mathcal E$,
apart from the family
which gives rise to the Liouvillian solution we found in subsection
\ref{1l}, where they take $M=\frac{1}{2}$.
It is highly likely that
the choice $M=\frac{1}{2}$
does not just simplify expressions,
but instead,
it is absolutely essential for the proof
given by C.H..

It is interesting and also useful for future applications of Kovacic's algorithm to  know if this is indeed the
case. In principle the proof of C.H.
can be given for one single value of
$M$, or for a range of values of $M$.
This is allowed by the quasi$-$invariance of the differential equation
$\mathcal P \mathcal E$
under the transformation (\ref{tran}).
A study of the results of C.H., in their present succinct form, does not
allow us to understand if this is indeed the case. For this to happen we need to repeat the analysis of C.H.\cite{C}  all
over again.






\it{

} \normalfont

It is also interesting and  useful for future applications of Kovacic's algorithm to clarify
why the method of proof
 given by C.H.\cite{C} fails, \it for any value of $M$, \normalfont  when we try
 to apply it to the families G3, E3, E7, and S3, which arise in our consideration of the problem when n=1.

All the remaining families in C.H., when
n=1, are associated with three$-$term recurrence relations which have the general form
\be
\label{recrel}
\mathcal C_{{\rm k}-1}(s,{\rm k},l,M)
{\rm P}_{{\rm k}-1}+
\mathcal C_{{\rm k}}(s,{\rm k},l,M)
{\rm P}_{{\rm k}} +
\mathcal C_{{\rm k}+1}(s,{\rm k},l,M)
{\rm P}_{{\rm k}+1}=0,  \ee
where the coefficients $\mathcal C_{{\rm k}-1}(s,{\rm k},l,M), \ \mathcal C_{{\rm k}}(s,{\rm k},l,M),$ and  $ \mathcal C_{{\rm k}+1}(s,{\rm k},l,M)$,
depend on the index ${\rm k}$,
on the parameter $s=2 \rm i \normalfont \sigma$, $\sigma$ is the frequency of the perturbations,  on the angular
harmonic index $l$, and on the mass $M$ of the black hole.

C.H. start their proof\cite{C} with the
three$-$term recurrence relation (\ref{recrel}). They take $M=\frac{1}{2}$, assume that the
recurrence relation terminates to give
a polynomial solution of degree n,
and they reach a contradiction.
In the following relations
we do not set $M=\frac{1}{2}$
only to bring at home
why the method of proof
of C.H.\cite{C} fails when we try to apply it
to the remaining families G3, E3, E7, and
S3, which arise when n=1, in our consideration of the problem.
Thus in the rest of the proof $M$ takes the
value $\frac{1}{2}$, although this is
not explicitly stated.

If the
three$-$term recurrence relation (\ref{recrel}) terminates to
give
a polynomial solution of degree n,
then, the coefficients $\mathcal C_{{\rm k}-1}(s,{\rm k},l,M), \ \mathcal C_{{\rm k}}(s,{\rm k},l,M),$ and  $ \mathcal C_{{\rm k}+1}(s,{\rm k},l,M)$, are such that, if we choose ${\rm P}_{{\rm n}}>0$,
the recurrence relation (\ref{recrel})
gives
\be
\label{an1}
{\rm P}_{{\rm n}-1}>0,  \quad  {\rm P}_{{\rm n}-1}   - {\rm P}_{{\rm n}}    >0.
\ee

The recurrence relation (\ref{recrel})
may be written as
\begin{eqnarray}
\label{recrelA}
&&\mathcal C_{{\rm k}-1}(s,{\rm k},l,M)
({\rm P}_{{\rm k}-1}-{\rm P}_{{\rm k}})=
\mathcal C_{{\rm k}+1}(s,{\rm k},l,M) ({\rm P}_{{\rm k}} - {\rm P}_{{\rm k}+1})-
\nonumber \\
&&(C_{{\rm k}-1}(s,{\rm k},l,M)
+ C_{{\rm k}}(s,{\rm k},l,M)
+ C_{{\rm k}+1}(s,{\rm k},l,M)
){\rm P}_{{\rm k}}.
\end{eqnarray}
What makes the
proof given by \cite{C}
tick
is
the structure of the
coefficients
$\mathcal C_{{\rm k}-1}(s,{\rm k},l,M)$, $\mathcal C_{{\rm k}}(s,{\rm k},l,M),$ and  $ \mathcal C_{{\rm k}+1}(s,{\rm k},l,M)$. 
In particular, the coefficients $\mathcal C_{{\rm k}-1}(s,{\rm k},l,M)$,  $\mathcal C_{{\rm k}}(s,{\rm k},l,M),$ and  $ \mathcal C_{{\rm k}+1}(s,{\rm k},l,M)$
are such that, when $M=\frac{1}{2}$, the
following relations hold
\begin{eqnarray}
\label{an2}
&&\mathcal C_{{\rm k}-1}(s,{\rm k},l,M)>0, \quad \mathcal C_{{\rm k}+1}(s,{\rm k},l,M)>0, \\
\label{an3}
&&
-(C_{{\rm k}-1}(s,{\rm k},l,M)
+ C_{{\rm k}}(s,{\rm k},l,M)
+ C_{{\rm k}+1}(s,{\rm k},l,M)
)>0,
\end{eqnarray}
\it{for every $s,{\rm k}, \  \rm and \ \it l $.}

\normalfont

Equations (\ref{an1}), (\ref{recrelA}), (\ref{an2}), and (\ref{an3}), imply
\be
\label{an4}
 {\rm P}_{{\rm n}-2}   - {\rm P}_{{\rm n}-1}    >0, \quad  {\rm P}_{{\rm n}-2}>0.
\ee
It follows by induction from equations
(\ref{an1}), (\ref{recrelA}), (\ref{an2}),  (\ref{an3}), and
(\ref{an4}), that
\be
\label{an5}
 {\rm P}_{{\rm k}-1}- {\rm P}_{{\rm k}}>0, \quad  {\rm P}_{{\rm k}-1}>0,
\ee
for all ${\rm k}=\rm n, \rm n -1, \rm n - 2,...,1$.

When $\rm k=1$ equation (\ref{an5}) reads
\be
\label{an6}
 {\rm P}_{{\rm 0}}- {\rm P}_{{\rm 1}}>0, \quad  {\rm P}_{{\rm 0}}>0.
\ee
Moreover,
when $\rm k=0$  the three$-$term recurrence relation
(\ref{recrel}) reads
\be
\label{an7}
\mathcal C_{{\rm 0}}(s,0,l,M)
{\rm P}_{{\rm 0}} +
\mathcal C_{1}(s,0,l,M)
{\rm P}_{1}=0.
\ee
We arrive at a contradiction,
because equations (\ref{an6}) and (\ref{an7}) are incompatible: Equation (\ref{an6}), and in particular the inequality
${\rm P}_{{\rm 0}}- {\rm P}_{{\rm 1}}>0$,  implies that
\be
\mathcal C_{{\rm 0}}(s,0,l,M)
{\rm P}_{{\rm 0}} +
\mathcal C_{1}(s,0,l,M)
{\rm P}_{1}>0.
\ee
This completes the proof given by C.H.\cite{C}.

\it{
This
proof
fails when we try to apply it
to the remaining families G3, E3, E7, and
S3, which arise when n=1, in our consideration of the problem. The reason is that in our consideration of the problem, for all the remaining families
G3, E3, E7, and
S3, equation (\ref{an2})
holds for every $s, \rm k, \it l, \ \rm and \ \it M,$
but equation (\ref{an3})
does not hold for any of the remaining families G3, E3, E7, and
S3. For all  the remaining families G3, E3, E7, and S3, and for every value of $M$, the sign of $-(C_{{\rm k}-1}(s,{\rm k},l,M)
+ C_{{\rm k}}(s,{\rm k},l,M)
+ C_{{\rm k}+1}(s,{\rm k},l,M)
)$ depends on $s, \rm k, \ \rm and \ \normalfont l.$ Consequently, in our
consideration of the problem, we cannot use, for any of the remaining families
G3, E3, E7, and S3,  equation
(\ref{recrelA}) in order to derive from it the sign of the difference ${\rm P}_{{\rm k}-1}-{\rm P}_{{\rm k}}$
from the signs of ${\rm P}_{{\rm k}}-{\rm P}_{{\rm k}+1}$ and ${\rm P}_{{\rm k}}$.
}

\normalfont

\subsection{Comments on the
implementation of Kovacic's algorithm}

\label{comments}

Kovacic's algorithm is based in the study of the structural properties of the Galois group of the differential equation $ y''=\nu y,
\nu \in C(x)$ (equation (\ref{katrouba})).
Our results and the results of C.H.\cite{C}
suggest that the implementation of
Kovacic's algorithm can become more
efficient when it is properly
combined with the search for other
types of symmetry.

In our consideration of the problem, in subsection \ref{elementarysym}, we showed
by using the
elementary symmetries of the confluent
Heun equation, that the examination of the two cases E3 and E7 can be reduced into the
examination of the case E7 only
(Corollary \ref{cor3}).

For every Fuchsian differential equation
with n singular points and its various
confluent forms there is a finite group
of transformations which acts on the
parameter space of the equation
and generates alternative expressions for its family of local solutions.
This group is of order n! $\rm 2^{\rm n -1}$ for every Fuchsian differential equation
with n singular points.

In particular, for the hypergeometric equation  this group is of order 24, and for the Heun's equation is of order
192 (see e.g. \cite{Maier}).
Moreover, for the confluent Heun equation
this group is of order 16\cite{El}.
It is precisely this group of elementary
symmetries we use in subsection \ref{elementarysym}
in order to reduce the examination
of the two cases E3 and E7  into the
examination of the case E7 only.
This group, in any case, can be derived in a systematic
fashion with the method expounded e.g. in
\cite{Maier}.


A key observation regarding the work
of C.H.
is that at the very beginning
of their study
they  use the transformation (\ref{tran}),
which leaves equation
$\mathcal P \mathcal E$
quasi$-$invariant.
In principle a proof, such that given by C.H.\cite{C},
can be given for one single value of
$M$, or for a range of values of $M$.
This is allowed precisely by the quasi$-$invariance of the differential equation
$\mathcal P \mathcal E$
under the transformation (\ref{tran}).

In general, if not in the problem under consideration,
transformations of the form (\ref{tran})
can be of crucial importance
when we study retained families,
and examine if they give rise to
Liouvillian solutions.
 C.H. do not explain how the transformation (\ref{tran})
can be  derived from first principles.
What is not well known, and is of
outmost importance for the implementation
of Kovacic's algorithm, is that
such transformations can be derived
in a systematic way\cite{Chris}
by using the theory of Lie groups
as this is applied to differential
equations (see e.g. \cite{Olver}).

Proving that a retained family does not give rise to a Liouvillian solution is
one of the most difficult problems we have to solve when we apply Kovacic's algorithm.
This in fact is an intriguing and difficult
problem, it
can be cast into different forms, and
it can be approached with a variety of
solution methods. Four such methods are the methods 1, 2, 3, and 4, given in subsection \ref{evidence}.

Another
method is the one we employ in subsection
\ref{sections3}.
This method is in general applicable
not only to second order ODEs,
which arise when n=1, but it is also applicable to  ODEs of any order,
and in particular it is applicable
to ODEs of order n+1
which arise when
n=2,4,6, and 12.

A concrete realization of the method 3  given in subsection \ref{evidence} is the method
of proof given by C.H.\cite{C}.
This method is in general applicable
not only to three$-$term recurrence
relations, but it is also applicable to  higher order recurrence relations
which may well arise
when n=2,4,6, and 12,
if not when n=1.

The previous remarks suggest that the
implementation of Kovacic's algorithm
becomes more efficient when we follow the
next steps
\begin{enumerate}
\item{By using the theory of Lie groups as this is applied to differential equations, we search for
    transformations of the form (\ref{tran}) which leave equation $ y''=\nu y, \
\nu \in C(x),$ \ quasi$-$invariant. In general equation $ y''=\nu y, \
\nu \in C(x),$ may contain accessory parameters. In the case of the perturbation theory of black hole geometries such accessory parameters are the mass $M$, the charge $Q$, and the angular momentum $J$  of the black hole. In this case,
if such a group of transformations
which leave equation $ y''=\nu y, \
\nu \in C(x),$ \ quasi$-$invariant
does exist,
then
equation  $ y''=\nu y$
will retain its
form under this group of transformations, whereas $M$, $Q$, and $J$,
will be replaced respectively by $\mathcal M$, $\mathcal Q$,
and $\mathcal J$, where
\begin{eqnarray}
\label{sub1}
\mathcal M &=& f_{\mathcal M}(M,Q,J), \\
\label{sub2}
\mathcal Q &=& f_{\mathcal Q}(M,Q,J),
\\
\label{sub3}
\mathcal J &=& f_{\mathcal J}(M,Q,J).
\end{eqnarray}

}

\item{We apply Kovacic's algorithm to equation $ y''=\nu y, \
\nu \in C(x).$}

\item{Every remaining family, which in principle could give rise to a Liouvillian solution is associated to an ODE, which for example when n=1 is of second order and has the form ${\rm P}^{''}+2\theta{\rm P}^{'}+ (
\theta^{2}+\theta^{'}-\nu  ) {\rm P}=0$ (equation (\ref{gfkkukku})). If this
ODE, e.g. if equation (\ref{gfkkukku}),
has a polynomial solution then a Liouvillian solution arises to
equation $ y''=\nu y, \
\nu \in C(x).$
For every such ODE we find a polynomial solution, or else we prove that it does
not admit a polynomial solution.
}

\item{When n=1 each family is associated with an ODE which has the form ${\rm P}^{''}+2\theta{\rm P}^{'}+ (
\theta^{2}+\theta^{'}-\nu  ) {\rm P}=0$ (equation (\ref{gfkkukku})). When n=2 each family is associated with an ODE which has the form ${\rm P}^{'''}
+3\theta{\rm P}^{''}+
(3\theta^{2}  +  3\theta^{'}  - 4 \nu               ){\rm P}^{'}+
(\theta^{''} + 3 \theta \theta^{'}
+ \theta^{3} - 4 \nu \theta - 2 \nu^{'}
) {\rm P}=0$ (as we easily find
from equation
(\ref{karokaro})
).
Similarly, from equation
(\ref{karokaro}),
we find the form of ODEs associated with the families which arise
when n=4, 6, and 12.
For each such ODE we find
the finite group of transformations which acts on the parameter space of the
equation and generates alternative expressions for its family of local solutions.
By using this group we can reduce the
number of remaining families which
could give rise to Liouvillian solutions to equation
$ y''=\nu y, \
\nu \in C(x).$
In the problem we
consider in this paper
it is precisely this finite group
of symmetries of the confluent Heun equation, of order 16,
we use in subsection \ref{elementarysym}
in order to reduce the examination
of the two families E3 and E7  into the
examination of the family E7 only.
}

\item{Proving that a retained family does not give rise to a Liouvillian solution is
one of the most difficult problems we have to solve when we apply Kovacic's algorithm, and
it can be approached with a variety of
solution methods. Four such methods are the methods 1, 2, 3, and 4, given in subsection \ref{evidence}. A concrete realization of the method 3  given in subsection \ref{evidence} is the method
of proof given by C.H.\cite{C}.}

\item{This method of proof,
as well as the method of proof
we employ in subsection
\ref{sections3} and the
methods of proof 1, 2, 3, and 4, we give in subsection \ref{evidence},
can be greatly facilitated if we take into account
the group of transformations, when such a group
exists, which
leaves equation $ y''=\nu y, \
\nu \in C(x)$   quasi$-$invariant,
and if we use the subsequent
relations
(\ref{sub1}), (\ref{sub2}), and (\ref{sub3}).
This group of transformations,
and the
subsequent relations
(\ref{sub1}), (\ref{sub2}), and (\ref{sub3})
allow us to give such proofs
for specific values, or for specific ranges of values, of the parameters $M$, $Q$, and $J$.
Even more, in general, we expect that some of these proofs
cannot be given
if we do not employ the quasi$-$invariance of $ y''=\nu y, \
\nu \in C(x)$, and the
subsequent  relations
(\ref{sub1}), (\ref{sub2}), and (\ref{sub3}). }

\item{In the search for polynomial solutions to the ODEs associated to the various values of n, and in particular in the search for Liouvillian pairs $\eta$ and $\eta \int \frac { e^{-\int a}}{\eta ^{2}}$, we can also employ the method of proof given in subsection \ref{G7}. }

\item{
Furthermore, in the search for polynomial solutions to the ODEs associated to the various values of n, both in
other black hole geometries,
4$-$dim and higher \cite{Ida}, and
in every other application of
Kovacic's algorithm,
we examine, if under certain conditions (corresponding to the $c=j \ \rm( =0, 1,2,...,fixed)$ condition in the Schwarzschild case)
the square matrix $\mathcal M$ of the linear homogeneous system associated to these ODEs
takes the form $
\mathcal M = \begin{pmatrix}
    \mathcal A & \bf O \\
\mathcal D & \mathcal B
  \end{pmatrix}.$ If this  is indeed the case then this will be an indication that a generalization of  Hautot's analysis can be applied to these ODEs, generalization similar to the generalization given in subsections \ref{extension}, \ref{truncon} and \ref{Laguerre1}.  As a result of this analysis, the polynomials contained in the derived
  Liouvillian solutions, will be
  written as  finite sums of special functions of Mathematical Physics.
We elaborate more on this and related issues in section \ref{D}
which follows.
}

\end{enumerate}

To the best of our knowledge, in the applications of Kovacic's algorithm so far, the steps 1, 4, 6,  7, and 8 which greatly facilitate its thorough and
complete implementation,
have not been employed.

\section{Discussion}

\label{D}

Regarding the results we obtained and the related open questions and problems the following remarks are now in order:

\begin{enumerate}


\item{Our results are apt to generalization to other black hole geometries, 4$-$dim and higher \cite{Ida}.
    It is an
    interesting open problem to apply Kovacic's algorithm to the master equation which governs the evolution of first order perturbations of other black hole geometries, 4$-$dim and higher \cite{Ida},  identify Liouvillian pairs, if they exist at all, as we did in section \ref{gl},  
     find the polynomial solutions to the  differential equations
    (\ref{karokaro}) 
    associated to the Liouvillian solutions   in ``closed form'', as we did in subsection \ref{G7},
    and express them as a finite sum
    of special functions of Mathematical Physics as we did in subsections \ref{truncon} and \ref{Laguerre1}.
    }

    \item{ As we proved in sections \ref{gl} and \ref{rem}, in the case of
    the perturbations of the Schwarzschild geometry,
    the  differential equation
(\ref{karokaro})
    associated to the Liouvillian solutions  turns out to be the confluent Heun equation (\ref{hautot1}). In the case of other black hole geometries, 4$-$dim and higher \cite{Ida}, equation (\ref{karokaro}) will turn out to be, in general, other type of equation,
    even of third or higher order.    }

    \item { Also, it will be interesting to examine, if in
     the case of other black hole geometries,
     Hautot's results can be
    generalized, and extended, if necessary,  as they were extended in the case of Schwarzschild geometry in  subsections \ref{extension}, \ref{truncon} and \ref{Laguerre1}, so that the polynomials  contained in the corresponding  Liouvillian solutions can be written as a finite sum of special functions
    of Mathematical Physics.
}

\item{The crucial starting remark made by Hautot upon which his whole analysis is based is that if $c=j \ \rm( =0, 1,2,...,fixed)$ in $z(z-1) {\rm P}^{''}(z) + (a z^{2}  + b z + c) {\rm P}^{'}(z) + (d + e z + f z^{2}){\rm P}(z)=0$ (equation (\ref{hautot})), then the square matrix $\mathcal M$ of the associated linear homogeneous system takes the form
    $
\mathcal M = \begin{pmatrix}
    \mathcal A & \bf O \\
\mathcal D & \mathcal B
  \end{pmatrix}
$, and consequently (equation (\ref{det1}))
$$
det(\mathcal M)=det(\mathcal A) \cdot det(\mathcal B).
$$
$det(\mathcal A)=0$ becomes one of Hautot's sufficient conditions (equation (\ref{con7})).
}

\item{Therefore, the starting point  in the generalization of Hautot's results to other black hole geometries,
4$-$dim and higher \cite{Ida},  will be to examine, if under certain conditions (corresponding to the $c=j \ \rm( =0, 1,2,...,fixed)$ condition in the Schwarzschild case)
the square matrix $\mathcal M$ of the linear homogeneous system associated to   the differential equation
(\ref{gfkkukku})
takes the form $
\mathcal M = \begin{pmatrix}
    \mathcal A & \bf O \\
\mathcal D & \mathcal B
  \end{pmatrix}.$
}

\item{In the case of the
gravitational perturbations of the Schwarzschild geometry,
and in particular in case G7 of Kovacic's algorithm, Hautot's sufficient condition $det(\mathcal A)=0$
is equivalent to condition (\ref{tersol}), i.e., to the
condition that the frequency $s$ of the perturbations takes the
values
$s=
\frac{l(l-1)(l+1)(l+2)}{6}, \ l=2,3,..., \ $ of the frequencies of the algebraically special perturbations\cite{Chandr,Couch} of the Schwarzschild geometry!
}

\item{As we proved in subsections \ref{extension}, \ref{truncon} and \ref{Laguerre1}, in the case
    G7 of the gravitational perturbations of the
    Schwarzschild geometry, which is the case of
    Chandrasekhar's Liouvillian solution $\chi \int \frac {{\rm d}r_{\!\ast}}{\chi^{2}}$, studied in subsection
    \ref{G7},
     when $s=
\frac{l(l-1)(l+1)(l+2)}{6}, \ l=2,3,..., \ $
 the differential equation (\ref{gfkkukku}),
 which is the confluent Heun equation, an equation
 with three singular points, admits polynomial solution
 which can be written as a finite sum of truncated confluent hypergeometric functions of the first kind, and also, as a sum of associated Laguerre polynomials; both the truncated confluent hypergeometric functions of the first kind, and the associated Laguerre polynomials  satisfy differential equations with two singular points!
 So, it appears, as Hautot remarks\cite{Hautot}, that one singular point disappears!
 }

\item{The most important aspect in Hautot's work\cite{Hautot,Hautot1,Hautot2,Hautot3}
    is his proof that certain ODEs
    of second order with polynomial coefficients
    and three singular points, when certain necessary conditions are satisfied, admit polynomial solutions
    which can be expressed as a finite sum of polynomials
    which satisfy differential equations with two singular
    points. On the first sight it appears that this is a
    purely ``mathematical'' fact which does not have any
    ``physical'' underpinnings, and/or connections with problems coming from Physics.}

\item    {One of the most important spin$-$offs of the  work we present in this paper is that Hautot's sufficient conditions, in the case of
    the gravitational perturbations of the Schwarzschild geometry, are nothing but the condition
 that the frequency  of the perturbations takes the
values
  of the frequencies of the algebraically special perturbations\cite{Chandr,Couch} of the Schwarzschild geometry. There is no reason whatsoever that we could
have anticipated this result.}

\item { This suggests that, at least in the case of
    the gravitational perturbations of the Schwarzschild geometry,  Hautot's
remarkable result, when certain sufficient conditions are satisfied
a singularity of an ODE
 be removed as a method of solving this differential equation,  is inextricably interwoven with the existence of  \it algebraically special \normalfont gravitational   perturbations in this geometry. Thus it is plausible to conjecture that the removal of the singularity has a ``physical origin''. }

\item{Thus the ``physical origin'', if it exists, is
expected to be related to the gravitational algebraically special perturbations of the Schwarzschild geometry. A   natural measure of the
complexity of a (confluent) Fuchsian equation is the
number of its singular points. This suggests that
a possible candidate for the ``physical origin''
of the removal of the singularity in the confluent Heun equation (\ref{hautot1}) is  the extremum of the correction  to the entropy of the Schwarzschild black hole due to the presence of a
gravitational algebraically special perturbation. Needless to say this is highly speculative.}

\item{ The extremum is understood with respect to a general, not algebraically special, gravitational perturbation of the Schwarzschild geometry, and/or respect with other type of perturbations of the Schwarzschild geometry.
     The correction to the entropy
of the Schwarzschild black hole due to the presence of
gravitational algebraically special perturbations, or due to the presence of other type of perturbations, can be calculated with the brick wall model of 't Hooft (see e.g. \cite{Ghosh}). }


\item{Hautot makes  his crucial observation about the confluent Heun equation, namely, when certain sufficient  conditions are satisfied, the confluent Heun equation admits polynomial solution
    which can be expressed as a sum of truncated confluent hypergeometric functions of the first kind in \cite{Hautot}. Moreover,  as we clarified in subsection \ref{Laguerre} the aforementioned polynomial solution can also be expressed as a sum of associated Laguerre polynomials. The important point, as Hautot points out \cite{Hautot} is that the confluent Heun equation has three singular points, whereas, hypergeometric functions (and Laguerre polynomials) satisfy differential equations with two singular points.
So it seems that a singular point has disappeared!
}

\item{Hautot in a series of papers \cite{Hautot1,Hautot2,Hautot3} gives other confluent Fuchsian equations of second order, with three singular points, where the same phenomenon occurs: They admit  polynomial solutions which are linear combinations of polynomials which satisfy confluent Fuchsian equations of second order with two singular points. However, what is  of outmost  significance for our study, is that Hautot does not give an explanation of the phenomenon.
    This phenomenon, is important on its own right, and irrespective of any possible ``physical origin'' it might have, it needs to be explained
    mathematically.}

\item{Not surprisingly, this phenomenon, after its initial observation by Hautot\cite{Hautot}, attracted further attention, and to the best
    of our knowledge, the most complete and thorough
    study with a mathematical explanation of it,
    was given by Craster and Shanin in \cite{Shanin}
    (see also \cite{Tak} for a subsequent study based on \cite{Shanin}). It is somehow surprising that
    it took  thirty years before a systematic
    theory about Hautot's observation could be developed. }

\item{    The key observation in \cite{Shanin}  is that a (confluent) Fuchsian
    ODE of second order may have \it false \normalfont singular points; i.e.,
    singular points which are singular points of the coefficients of the ODE but not of the solutions of the ODE. This suggests that when a (confluent) Fuchsian ODE has false singular points it may have solutions
    which are expressed as linear combinations of functions which solve ODEs with coefficients which do not have the false singular points as singular points, and in this sense they are simpler than the initial ODE.}

\item{In \cite{Shanin} it is proved that this is indeed the case when the false singular points have roots of the associated indicial equations
    0 and $N$, where $N$ is a positive integer, and the local expansions around these false singular points contain no logarithmic terms in their developments. Let us call these false singular points of type $\mathcal A.$ In fact Shanin and Craster prove that  false singular points of type $\mathcal A$  can be removed
    by using isomonodromy mappings of the solution space of the ODE and show that the same isomonodromy mappings can be used in order to find
    solutions of the ODE which are expressed as linear combinations of functions which solve simpler ODEs 
    which do not have in their coefficients the
    false singular points of type $\mathcal A$ as singular points.}

\item{    Interestingly enough
the singular point 1 of the
confluent Heun equation  (\ref{hautot1}) is a false singular point of type $\mathcal A$.
Even more what lies behind Hautot's observation
\cite{Hautot,Hautot1,Hautot2,Hautot3}
is the theory  developed by  Shanin and Craster in
 \cite{Shanin}. It will be interesting to derive Hautot's results, and in particular expansion (\ref{sum}) with coefficients $A_{k}$ given by the linear homogeneous system (\ref{recrel}), when condition (\ref{conhau}) is satisfied, by using the theory developed in \cite{Shanin}.  Moreover, it
 will be interesting to investigate if the extension of Hautot's results, given  in subsections \ref{extension}, \ref{truncon} and \ref{Laguerre1}, can be derived by using the theory developed in \cite{Shanin}, or if an extension and/or modification of this theory is needed, in order to derive the extension of subsections \ref{extension}, \ref{truncon} and \ref{Laguerre1}, by using this theory. }

\item{Hautot's results, in conjunction with the
underlying theory developed in \cite{Shanin},
provide the means in order to address a whole
new area of research pertaining to black hole physics
and to the theory of solvability of ODEs. The objective of the  part of research related to black hole physics will be to find solutions in
quadratures of the first order perturbation equations
of the black hole geometries in 4$-$dim and higher \cite{Ida}. The aim of the  part of research related to the theory of solvability of ODEs will be to find
solutions of (confluent) Fuchsian ODEs by removing false singular points from these equations and explore
the connections with the solvability theory of other types of equations.}

\item{Guiding problems in the part of this research
    which pertains to black hole physics are:
\begin{enumerate}
\item{The application of Kovacic's algorithm to the first order perturbation equations of black hole space$-$times, 4$-$dim and higher, in order to find  all solutions in quadratures of these equations.}
\item{The application of Hautot's observation, in the way indicated in the remarks 4 and 5 in this section, with the simultaneous  application of the theory developed in \cite{Shanin}, in order to solve the associated ODE, the associated ODE in the problem under consideration is the ODE given in equation  (\ref{gfkkukku}),   in each case.   }
\end{enumerate}

Verifying that the theory in  \cite{Shanin}
gives Hautot's results\cite{Hautot,Hautot1,Hautot2,Hautot3} is the first natural step for the materialization of this research program.
}

\item{Guiding problems in the part of this research
    which pertains to the solvability theory of ODEs are:
\begin{enumerate}
\item{The extension of the theory developed by Shanin and Craster in \cite{Shanin} to (confluent) Fuchsian equations of third order and higher. Shanin and Craster  developed their theory for second order (confluent) Fuchsian equations. }
\item{The study of the connections of the solvability theory of (confluent) Fuchsian equations of second order developed in \cite{Shanin} with the solvability theory of other type of equations, e.g., with the solvability theory of Painlev$\acute{\rm e}$  equations, and also, as Shanin and Craster point out\cite{Shanin}, with generalizations of the    Darboux$-$Halphen system.}
\end{enumerate}
}

\item{A different mathematical explanation of Hautot's observation can be given with the use of group theory, and in particular with the use of tridiagonalization\cite{Gr}. Tridiagonalization is a cogent algebraic framework via which the Heun differential operator, and two of its confluent forms,
    the confluent Heun operator and the double confluent Heun operator can be obtained from the hypergeometric operator, the confluent hypergeometric operator, and the Hermite operator respectively.   }

\item{  It will be interesting to derive Hautot's results, and in particular expansion (\ref{sum}) with coefficients $A_{k}$ given by the linear homogeneous system (\ref{recrel}), when condition (\ref{conhau}) is satisfied, by using the
algebraic framework of tridiagonalization\cite{Gr}.
Moreover, it
 will be interesting to investigate if the extension of Hautot's results, given  in subsections \ref{extension}, \ref{truncon} and \ref{Laguerre1}, can be derived by using tridiagonalization,
 or if an extension and/or modification of this
algebraic framework
 is needed in order to derive this extension.
  }

\item{
The theory developed by Shanin and Craster in \cite{Shanin},  is based on the notion of the false singular point of a (confluent) Fuchsian equation of second order, and uses isomonodromy mappings of
its solution space
in order to remove the false singular point from its coefficients,  and derive solutions of
this equation
which are expressed as linear combinations of functions which solve simpler ODEs, i.e. ODEs which do not have the false singular point in their coefficients.}

\item{ By doing so Shanin and Craster
effectively construct more involved operators from simpler ones. They
use the Heun equation, which is the more general Fuchsian equation of second order with four regular singular points to develop their theory. In the  subsection 4.3 of their paper they also consider the confluent Heun equation.
They effectively construct the Heun operator from the hypergeometric operator and the confluent Heun operator from the confluent hypergeometric operator.
}

\item{     On the other hand tridiagonalization constructs explicitly\cite{Gr}  more involved operators from simpler operators, and in particular, it constructs the Heun operator from
    the hypergeometric operator, the  confluent Heun operator from the confluent hypergeometric operator, and the double confluent Heun operator from the Hermite operator. It will be interesting
to compare the results of the two approaches, that of Shanin and Craster\cite{Shanin} and that of tridiagonalization\cite{Gr}, and see if their results coincide in the cases of the Heun operator,  of the confluent Heun operator, and of the double confluent Heun operator.}

\item {  Moreover, it will be interesting to examine the importance of the notion of the false singular point in tridiagonalization, and also it will be interesting to investigate how the two approaches complement each other and if they can merge in a coherent framework. Hautot's results\cite{Hautot,Hautot1,Hautot2,Hautot3} and their extension, given in subsections \ref{extension}, \ref{truncon} and \ref{Laguerre1}, provide guidance and a test bed for this investigation.  }

\item{It  is an open problem to find the  whole set of polynomial solutions to the confluent Heun equation; some partial results have appeared in the literature \cite{Hautot,Hautot1,Hautot2,Hautot3,Ish,Ish1} but the whole set of polynomial solutions to this equation is missing.   Hautot's results  and their extension, given in subsections \ref{extension}, \ref{truncon} and \ref{Laguerre1}, give only a subclass of the polynomial solutions to the confluent Heun equation, those solutions which can be expressed   as a finite sum of truncated confluent hypergeometric functions of the first kind, and also, as a finite sum of associated Laguerre
     polynomials.}


\end{enumerate}


\newpage

\appendix
\section{The case n=2}
\label{app}
Here  the ${\rm n}=2$ case is presented. Having set ${\rm n}=2$ we go
to step \textbf{2}.


\noindent
{\bfseries  Second step:}


\noindent
{\bfseries 2a.}
\hspace{0.5cm}
It is not applicable since ${\rm n}=2.$

\noindent
{\bfseries 2b.}
\hspace{0.5cm}
From equations (\ref{gyretaate}), (\ref{koloskalos}) it follows that
\begin{eqnarray}
{\rm E}_{0} & = &
\label{dasiisad}
\left \{ 2-4\sqrt{1-\beta}, \; 2, \;2+4\sqrt{1-\beta} \right \} \cap Z.
\end{eqnarray}
Equation (\ref{dasiisad}) gives
\begin{eqnarray}
\label{huroihyroi}
{\rm For} \;\;\; & \beta=-3 & \;\;\; {\rm E}_{0}=\left \{ -6,2,10 \right \} \\
\label{frodossodorf}
{\rm For} & \beta=0 & {\rm E}_{0}=\left \{ -2,2,6 \right \} \\
\label{gufesaas}
{\rm For} & \beta=1 & {\rm E}_{0}=\left\{ 2 \right \}
\end{eqnarray}
For the second root $r=2$ we  obtain \be \label{gdnkk} {\rm
E}_{2}= \left \{ 2-4s, \; 2, \; 2+4s \right \} \cap Z. \ee
Equation (\ref{gdnkk}) implies \be \label{boukibouko} {\rm
E}_{2}=\left \{ 2-4s, \;2, \; 2+4s \right \} \qquad {\rm where}
\quad s=\frac{{\rm k}}{4}-\frac{1}{2},\; {\rm k} \in Z. \ee


\noindent
{\bfseries 2c.}
\hspace{0.5cm}
It is not applicable since ${\rm n}=2.$


\noindent
{\bfseries 2d.}
\hspace{0.5cm}
Since $\Gamma_{4}=\left \{ \infty \right \}$ (see equation (\ref{kolossss})),
equation (\ref{nugehamm}) implies
\be
\label{nubonobu}
{\rm E}_{\infty}=\left \{ 4 \right \}.
\ee


\noindent
{\bfseries  Third step: }


\noindent
{\bfseries 3a-3b.}
\hspace{0.5cm}
Making use of equations
(\ref{huroihyroi}), (\ref{frodossodorf}), (\ref{gufesaas}),
(\ref{boukibouko}), and (\ref{nubonobu}), one can readily find the families
$\underline{\rm e}= \left ( {\rm e}_{c} \right )_{ c \in \Gamma}$ of elements
${\rm e}_{c} \in {\rm E}_{c}$. Since ${\rm E}_{0}$ depends on the value of
$\beta$ we distinguish three cases.
\vspace{0.2cm}

\centerline{\boldmath${\beta=-3}\;\;\;$  \qquad
$\bf (Gravitational \; \; case)$}
\vspace{-1cm}
\begin{eqnarray}
\mathbf{ e_{0}} \qquad \qquad \qquad \qquad \qquad
& \quad \mathbf{ e_{2}} \qquad \qquad \qquad \qquad \qquad &
\mbox{\boldmath${{\rm e}_{\infty}}$} \nonumber \\
\!\!\!\! -6 \qquad \qquad \qquad \qquad  \qquad &   2-4s  \qquad \qquad
\qquad \qquad \qquad & \;  4 \nonumber \\
\!\!\!\! -6 \qquad \qquad \qquad \qquad \qquad  &   2 \qquad \qquad \qquad \qquad
\qquad  & \;  4\nonumber \\
\!\!\!\! -6 \qquad \qquad \qquad \qquad \qquad &   2+4s \qquad \qquad \qquad \qquad
\qquad & \;   4 \nonumber \\
 2 \qquad \qquad \qquad \qquad \qquad &   2-4s
\qquad \qquad \qquad \qquad \qquad & \;  4 \nonumber \\
2 \qquad \qquad \qquad \qquad \qquad &  2
\qquad \qquad \qquad \qquad \qquad & \;  4 \nonumber \\
 2 \qquad \qquad \qquad \qquad \qquad &   2+4s
\qquad \qquad \qquad \qquad \qquad& \; 4 \nonumber \\
\!\! 10 \qquad \qquad \qquad \qquad \qquad &2-4s
\qquad \qquad \qquad \qquad \qquad & \;  4 \nonumber \\
\!\! 10 \qquad \qquad \qquad \qquad \qquad&   2
\qquad \qquad \qquad \qquad \qquad& \;   4 \nonumber \\
\!\! 10 \qquad \qquad \qquad \qquad \qquad &   2+4s
\qquad \qquad \qquad \qquad \qquad & \;   4   \nonumber
\end{eqnarray}



\centerline{\boldmath${\beta=1}\;\;\;$  \qquad
$\bf (Scalar  \; \; case)$}
\vspace{-1cm}
\begin{eqnarray}
\mathbf{ e_{0}} \qquad \qquad \qquad \qquad \qquad
& \quad \mathbf{ e_{2}} \qquad \qquad \qquad \qquad \qquad &
\mbox{\boldmath${{\rm e}_{\infty}}$} \nonumber \\
 2 \qquad \qquad \qquad \qquad  \qquad &   2-4s  \qquad \qquad
\qquad \qquad \qquad & \;  4 \nonumber \\
 2 \qquad \qquad \qquad \qquad \qquad  &   2 \qquad \qquad \qquad \qquad
\qquad  & \;  4\nonumber \\
 2 \qquad \qquad \qquad \qquad \qquad &   2+4s \qquad \qquad \qquad \qquad
\qquad & \;   4 \nonumber
\end{eqnarray}


\centerline{\boldmath${\beta=0}\;\;\;$  \qquad $\bf
(Electromagnetic \; \; case)$}
\vspace{-1cm}
\begin{eqnarray}
\nopagebreak
\mathbf{ e_{0}} \qquad \qquad \qquad \qquad \qquad
&\quad \mathbf{ e_{2}} \qquad \qquad \qquad \qquad \qquad &
\mbox{\boldmath${{\rm e}_{\infty}}$} \nonumber \\
\!\!\!\! -2 \qquad \qquad \qquad \qquad  \qquad &   2-4s  \qquad \qquad
\qquad \qquad \qquad & \;  4 \nonumber \\
\!\!\!\! -2 \qquad \qquad \qquad \qquad \qquad  &   2 \qquad \qquad \qquad \qquad
\qquad  & \;  4\nonumber \\
\!\!\!\! -2 \qquad \qquad \qquad \qquad \qquad &   2+4s \qquad \qquad \qquad \qquad
\qquad & \;   4 \nonumber \\
 2 \qquad \qquad \qquad \qquad \qquad &   2-4s
\qquad \qquad \qquad \qquad \qquad & \;  4 \nonumber \\
2 \qquad \qquad \qquad \qquad \qquad &  2
\qquad \qquad \qquad \qquad \qquad & \;  4 \nonumber \\
 2 \qquad \qquad \qquad \qquad \qquad &   2+4s
\qquad \qquad \qquad \qquad \qquad& \; 4 \nonumber \\
 6 \qquad \qquad \qquad \qquad \qquad &2-4s
\qquad \qquad \qquad \qquad \qquad & \;  4 \nonumber \\
 6 \qquad \qquad \qquad \qquad \qquad&   2
\qquad \qquad \qquad \qquad \qquad& \;   4 \nonumber \\
\nopagebreak
6 \qquad \qquad \qquad \qquad \qquad &   2+4s \qquad \qquad \qquad
\qquad \qquad & \;   4   \nonumber
\end{eqnarray}
\noindent
In none of these families one is to find two or more ${\rm e}_{c}$ which are
odd numbers. Hence all of them are discarded.

\noindent \bf{Output:} \normalfont

\noindent \underline{CONTINUATION}: It is not applicable since
${\rm n}=2$ is the largest element of L.

\noindent
OUTPUT2: The equation (\ref{kooala}) does not have any Liouvillian
solutions.


\clearpage



\end{document}